\documentclass[pdflatex,epsf,useAMS,usenatbib,usegraphicx]{mn2e}
\usepackage{pdfpages}
\usepackage{fixltx2e}
\usepackage{epstopdf}
\usepackage{epsfig}
\usepackage{amssymb, amsmath, graphics}
\usepackage{ulem}
\usepackage{longtable, lscape}
\usepackage{subfloat}
\usepackage{graphicx}
\usepackage{booktabs,caption}
\usepackage{mathrsfs}
\usepackage{verbatim}
\usepackage{aas_macros}  % required to get journal abbreviations
\usepackage{longtable}
\usepackage{ltablex}
\usepackage{pdflscape}
\usepackage{booktabs}
\usepackage[flushleft]{threeparttable}
\usepackage[percent]{overpic}
\title[`Peculiar' RRd Stars Observed by {\it K2}]{Four `Peculiar' RRd Stars Observed by {\it K2}}

\author[James M.\,Nemec and Pawe{\l} Moskalik]{James M. Nemec$^{1,2}$\thanks{E-mail: nemec@camosun.ca (JMN)} and Pawe{\l} Moskalik$^{3}$
\\
\\
$^{1}$Department of Physics \& Astronomy, Camosun College, Victoria, British Columbia, Canada\\
$^{2}$International Statistics \& Research Corporation, Brentwood Bay, British Columbia, Canada \\
$^{3}$Nicolaus Copernicus Astronomical Center, Polish Academy of Sciences, ul. Bartycka 18, 00-716
Warszawa, Poland  \\
}
\date{Accepted 2021 July 2.  Received 2021 July 2;  in original form 2021 February 25}

\pubyear{2021}

\begin{document}
\label{firstpage}
\pagerange{\pageref{firstpage}--\pageref{lastpage}}
\maketitle

\begin{abstract}

% see C7p469 for Pawel's summary page

\noindent Four stars pulsating simultaneously with a {\it dominant} period
$P_D$$\in$(0.28,\,0.39)\,d and an {\it additional} period
$P_A$$\in$(0.20,\,0.27)\,d have been identified from among the more than 3000
RR~Lyrae stars observed by the {\it Kepler} space telescope during NASA's {\it
K2} Mission. All four stars are located in the direction of the Galactic Bulge
and have period ratios, $P_A$/$P_D$, significantly smaller than those of most
double-mode RR~Lyrae (RRd) stars: $P_A$/$P_D$$\in$(0.694,\,0.710) vs.
$P_1$/$P_0$$\in$(0.726,\,0.748). Three of the stars are faint
($<$$V$$>$=18--20\,mag) and distant and are among the `peculiar' RRd (pRRd)
stars discovered by Prudil et al. (2017); the fourth star, EPIC\,216764000
(=\,V1125\,Sgr), is a newly discovered pRRd star several magnitudes brighter
than the other three stars. In this paper the high-precision long-cadence {\it
K2} photometry is analyzed in detail and used to study the cycle-to-cycle light
variations.  The pulsational characteristics of  pRRd stars are compared with
those of `classical' and `anomalous' RRd (cRRd, aRRd) stars.   The conclusion
by Prudil et al. that pRRd stars form a separate group of double-mode pulsators
and are not simply very-short-period cRRd stars is confirmed.  V1127\,Aql and
AH\,Cam are identified as other probable members of the class of pRRd stars.

\end{abstract}


%&&&&&&&&&&&&&&&&&&&&&&&&&&&&&&&&&&&&&&&&&&&&&&&&&&&&&&&&&&&&&&&&&&&&&&&&&&&&&&&
% Section 1
\section{INTRODUCTION}

% see C17V17p494-502 for Alcock et al. 1997 paper
% See Soszy\'nski et al. 2016b for anomd RRd in  MCs

For most of the 20th century RR~Lyrae stars were thought to be
pulsating either in the radial fundamental mode (RRab stars) or
radial first-overtone mode (RRc stars), often with slowly increasing
or decreasing pulsation periods, and sometimes with amplitude and/or
frequency modulations (Blazhko effect). Since the discovery that
AQ~Leo pulsates simultaneously in the fundamental {\it and}
first-overtone modes (Jerzykiewicz \& Wenzel 1977, Cox, King \&
Hodson 1980) more than 3000 double-mode RR~Lyrae (RRd) stars have
been found in various Galactic and extragalactic environments: in
globular clusters (Cox et al. 1980; Cox, Hodson \& Clancy
1983; Sandage, Katem \& Sandage 1981; Nemec 1985b;  Clement et al.
1986; Nemec \& Clement 1989), in nearby dwarf galaxies (Nemec 1985a;
Kov\'acs 2001a; Cseresnjes 2001; Bernard et al. 2009), in the
Magellanic Clouds (Alcock et al. 1997; Nemec, Walker \& Jeon
2009; Soszy\'nski et al. 2009, 2010, 2016a,b, 2017a), and in the
Galactic Bulge and Disk (Soszy\'nski et al. 2011, 2014b, 2017b,
2019).

% FIGURE 1   Petersen Diagram for the {\it K2} and Prudil RRd stars
% See  C17V6p1030 (replaces  C17V3p567,580,619;   C17p535 Sept 21, 2018)
% Removed:  A dotted vertical line joins the two observed pulsation modes for \texttt{OGLE-BLG-RRLYR-14135}.

\begin{figure}
\begin{center}
\vskip0.1truecm
\includegraphics[width=8.6cm]{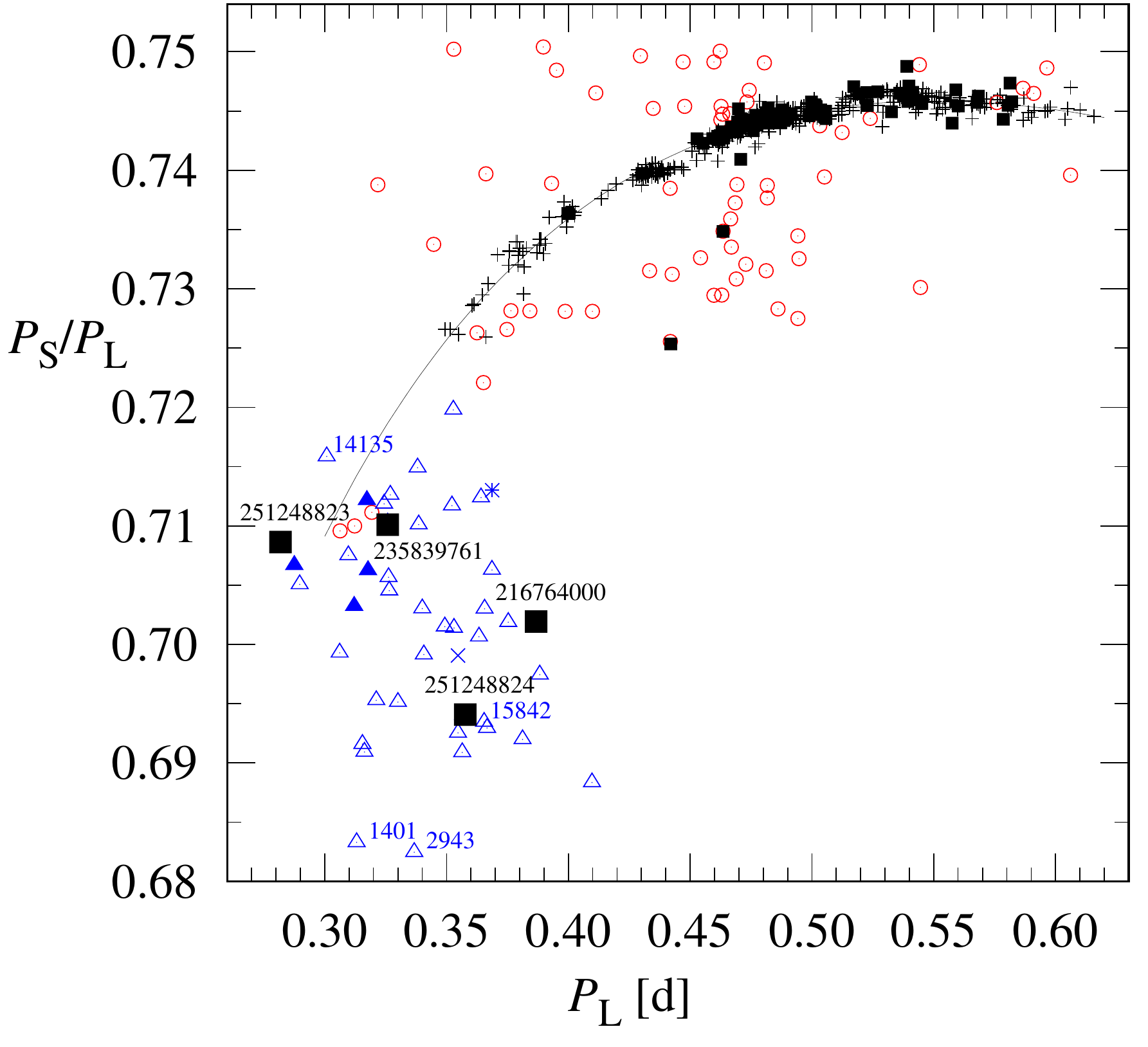}               % renamed Prudil_P0_2020Oct11.pdf --> Prudil_P0_2021Apr28.pdf (removed pRRd curve)
\end{center}

\caption{Petersen diagram identifying the four `peculiar' RRd (pRRd) stars
observed by {\it K2} (large black squares labelled with \texttt{EPIC} numbers).
Also plotted are the 42 Galactic Bulge pRRd stars studied by Prudil et al.
(2017; blue triangles),  the Galactic Bulge and Disk RRd stars from Soszy\'nski
et al. (2019; black plus signs for 458 classical RRd stars, and red open
circles for 63 anomalous RRd stars), and the 76 Ecliptic Plane RRd stars
observed by {\it K2} (Nemec, Moskalik et al., in preparation; small black
squares).  The four pRRd stars with the smallest $A_{\rm S}$/$A_{\rm L}$
amplitude ratios are labelled (in blue) with their \texttt{OGLE-BLG-RRLYR} star
numbers (see Fig.\,2). The blue filled triangles are the four low-amplitude
pRRd stars classified `RRc' by the \texttt{OGLE} survey. V1127\,Aql is
represented by the blue cross and AH\,Cam by a blue asterisk. The mean curve
for the `classical' RRd stars is derived in $\S$4.1. }

\label{fig:PetersenDiagram}
\end{figure}

% \vfill \eject

Most RRd stars are of the `classical' variety, pulsating
simultaneously with a dominant shorter period ($P_{\rm S}$) and a
secondary longer period ($P_{\rm L}$). The overall light curves
phased with $P_{\rm S}$ are of relatively low amplitude and roughly
sinusoidal with significant scatter about the mean curve; in fact,
observed excess scatter is the usual way in which RRd stars are
discovered.  In period histograms (see, for example, fig.\,32 of
Nemec 1985a and fig.\,2 of Soszy\'nski et al. 2016b) the two periods
straddle the long-period side of the period distribution for RRc
(first-overtone) stars and the short-period side of the period
distribution for RRab (fundamental-mode) stars.  Perhaps the
strongest argument that  $P_{\rm L}$ is the fundamental-mode period
and $P_{\rm S}$ is the first-overtone period comes from the observed
very-narrow range of period ratios, 0.726 $<$ $P_{\rm S}$/$P_{\rm
L}$ $<$ 0.750, the range amounting to only 3.3 per cent of the
mean period ratio. For a given pair of radial modes the period ratio
is always strongly constrained by pulsation models.

\vfill \eject

In a Petersen (1973) diagram (see Figure~1) the classical RRd
(hereinafter `cRRd') stars trace out a smooth curve, the period
ratios decreasing as the periods decrease.  The earliest such
diagrams revealed that in low-metallicity Oosterhoff~II (OoII)
globular clusters, such as M15 and M68 with [Fe/H]=--2.1\,dex (i.e.,
$Z$$\sim$0.0001, assuming [Fe/H]=$\log\,Z$+1.9), the RRd stars have
fundamental-mode periods, $P_0$\,(=$P_L$), greater than 0.51\,d, and
first-overtone periods, $P_1$\,(=$P_S$) $\in$(0.37\,d,0.45\,d). And,
in less metal poor (i.e., intermediate-metallicity)
Oosterhoff~I (OoI) globular clusters, such as IC\,4499 and M3 with
[Fe/H]$\sim$--1.5\,dex (i.e., $Z$$\sim$0.0004), the RRd stars have
shorter fundamental-mode periods, $P_0$$\in$(0.45\,d,\,0.51\,d) and
minimum period ratios ($P_1$/$P_0$)$_{\rm min}$=0.742 (see, for
example, fig.\,10 of Clement et al. 1986).  These observations
clearly established that the location of a cRRd star on the
`classical' curve mainly depends on [Fe/H]: the shorter the periods
the smaller the period ratio and the higher the metallicity.

% \footnote{The old stellar population of the SMC consists mainly of stars with
% [Fe/H] values between --1.6 and --0.4\,dex (see fig.\,1 of Mucciarelli 2014).}
% For example, the Small Magellanic Cloud (SMC) has a longer
% minimum fundamental-mode period, $P_{\rm 0,min}$=0.478\,d, and a larger minimum
% period ratio, ($P_1$/$P_0$)$_{\rm min}$=0.744 than the Large Magellanic Cloud
% (LMC), where $P_{\rm 0,min}$=0.457\,d and ($P_1$/$P_0$)$_{\rm min}$=0.742

In environments consisting of mixtures of stars exhibiting a range
of metallicities, such as dwarf galaxies and the Magellanic Clouds
(Nemec 1985a; Alcock et al. 1997, 2000; Cseresnjes 2001; Kov\'acs
2001a; Soszy\'nski et al. 2016b), the RRd stars exhibit a wide range
of periods and period ratios. In such systems significant
differences in the periods of the shortest-period RRd stars are
observed. For example, the LMC has a shorter minimum
fundamental-mode period, $P_{\rm 0,min}$, and a smaller minimum
period ratio, ($P_1$/$P_0$)$_{\rm min}$, than the SMC (Soszy\'nski
et al. 2010; see fig.\,1 of Soszy\'nski et al. 2016b). If
metallicity is the main cause of these differences then it follows
that the most metal-rich RRd stars in the LMC, those with the
smallest period ratios, are more metal-rich than those in the SMC,
a result consistent with the LMC having a higher mean metallicity
than the SMC (see fig.\,2 of Kov\'acs 2009, and fig.\,1 of
Mucciarelli 2014). Since the shortest-period RRd stars in the LMC
also tend to be more centrally concentrated than the less metal-rich
RRd stars in the LMC (see figs.\,3-4 of Soszy\'nski et al. 2016a)
this also argues for a metallicity gradient. In our own Galaxy the
most metal-rich RR~Lyrae stars (see Soszy\'nski et al. 2019) are
more metal-rich than the most metal-rich RR~Lyrae stars in the
Magellanic Clouds, and have kinematics consistent with membership in
the thick- and thin-disk and bulge stellar populations, possibly
with quite young ages (Preston 1959, Layden 1995, Kinemuchi et al.
2006, Chadid, Sneden \& Preston 2017, Sneden et al. 2018).

% see C19p327 for Popielski+ Fig.3
Clear illustrations of how the periods and period ratios of RRd stars depend on
mass ($M$), luminosity ($L$), effective temperature ($T_{\rm eff}$) and
chemical composition are given by Popielski, Dziembowski \& Cassisi (2000,
fig.\,3) and by Soszy\'nski et al. (2011, fig.\,9; 2014a, fig.\,2).  Several
thousand RRd stars are now known, over 2600 of them in the Magellanic Clouds.
More than 500 have been identified in our Galaxy, many of which reside in the
Galactic Bulge and tend to have shorter periods and smaller period ratios than
do the RRd stars located in  globular clusters or in the Magellanic Clouds.
Varying one or more of the above physical characteristics introduces spread
into the distribution of stars about the classical RRd curve.  Such variations
might help to explain some of the unusual period ratios of the `anomalous' RRd
stars (see below), however, at present they have little bearing on the question
of amplitude ratios or the Blazhko effect.  Still not well understood is why a
particular pulsation mode is dominant (i.e., mode selection), why some RRd
stars switch modes on short time-scales (Clement et al. 1997; Soszy\'nski et
al.  2014a, 2019), the cause(s) of Blazhko modulations, and the non-linear
properties of multi-mode pulsators (e.g., amplitude ratios).

% Anomalous RRd stars
Large-scale surveys and high-precision time-series photometry from
the ground and with space telescopes have led to the discovery of
several new types of multi-mode RR~Lyrae stars (see Moskalik 2014,
and Smolec et al. 2017b for recent reviews). The so-called
`anomalous' RRd stars (hereinafter `aRRd' stars, the notation
adopted in the on-line `\texttt{OGLE} Collection of Variable Stars')
differ from cRRd stars in one or more of the following ways (see
Jurcsik et al. 2015; Soszy\'nski et al. 2014b, 2016b, 2017a,b, 2019;
Smolec et al. 2015, 2017a): (1) in the Petersen diagram they lie
below or above the curve defined by the cRRd stars; (2) in the
majority of cases the dominant pulsation mode is the (longer-period)
fundamental mode rather than (shorter-period) first-overtone mode,
i.e., $A_1$$<$$A_0$; and  (3) the Blazhko effect (i.e. long-term
modulation of the amplitudes, frequencies and period ratios) is
often present, with several Blazhko stars having been observed to
switch between double-mode and single-mode RRab pulsation and to
have undergone an abrupt change of the fundamental-mode period
(e.g., Jurcsik et al. 2015). Sixty-three aRRd stars (from
Soszy\'nski et al. 2019) are present in Fig.\,1 (red open circles);
five are triple-mode pulsators; at least one has switched pulsation
mode from double-mode (Soszy\'nski et al. 2017b, 2019) to RRab (with
Blazhko modulations); and at least three were observed by {\it K2}
(Nemec, Moskalik et al., in preparation). To explain the anomalous
locations in the Petersen diagram of several RRd stars in the
globular cluster M3, the possibilities of mass transfer in a binary
system and unusual helium compositions have been considered by
Clementini et al. (2004).  More recently, Soszy\'nski et al. (2016b)
attribute the behaviour to a resonance instability involving the
fundamental, first-overtone and second-overtone frequencies
(specifically, $2f_1$$\sim$$f_0$+$f_2$).

% amplitude ratios
In Figure~2 amplitude ratios for the stars included in Fig.\,1
are plotted against the longer periods.  A similar graph, apart from
the logarithmic scaling and the labelled groupings, was made by
Soszy\'nski et al. (2009) for the RRd stars in the LMC.  The longer
period is seen to be dominant (i.e., $A_{\rm L}$$>$$A_{\rm S}$) for
all of the pRRd stars, most of the aRRd stars, and for 14 of the 39
short-period cRRd stars, whilst the first-overtone dominates for
almost all of the intermediate- and long-period cRRd stars (OoII-
and OoI-like) and for 25 of the short-period cRRd stars. The
amplitude ratios appear to trend upward as the periods increase,
with an apparent discontinuity occuring around $P_L$$\sim$0.45\,d,
resembling somewhat the jump of $P_S$/$P_L$ occuring for the cRRd
stars at the same period (see Fig.\,1).   Figs.\,1-2, which are
discussed further in $\S$4, provide valuable observational
constraints for models that aim to understand the pulsational and
evolutionary states of RRd stars (Kov\'acs 2001b; Koll\'ath et al.
2002; Szab\'o, Koll\'ath \& Buchler 2004; Smolec et al. 2016).

% FIGURE 2 - Amplitude ratios for the {\it K2} and Prudil RRd stars
% originally C17p535 (Sept 21, 2018) but changed C17V3p564-565 (May 30, 2019)
% Improved even further on June 1,2019 - see C17V3p568-571;  and again on Aug.11,2019 -
% see C17V3p577; and again on Nov29,2019; see p.581-582;  C17p621; C17p934
\begin{figure}
\begin{center}
\includegraphics[width=8.1cm]{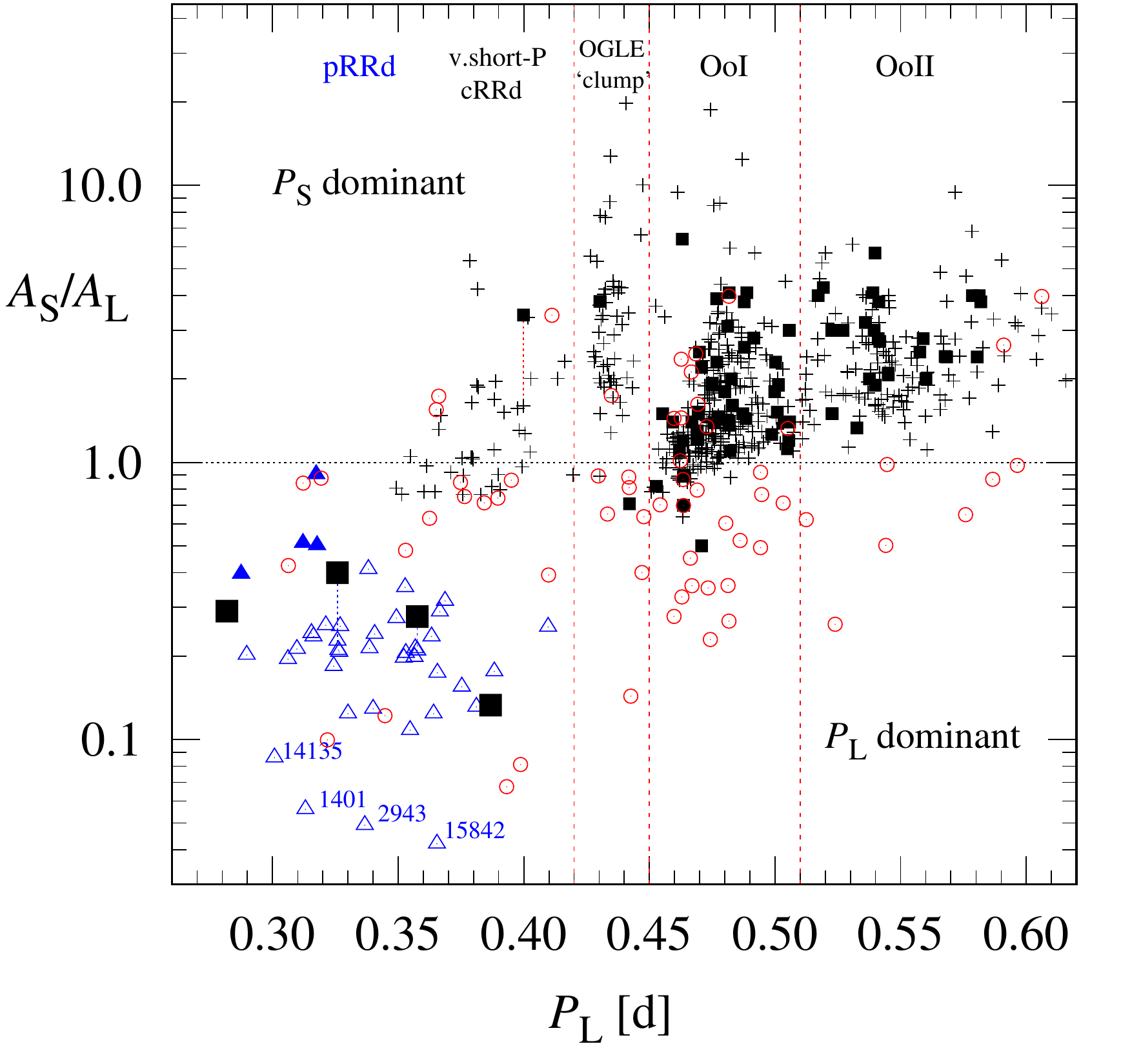}                 % renamed Prudil_ampl_2020Sept4.pdf
\end{center}

\caption{Amplitude ratio $A_{\rm S}$/$A_{\rm L}$ vs. longer period $P_{\rm L}$
for the RRd stars in Fig.\,1 (same symbols). Below $A_{\rm S}$/$A_{\rm L}$=1.0
the longer period is dominant, which is the case for all the pRRd stars and
most of the aRRd stars. Except for EPIC\,216764000, whose amplitude ratio is
based on {\it K2} photometry, the amplitude ratios were derived from Fourier
first-term $I$-passband photometry.  The short red and blue vertical lines join
the \texttt{OGLE-IV}-$I$ and {\it K2}-{\it Kp} amplitude ratios.  For
EPIC\,251248823 (=\,\texttt{OGLE-BLG-RRLYR-00595}) the $I$ and {\it Kp}
amplitude ratios are practically identical. The red vertical dashed lines at
$P_L$=0.42\,d, 0.45\,d and 0.51\,d identify the \texttt{OGLE} `clump', OoI and
OoII stars.}

\label{fig:AmplRatios}
\end{figure}

Almost all cRRd stars that have been observed from space have been
found to exhibit, in addition to the two main pulsation modes,
low-amplitude oscillations with periods $P_x$$\sim$0.61$P_1$
(Moskalik et al. 2015; Nemec, Moskalik et al., in preparation). Such
period ratios, which are also observed in most RRc stars, can be
interpreted as ``a manifestation of nonradial mode excitation''
(Dziembowski 2012, 2016).  Some examples include AQ~Leo (Gruberbauer
et al. 2007), \texttt{CoRoT}\,101368812 (Chadid 2012, Szab\'o et al.
2014), EPIC\,60018653 and 60018662 (Moln\'ar et al. 2015), and
EPIC\,201585823 (Kurtz et al. 2016).

Other kinds of multi-mode RR~Lyrae stars are summarized by Netzel \&
Smolec (2019). These include the so-called R$_{\rm 0.61}$ and
R$_{\rm 0.68}$ RRc stars (see Netzel, Smolec \& Moskalik
2015a,b), long-period RRab stars ($P_0$$>$0.6\,d) exhibiting
additional shorter-period low-amplitude variability where the period
ratios are not unlike those of cRRd and aRRd stars (Smolec et al.
2016), and RRd stars pulsating in the fundamental and
second-overtone modes with period ratios $\sim$0.58 (Benk\H o et al.
2010, 2014). Amplitude and phase modulations (Blazhko effect) are
also present in many RR~Lyrae stars (Jurcsik et al. 2009, 2017;
Benk\H o et al. 2014). Recent developments concerning such
modulations are discussed by Jurcsik et al. (2018), Kov\'acs
(2018a), Koll\'ath (2018) and Netzel et al. (2018).

Of particular interest for the current paper, Prudil et al. (2017; hereinafter
P17) identified 42 doubly periodic RR~Lyrae stars that they described as
`peculiar'. The stars were found as a result of reanalyzing 2010-2013
\texttt{OGLE-IV} photometry for over 38000 Galactic Bulge RR~Lyrae stars
(Soszy\'nski et al. 2011, 2014b; Udalski, Szyma\'nski \& Szyma\'nski 2015).
Fourteen of the stars had been seen to have secondary pulsations by the
\texttt{OGLE} survey but had not been discussed.  The 42 `peculiar' RRd (or
`pRRd') stars were found to have {\it dominant} pulsation periods $P_{\rm
D}$$\sim$0.34\,d, and shorter {\it additional} periods $P_{\rm
A}$$\sim$0.24\,d, with period ratios $P_{\rm A}$/$P_{\rm D}$ between 0.68 and
0.72. Such period ratios are significantly smaller than the observed period
ratios of `classical' and `anomalous' RRd stars (see Fig.\,1). Since the light
curves of the dominant mode for all 42 stars are asymmetric when phased with
$P_D$ (see fig.\,4 of P17), and resemble the light curves of low-amplitude RRab
stars, P17 concluded that the {\it dominant} (longer-period) pulsation mode
most likely is the radial fundamental mode for all 42 stars. The nature of the
{\it additional} shorter-period pulsation was considered uncertain and of
unknown origin, leading P17 even to question whether or not the stars are
RR~Lyrae stars, that is to say, old low-mass (0.5--0.9\,M$_{\odot}$),
high-luminosity (30$<$L/L$_{\odot}$$<$60), horizontal branch stars evolving
through the Instability Strip with mean surface temperatures 6200$<$T$_{\rm
eff}$$<$7500\,K.

% See C17V3p570 for table of K2 RRd stars
The present paper concerns four such `peculiar' RRd stars, all located in the
direction of the Galactic Bulge and observed during the {\it K2} Mission
(Howell et al. 2014). The stars are among $\sim$80 RRd stars that were found
when more than 3000 RR~Lyrae stars observed by {\it K2} were screened in order
to identify and study the RRd stars (Nemec, Moskalik et al., in preparation;
see Moskalik et al. 2018a,b for a preliminary report). The brightest of the
four stars, EPIC\,216764000 (=\,V1125\,Sgr), was observed during {\it K2}
Campaign\,7 and is located in the foreground of the Sagittarius dwarf galaxy.
It is a newly discovered pRRd star, and although previously observed, the
additional pulsation detected in the {\it K2} photometry is a new discovery.
The three fainter stars were observed during Campaign\,11 and are considerably
more distant.  When the four stars were first plotted in a Petersen diagram
they stood out from the other classical and anomalous RRd stars (see Fig.\,1).
An overlay of RRd stars in the Galactic Bulge (Soszynski et al.  2011, 2014b,
2019) and the 42 pRRd stars discovered by P17 showed that the three Campaign 11
variables are among the stars studied by P17. The fourth star,
EPIC\,216764000, almost certainly belongs to his group as well.

% Section 2
\section{PHOTOMETRY}

The main data analyzed were the brightness measurements made with
the {\it Kepler} telescope (Borucki et al. 2010) during the {\it K2}
Mission (Howell et al. 2014). The high precision and relatively-long
duration of the {\it Kepler} photometry ({\it Kp} filter) offers the
opportunity for improving our understanding of pRRd stars. The
integration times were 29.4\,min (i.e., long cadence), corresponding
to a sampling rate of 48.9 measurements per day. Thus the
highest discernable frequencies are near the Nyquist frequency,
$f_N$=24.47\,d$^{-1}$, corresponding to minimum periods of
$\sim$1\,h. The {\it K2} data were subject to continuous
drifting of the telescope, with thrusters fired every six hours to
maintain pointing. Owing to this instrumental effect the photometry
often has an extraneous trend due to the drift and imposed
thruster-firing frequencies near 4.08\,d$^{-1}$ and its multiples
(seen in Fig.\,6g). Once detrended the photometry offers almost
continuous brightness measurements with milli-magnitude precision.
Since all four stars were observed by {\it K2} for $\sim$11 weeks
the high-precision photometry also offers the opportunity to observe
directly cycle-to-cycle light variations. The source files are
available at the \texttt{MAST} website (https://archive.stsci.edu).

Data from other surveys undertaken prior to {\it K2} were available
for all four stars.  These earlier observations provide valuable
auxiliary information on such quantities as mean magnitudes and
colours, distances, pulsation frequencies at different epochs,
high-resolution finding charts, etc. (see $\S$2.3).

% Subsection 2.1
\subsection{{\it K2} Campaign\,7 photometry}

The {\it K2} observations of EPIC\,216764000, the brightest of the
four stars, were made in 2015 during Campaign\,7, which began on
October\;4 and ended 81.37 days later on December 26. A total of
more than 3700 long-cadence (i.e., 30-min) sequential brightness
measurements were made and are shown in Figure~3. The mean
magnitude, $<${\it Kp}$>$=13.741\,mag (Huber et al. 2016), makes
EPIC\,216764000 the brightest pRRd star yet discovered\footnote{The
brightest star in the P17 sample is \texttt{OGLE-BLG-RRLYR-15842}
with $I$=15.00\,mag, which is more than two magnitudes fainter than
the $I$=12.63\,mag for EPIC\,216764000.} and an ideal candidate for
spectroscopy. On the sky it lies in a direction of the Sagittarius
constellation that is relatively free of bright stars, with only
minor contamination from a faint neighbour 3.8\,arcsec to the
south-west\footnote{The \texttt{Gaia} (DR2) equatorial coordinates
for EPIC\,216764000 and its close neighbour are RA,\,DEC
\,= 19$^{\rm h}\,\!$17$^{\rm m}$33${^{\rm s}\;\!\!\!\!.\;\!}$344,
--21${^\circ\;\!\!}$46${\;\!'}$ 02$\,\!'\!'\,\!\!\!\!.\;\!$79 and
19$^{\rm h}\,\!$17$^{\rm m}$33${^{\rm s}\;\!\!\!\!.\;\!}$141,
--21${^\circ\;\!\!}$46${\;\!'}$05$\,\!'\!'\,\!\!\!\!.\;\!$11,
respectively.
%(19:17:33.331,\;--21:46:2.78) and (19:17:33.141,\;--21:46:5.11), respectively.
The neighbour, fainter than EPIC\,216764000 by
$\Delta$m$_g$=3.6\,mag, contributes negligibly to the {\it K2} flux.
\texttt{Gaia} (DR2) gives for the parallax
$\pi$=0.235$\pm$0.031\,mas (see Table~1), corresponding to
the distance 4.26$\pm$0.57\,kpc, which is considerably nearer
than the 20$\pm$2\,kpc distance to the background Sagittarius
dwarf galaxy, and roughly half-way to the Galactic Centre.
\texttt{Gaia} (DR2) also finds for the effective temperature $T_{\rm
eff}$\,=\,6135\,K, which is $\sim$900\,K smaller than the 7046\,K value
given in the \texttt{EPIC} catalog;  both values are within the
observed range for RR~Lyrae stars.}.

% FIGURE 3
% Prudil star See C7p247 for the creation of this figure;  see also C7p414-426; C7p572
% See C7p242     for the creation of this figure;   See C7p398 for EAP most recent analysis;
%  3728 comes from K2C7p626
\begin{figure}
% \begin{center}
% \includegraphics[width=8.5cm]{RRd_C7_P04fits_odd_2017July25.pdf}
% \end{center}
%\begin{overpic}[width=8.5cm]{RRd_C7_P04fits_odd_2017July25.pdf} \put(23,55){V1125 Sgr  = }  \put(63,55){= ASAS\,191734-2146.2} \end{overpic}
%\begin{overpic}[width=8.5cm]{RRd_C7_P04fits_odd_2017July25.pdf}  \end{overpic}
% \begin{overpic}[width=8.3cm]{RRd_C7_P04fits_odd_2019Feb28.pdf}  \end{overpic}    % C7p572
% \begin{overpic}[width=8.3cm]{RRd_C7_P04fits_odd_2019Feb28.pdf}  \end{overpic}    % C7p572
\begin{overpic}[width=8.3cm]{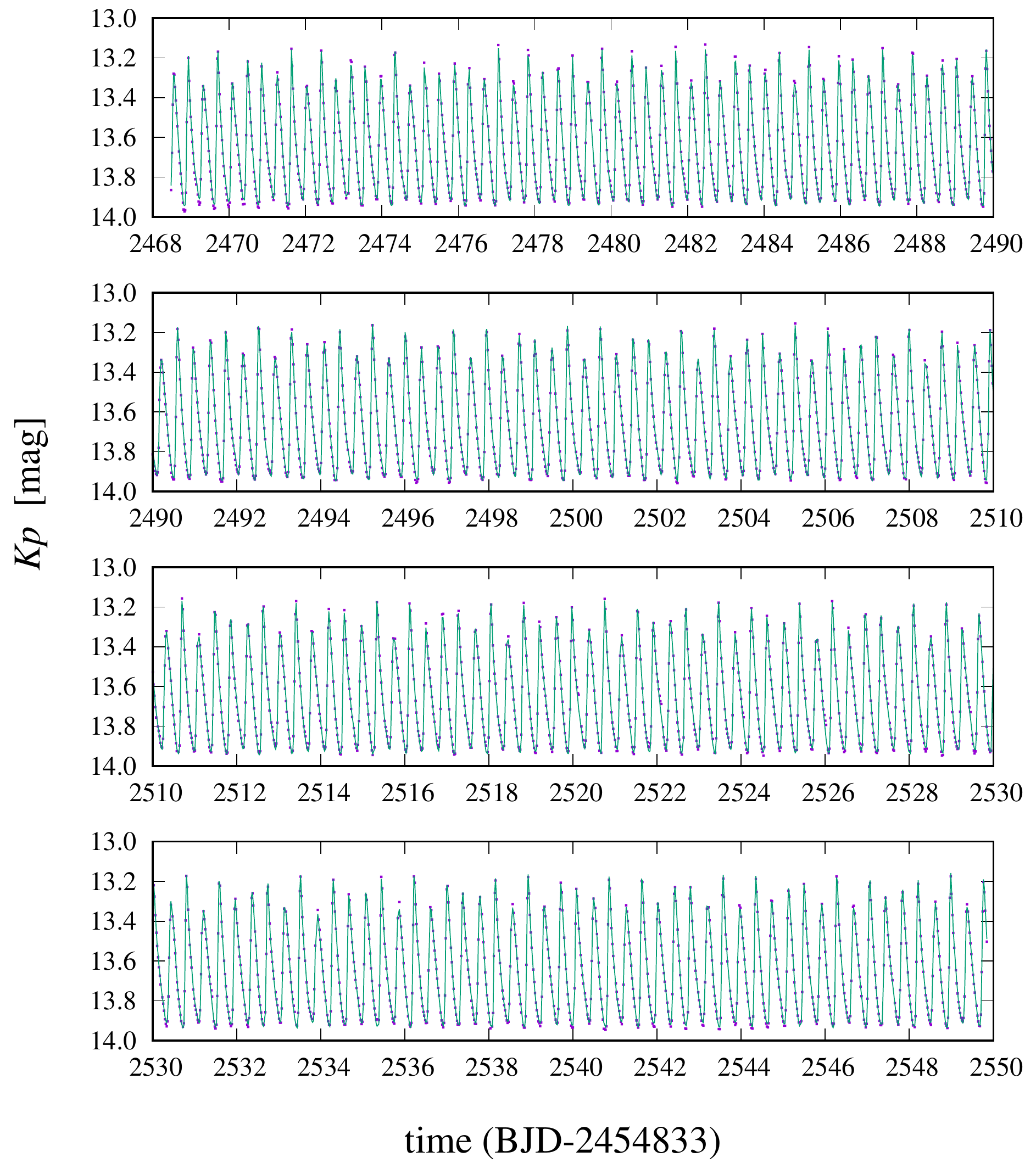}   \end{overpic}

\caption{Observed (and fitted) light curve for the `peculiar' RRd
star EPIC\,216764000 (=\,V1125\,Sgr = ASAS\,191734-2146.2) observed
during Campaign\,7.  The four panels show all 81.37 days of {\it K2}
photometry, a total of 3728 long-cadence measurements.  The {\it
dominant} variability is asymmetric,  periodic with period
0.38675\,d, and clearly exhibits amplitude variability with most of
the variation near maximum light.  }

\label{fig:216764000-c7}
\end{figure}

% \subsubsection{Raw Fluxes to Detrended Light Curves}
Various approaches were tried to obtain optimally-detrended,
flat-field-corrected, outlier-free photometric fluxes.  Kurtz et al.
(2016) in their analysis of the RRd star EPIC\,201585823,
recommended that {\it K2}-Pixel Photometry (\texttt{K2P$^2$}, see
Lund et al. 2015; Handberg \& Lund 2017) be used for RR~Lyrae stars;
however, such detrendings were not made for the Campaign\,7 (and 11)
data sets.

Our analysis of EPIC\,216764000 started with inspection of the
`Pre-search Data Conditioning Simple Aperture Photometry'
\texttt{PDCsap} (see Smith et al. 2012) photometric measurements
provided by the NASA-Ames {\it Kepler}/{\it K2} Science Center. The
\texttt{PDCsap} data comprise 3728 flux measurements ranging from
33966\;electrons per second (minimum light) to
73552\;electrons per second (maximum light). Even better light
curves were obtained with the \texttt{Everest} (Luger et al.
2016) and \texttt{EAP} (Plachy et al. 2019) pipelines, and with our
own reductions using the \texttt{Pyke} software (Still \& Barclay
2012), all of which gave similar results.  Fluxes were transformed
to {\it Kepler} magnitudes with the Lund et al. (2015) relation,
i.e., {\it Kp} = 25.3 -- 2.5$\log$(flux), and all the photometric
results are based on this transformation.

% See C7V5p825 for PDCsap light curve showing constancy at minimum light
The {\it K2} photometry for EPIC\,216764000 shown in Fig.\,3
includes a fitted (\texttt{Period04}) light curve (see $\S$3.1
below). The dominant period 0.386748\,d is slightly longer than the
period 0.3867145($\pm$13)\,d first determined by Uitterdijk (1949),
with a trough-to-peak total amplitude $\sim$0.7\,mag and an
asymmetric light curve shape for each cycle. Cycle-to-cycle
amplitude variations are seen with most of the variation occuring at
the brightest phases, i.e., the magnitudes at minimum are
approximately constant (at 13.95\,mag) while those at maximum light
vary from 13.10 to 13.35\,mag. The mean magnitude derived using the
Lund et al. formula, {\it Kp}=13.680\,mag, agrees well with the
13.741\,mag given in the \texttt{EPIC} catalog.

% Subsection 2.2
\subsection{{\it K2} Campaign\,11 photometry}

% K2 data - see C11p669 for PM comments
The three pRRd stars observed during Campaign\,11, in order of
decreasing brightness, are EPIC\,235839761, 251248823, and
251248824.  All are relatively faint, the brightest being almost
four magnitudes fainter than EPIC\,216764000, and EPIC\,251248823
(=\,\texttt{OGLE-BLG-RRLYR-00595}) is the shortest-period pRRd star
presently known.  All were observed in 2016 over 74.2 days (from
September\;24 to December\;8). The Campaign\,11 field lies in
the direction of the Galactic Bulge and suffers from considerable
crowding.  Owing to ``an error in the initial roll-angle used to
minimize solar torque on the spacecraft", and corrective action, a
three-day break occurred in the observations (between BJD\,2457679
and 2457682), necessitating a division of the data into two
subcampaigns, C111 and C112, the former consisting of 1060
long-cadence measurements and the latter 2188 long-cadence
measurements.

After evaluating several different reduction pipelines the
\texttt{PDCsap} photometry from the NASA-Ames {\it K2} pipeline (see
https://keplerscience.arc.nasa.gov/pipeline.html) was adopted for
all three stars. In general the mean levels of the C111 fluxes were
found to be smaller by about 10 per cent than the fluxes for
the C112 fluxes, and for EPIC\,251248823 the \texttt{PDCsap} C111
data show an upward trend.  To correct for these effects it was
necessary to level the fluxes by fitting polynomials and then to
bring the C111 fluxes up to the C112 levels (after removing
5$\sigma$ outliers and other points flagged in the headers as
possibly in error) by multiplying by an appropriate constant. In
this way the mean flux levels of the optimum photometry for the
three stars were found to be 1212, 163 and 70\;electrons per
second. After transforming the observed fluxes to {\it Kp}
magnitudes using the Lund et al. (2015) formula, the derived mean
{\it Kp} magnitudes were found to be 17.59, 19.77 and 20.69\,mag,
respectively, all of which are significantly fainter than the
\texttt{EPIC} catalog (Huber et al. 2016) values: 16.90, 18.80 and
19.80\,mag.

To resolve these differences the mean magnitudes given in the
\texttt{OGLE} and \texttt{Gaia} catalogs were considered. The mean
$V$ magnitudes given in the latest \texttt{OGLE-IV} catalog are
17.51, 18.84 and 19.87 (Soszy\'nski et al. 2017b), respectively, and
the $m_{\rm bp}$ magnitudes given in the \texttt{Gaia} (DR2) catalog
are 17.64$\pm$0.06, 18.94$\pm$0.08, and 19.65$\pm$0.09. Note that
for each star the magnitudes are similar. Recalling also that the
mean $V$ magnitudes for the non-Blazhko RR~Lyrae stars observed
during the original {\it Kepler} Mission are usually within
0.2\,mag of the mean {\it Kp} magnitudes (see Nemec et al. 2011)
suggested that the measured {\it Kp} magnitudes should be similar to
these \texttt{OGLE-IV} and \texttt{Gaia} magnitudes.

For EPIC\,235839761 the measured mean {\it Kp}=17.59\,mag (\texttt{PDCsap}+Lund
transformation) is consistent with the \texttt{OGLE-IV} and \texttt{Gaia}
magnitudes but not with the brighter \texttt{EPIC} value ({\it Kp}=16.90\,mag).
For this reason the \texttt{EPIC} mean magnitude was rejected in favour of the
magnitudes derived using Equation 2.1 of Lund et al. (2015). For
EPIC\,251248823 the derived mean ({\it Kp}=19.77\,mag) was found to be a full
magnitude fainter than the \texttt{EPIC}, \texttt{OGLE} and \texttt{Gaia} means
({\it Kp}=18.80\,mag, $V$=18.84\,mag, $m_{\rm bp}$=18.94\,mag) which are all
consistent; so for it the \texttt{PDCsap} mean flux level was increased (by
multiplication) from 163 to 400\;electrons per second (corresponding to {\it
Kp}=18.8\,mag). For EPIC\,251248824, since the \texttt{EPIC}, \texttt{OGLE-IV}
and \texttt{Gaia} catalogues consistently give {\it Kp}$\sim$19.8\,mag, the
unrealistic  {\it Kp}=20.69\,mag was rejected in favour of normalizing the
fluxes so that {\it Kp}$\sim$19.8\,mag. The resulting time-series are shown in
Figure~4.  Apart from minor changes to Fourier amplitudes, these adjustments
had  little effect on the results of the frequency analyses discussed below
($\S$3).

% FIGURE 4
% See C11V5p698-702 for OGLE Web page information.
% See C11V5p819 for creation of file 2019Mar27; replaced p895, 959, 1049
\begin{figure}
\begin{center}
\begin{overpic}[width=7.8cm]{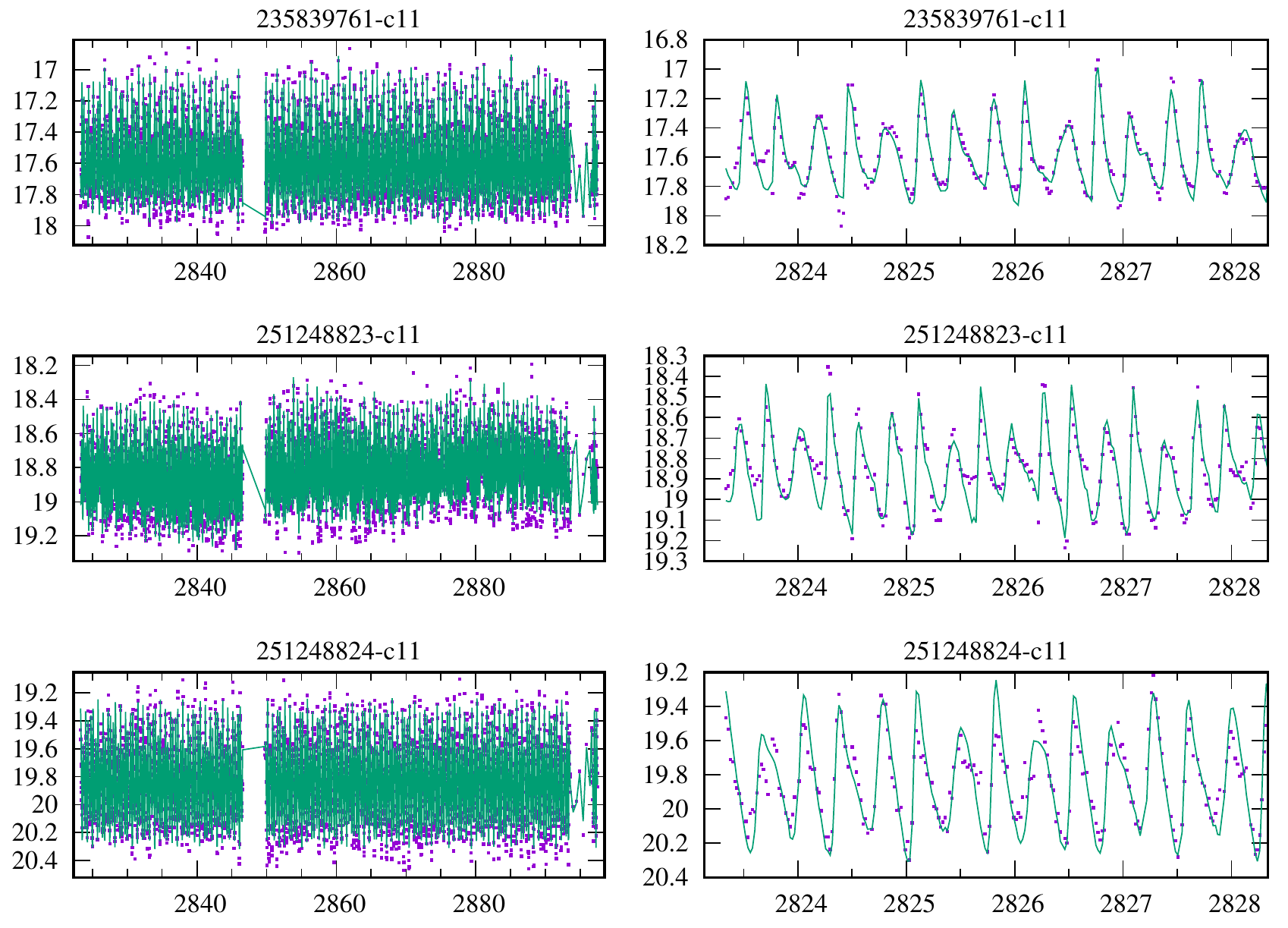} \put(30,-5){time\,(BJD-2454833)}    \put(-5,33){\rotatebox{90}{\it Kp {\rm [mag]}}}   \end{overpic}  % renamed Fig4_pecRRd_C11_2019July21.pdf  (see C11p1049)
% \begin{overpic}[width=7.9cm]{Fig4_pecRRd_C11_2019July21.pdf} \put(30,-5){Time (BJD-2454833)}    \put(-7,37){\it Kp}    \end{overpic}
% \begin{overpic}[width=7.9cm]{pecRRd_C11_p04fits_2019July15.pdf} \put(30,-5){Time (BJD-2454833)}    \put(-7,37){\it Kp}    \end{overpic}
\end{center}
\vskip0.4truecm

%  pRRd         Fig.4       MAST   see:
% 235839761-c11  17.591  vs  16.90  C11V5p743-770, see p.745 for MAST rtlc --> 17.59=25.3-2.5log(1212),   16.90=25.3-2.5log(2291)
% 251248823-c11  18.841      18.80  C11V5p781-802, see p.791 for MAST rtlc --> 18.80=25.3-2.5log(398),   19.53=25.3-2.5log(204)
% 251248824-c11  21.134      19.80  C11V5p803-815, see C11V6p910 for MAST rtlc (DR30);   20.7=25.3-2.5log(70);  for MAST: 19.80=25.3-2.5log(158); see p910,912

\caption{Observed (and fitted)  light curves for the three `peculiar'
RRd stars observed during Campaign\,11.   The left panels show all
the ${\it K2}$ photometry, and the right panels show the first
five days of data. The observations before time=2848 are from the
aborted data segment `C111' and the later observations are from the
corrected data segment `C112'.  The photometry (purple points) is
\texttt{PDCsap} data with additional detrending and the fitted light
curves (green) are from \texttt{Period04} analyses.}

\label{fig:FT}
\end{figure}

% TABLE 1 - Prior information
% see C7p691
\begin{table}
%\begin{small}   <-- does not change from normal size
{\fontsize{6}{7.2}\selectfont  %  <--- works and better; 2nd (skip) should be 1.2 times 1st (size)
% {\fontsize{7}{8.4}\selectfont  %  <--- works and better; 2nd (skip) should be 1.2 times 1st (size)
% {\fontsize{5}{6.5}\selectfont   % <-- works but too small
\centering

\caption{Previous observations of EPIC\,216764000 }

\label{tab:five}
\begin{tabular}{lc}
\toprule
\multicolumn{1}{c}{Observable} &  \multicolumn{1}{c}{Value}    \\
\multicolumn{1}{c}{(1)} & \multicolumn{1}{c}{(2)}    \\
\midrule
% \multicolumn{2}{l}{\underline{Coordinates:}} \\ [0.04cm]
% GCVS name                                  & V1125\,Sgr  \\

RA (J2000)                                   &    19$^{\rm h}\,\!\!$17$^{\rm m}\!$33$^{\rm s}\;\!\!\!\!.\;\!$34375  \\
DEC (J2000)                                  &  --21$\;\!\!^\circ\,\!\!$46$\,\!'\;\!\!$02$\,\!'\!'\,\!\!\!\!.\;\!$7859    \\
RA (J2000)                                   &   289$^\circ\,\!\!\!\!.\;\!$388932      \\
DEC (J2000)                                  &  --21$^\circ\,\!\!\!\!.\;\!$767441      \\
Galactic longitude (J2000), $l$              & ~\,15$^\circ\,\!\!\!\!.\;\!$9433        \\
Galactic latitude (J2000), $b$               &  --15$^\circ\,\!\!\!\!.\;\!$2531        \\ [0.1cm]

% See C7V6pxxx
\multicolumn{2}{l}{\underline{Johannesburg plates (1928-1937):} Uitterdijk (1949)} \\ [0.04cm]
Identification number (van Gent)             &    130 (38)   \\
No. photographic plates (epoch)              &     334 (1928-1937)  \\
Period, $P$ ($\pm$\,mean error) [d]          &   0.3867145\,($\pm$13)   \\
Time of max. light ($\pm$\,mean error), $t_0$ [BJD] &   2426568.308\,($\pm$3)   \\
$\Delta$phase (=\,risetime, RT)              &     0.20    \\  [0.1cm]

% See C7p851 for spreadsheet with info
\multicolumn{2}{l}{\underline{\texttt{EPIC} Catalog:} Huber et al. (2016)} \\ [0.04cm]
Identification number                        &     216764000    \\
    {\it Kepler} magnitude, {\it Kp} [mag]   &    13.741        \\
\texttt{UCAC4} proper motion (RA), $\mu_{\alpha}$\,[mas\,yr$^{-1}$]  & \;\,5.6 $\pm$ 2.2   \\
\texttt{UCAC4} proper motion (DEC), $\mu_{\delta}$\,[mas\,yr$^{-1}$] & \;\,1.0 $\pm$ 3.7   \\
$B$ [mag]                                    &  14.20 $\pm$ 0.34   \\
$V$ [mag]                                    &  13.79 $\pm$ 0.32   \\
$g$\; [mag]                                  &  13.91 $\pm$ 0.41   \\
$r$\; [mag]                                  &  13.69 $\pm$ 0.18   \\
$i$\;\, [mag]                                &  13.48 $\pm$ 0.30   \\
Effective temperature, $T_{\rm eff}$\,[K]    &     7046             \\
Surface gravity, $\log g$ [cgs]              &     4.11       \\
Metal abundance, $[{\rm Fe/H}]$ [dex]        &     +0.098    \\
Radius, R/R$_\odot$                          &    1.797     \\
Mass, M/M$_\odot$                            &    1.582   \\
Density, $\rho$ [g\,cm$^{-3}$]               &   0.257   \\
Distance [pc]                                &  1400   \\
Reddening, $E(B\!-\!V)$\,[mag]               &  0.227   \\   [0.1cm]

%  \multicolumn{2}{l}{\underline{\texttt{APASS} Survey}} \\ [0.04cm]
%  $V$ [mag]                                  &  13.825 $\pm$ 0.271  \\
%  $B$--$V$ [mag]                             &  0.472 $\pm$ 0.431  \\  [0.1cm]

\multicolumn{2}{l}{\underline{\texttt{ASAS-3} Survey (2001-2009):} Szczygie{\l} et al. (2009)} \\ [0.04cm]
Identification                               &   191734-2146.2  \\
Type of variable star                        & RRab   \\
Period, $P$ [d]                              &    0.3867075       \\
Time of maximum light, $t_0$ [BJD]           &   2452026.41   \\
% $V$(maximum light)                         &   13.19  \\
Total amplitude, $A_{\rm tot}$($V$), $A_{\rm tot}$($I$) [mag] &     0.439, 0.299    \\
$V$\,min,\,max,\,avg [mag]                   &    13.523, 13.104, 13.369      \\
% Total amplitude, $A_{\rm tot}$($I$) [mag]  &     0.299     \\
$I$\, min,\,max,\,avg [mag]                  &    12.712, 12.522, 12.626      \\
${\rm [Fe/H]}$\,JKZW [dex]                   &   0.14 $\pm$ 0.31  \\
${\rm [Fe/H]}$\,S04 [dex]                    &      0.50    \\  [0.1cm]

\multicolumn{2}{l}{\underline{\texttt{2MASS} Survey:} Skrutskie et al. (2006)} \\ [0.04cm]
Identification                               &   J19173334-2146027   \\
$J$\; [mag]                                  &  12.749 $\pm$ 0.023   \\
$H$ [mag]                                    &  12.601 $\pm$ 0.027   \\
$K$ [mag]                                    &  12.524 $\pm$ 0.029   \\ [0.1cm]

\multicolumn{2}{l}{\underline{\texttt{Pan-STARRS} (DR1) Survey:} Chambers et al. (2019),}  \\ [0.04cm]
g [mag]                                      & 14.1612 $\pm$ 0.1087  \\
r\;\! [mag]                                  & 13.6603 \\
i\hskip 4pt [mag]                            & 13.4406 \\
z\hskip 3pt [mag]                            & 13.6128 $\pm$ 0.0064  \\  [0.1cm]

% \multicolumn{2}{l}{\underline{\texttt{UCAC4} Survey}}  \\ [0.04cm]
% Id                                          &  342-180797    \\
% RA proper motion, $\mu_{\alpha}$\,[mas\,yr$^{-1}$]   &  5.6 $\pm$ 2.2   \\
% DEC proper motion, $\mu_{\delta}$\,[mas\,yr$^{-1}$]  &  1.0 $\pm$ 3.7   \\  [0.1cm]

\multicolumn{2}{l}{\underline{\texttt{WISE} mid-IR Survey:}  Gavrilchenko et al. (2014)}  \\ [0.04cm]
No.                                          &          2973         \\
Identification                               & J191733.34-214602.9   \\
W1 [mag]                                     &     12.593 $\pm$ 0.026   \\
W2 [mag]                                     &     12.604 $\pm$ 0.030    \\
W3 [mag]                                     &     12.284 $\pm$ 0.383   \\
W4 [mag]                                     &      8.778       \\  [0.1cm]

%sourceid  ra        dec    hjdstart         hjdstop                                       wasp_mag npts tm_statnpts tm_median  tm_stddevwrtmed   tm_range595
%1SWASP J191731.08-214641.1 289.379508  -21.778092  2453868.507 2454603.685 13.8336 5657    5657    1.40E+01    1.90E-01    4.94E-01
\multicolumn{2}{l}{\underline{\texttt{SuperWASP} Survey (2006-2008):} Pollacco et al. (2006)}  \\ [0.04cm]
Identification                               &  J191731.08-214641.1 \\
% HJDstart                                   &  2453868.507  \\
% HJDstop                                    &  2454603.685  \\
% No. points in light curve                  &  5657  \\
m(WASP) [mag]                                &  13.8336   \\
Period, $P$ [d]                              & 0.38671815       \\ [0.1cm]
% Pulsation amplitude, A [mag]               &  0.3 $\pm$ 0.4     \\  [0.1cm]
% Median magnitude [mag]                     & 14.0 $\pm$ 0.2    \\  [0.1cm]

% see K2C19p444 for 1st page of Clementini et al. 2019
\multicolumn{2}{l}{\underline{\texttt{Gaia} (DR2) Survey:} Brown et al. (2018); Clementini et al. (2019)} \\ [0.04cm]
Identification number                        & 4081034120298848256       \\
$m_{\rm g}$ [mag]                            &   13.782 $\pm$ 0.022  \\
$m_{\rm bp}$ [mag]                           &   14.099 $\pm$ 0.072     \\
$m_{\rm rp}$ [mag]                           &   13.387 $\pm$ 0.051    \\
Colour, bp$\_$rp [mag]                       &   0.711               \\
Effective temperature, $T_{\rm eff}$\,[K]    &     6135 $\pm$ 130            \\
Parallax, $\pi$ [mas]                        &    \;\,0.235 $\pm$ 0.031   \\
Proper motion (RA), $\mu_{\alpha}$\,[mas\,yr$^{-1}$]  & \;\,2.155 $\pm$ 0.048    \\
Proper motion (DEC), $\mu_{\delta}$\,[mas\,yr$^{-1}$] &   --2.899 $\pm$ 0.041   \\

\bottomrule
\end{tabular}
}
\end{table}

% TABLE 2 - Prior information
% see C7p691
\begin{table*}
%\begin{small}   <-- does not change from normal size
{\fontsize{6}{7.2}\selectfont  %  <--- works and better; 2nd (skip) should be 1.2 times 1st (size)
% {\fontsize{7}{8.4}\selectfont  %  <--- works and better; 2nd (skip) should be 1.2 times 1st (size)
% {\fontsize{5}{6.5}\selectfont   % <-- works but too small
\centering

\caption{Previous observations for the three `peculiar' RRd stars observed by {\it K2} during Campaign\,11.   }

\label{tab:five}
\begin{tabular}{lccc}
\toprule

\multicolumn{1}{c}{Observable}&  \multicolumn{1}{c}{EPIC\,235839761}             & \multicolumn{1}{c}{EPIC\,251248823}             & \multicolumn{1}{c}{EPIC\,251248824} \\
                      &  \multicolumn{1}{c}{=\,\texttt{OGLE-BLG-RRLYR-16999}}    & \multicolumn{1}{c}{=\,\texttt{OGLE-BLG-RRLYR-00595}} & \multicolumn{1}{c}{=\,\texttt{OGLE-BLG-RRLYR-19121}} \\
                      &  \multicolumn{1}{c}{=\,V2199\,Oph }                      &                                                    &                                       \\
\multicolumn{1}{c}{(1)} & \multicolumn{1}{c}{(2)} &\multicolumn{1}{c}{(3)}   & \multicolumn{1}{c}{(4)}   \\
\midrule
%                                              &    EPIC 235839761                &      EPIC 251248823              &   EPIC 251248824  \\

% \multicolumn{4}{l}{\underline{Coordinates:}} \\ [0.04cm]
RA (J2000)                                     &   17$^{\rm h}\,\!\!$16$^{\rm m}\!$22$^{\rm s}\;\!\!\!\!.\;\!$732   &
                                                   17$^{\rm h}\,\!\!$25$^{\rm m}\!$53$^{\rm s}\;\!\!\!\!.\;\!$310   &
                                                   17$^{\rm h}\,\!\!$30$^{\rm m}\!$19$^{\rm s}\;\!\!\!\!.\,\!$106   \\  % C7V5p829   289.389 $\pm$ 0.027;
DEC (J2000)                                    & --29$\,\!^\circ\,\!\!$11$\,\!'\,\!\!$48$\,\!'\!'\,\!\!\!\!.\;\!$65  &
                                                 --29$\,\!^\circ\,\!\!$15$\,\!'\,\!\!$46$\,\!'\!'\,\!\!\!\!.\;\!$80  &
                                                 --28$\,\!^\circ\,\!\!$58$\,\!'\,\!\!$12$\,\!'\!'\,\!\!\!\!.\;\!$89  \\  % C7V5p829    --21.767 $\pm$ 0.024
Galactic longitude (J2000), $l$                & ~356$^\circ\,\!\!\!\!.\;\!$27568    &
                                                 ~357$^\circ\,\!\!\!\!.\;\!$39955    &
                                                 ~358$^\circ\,\!\!\!\!.\,\!$179856   \\
Galactic latitude (J2000), $b$                 &   ~5$^\circ\,\!\!\!\!.\;\!$20103    &
                                                   ~3$^\circ\,\!\!\!\!.\;\!$45364    &
                                                   ~2$^\circ\,\!\!\!\!.\;\!$80965    \\  [0.1cm]

\multicolumn{4}{l}{\underline{\texttt{EPIC} Catalog} } \\ [0.04cm]
Mean {\it Kepler} magnitude, $<${\it Kp}$>$ [mag]   &     (16.895)                &     18.800                       &   19.802                      \\ [0.1cm]
% 2MASS number                                  &  17162273-2911486                &   \dots                          &   \dots                        \\
% $J$ magnitude, $J$ (2MASS)                    &     15.440 $\pm$ 0.062           &    \dots                         &    \dots                       \\ [0.1cm]     % See C11p885

\multicolumn{4}{l}{\underline{\texttt{OGLE-IV} Survey:} Soszy\'nski et al. (2017b)}  \\ [0.04cm]

Period, $P$ [d]                                 &    0.32582139\,($\pm$42)        & 0.28183106\,($\pm$19)            &  0.35759693\,($\pm$28)        \\  % Soszy\'nski et al. 2017
Frequency, $f$ [d$^{-1}$]                       &    3.0691666\,($\pm$44)         & 3.5482250\,($\pm$24)             &  2.7964446\,($\pm$22)         \\   %  calculated from line above
Time of first observation, $t_1$\!($I$) [BJD]   &     2455292.79811               & 2455265.81879                    &   2455265.81563               \\   %  see C11V5p698,700,702
Time of last observation, $t_N$\!($I$) [BJD]    &     2457971.67161               & 2457975.65283                    &   2457975.62104                \\  %  see C11V5p698,700,702
No. of $I$ observations (2010-2017), N($I$)     &     223                         &    827                           &     815                        \\   %  see C11V5p698,700,702
No. of $V$ observations (2010-2017), N($V$)     &         13                      &    44                            &     44                         \\   %  see C11V5p698,700,702
Time of maximum light, $t_0$  [BJD]             &     2457000.29828               &  2457000.05823                   &   2457000.13778                \\    % Soszy\'nski et al. 2017
Intensity-mean $I$ magnitude, $<$$I$$>$ [mag]   &       16.219                    &  16.764                          &   17.456                    \\   % Soszy\'nski et al. 2017
Intensity-mean $V$ magnitude, $<$$V$$>$ [mag]   &       17.508                    &  18.837                          &   19.874                      \\  % Soszy\'nski et al. 2017
$V$--$I$ colour [mag]                           &       1.289                     &   2.073                          &   2.418             \\
Total amplitude, $A_{\rm tot}$($I$) [mag]        &    0.446                       & 0.265                            &   0.360          \\
Fourier amplitude ratio, R$_{\rm 21}$($I$)       &      0.327                     &  0.380                           &   0.338      \\
Fourier amplitude ratio, R$_{\rm 31}$($I$)       &       0.187                    &  0.161                           &   0.144      \\
Fourier phase difference, $\phi^c_{\rm 21}$($I$) &      4.553                     &  4.450                           &   4.598      \\
Fourier phase difference, $\phi^c_{\rm 31}$($I$) &      2.631                     &  2.346                           &   2.876      \\ [0.1cm]

\multicolumn{4}{l}{\underline{\texttt{OGLE-IV} Survey:} Prudil et al. (2017)  }  \\ [0.04cm]
% Prudil et al.(2017):    I       P_D      P_A    P_A/P_D    A_D    A_A/A_D  Remarks
% OGLE-BLG-RRLYR-16999  16.216  0.32582  0.23142  0.71026  0.17399  0.227     blA
% OGLE-BLG-RRLYR-00595  16.770  0.28183  0.19971  0.70863  0.11414  0.287
% OGLE-BLG-RRLYR-19121  17.455  0.35760  0.24818  0.69402  0.15869  0.209     blD, f
% blD - peak(s) of the Blazhko effect in the vicinity of the dominant mode;
% blA - peak(s) for the Blazhko effect in the vicinity of the additional mode;
% f - listed in remarks file in OGLE archive
Dominant period, $P_D$ [d]                      &     0.32582                     & 0.28183                          &  0.35760 ($P_B$=482$\pm$5\,d)  \\
Additional period, $P_A$ [d]                    & 0.23142 ($P_B$=507$\pm$11\,d)   & 0.19971                          &  0.24818                       \\
Period ratio, $P_A$/$P_D$                       &    0.71026                      & 0.70863                          &  0.69402                       \\
Dominant amplitude, $A_{1,D}$($I$) [mmag]       &     173.99                      & 114.14                           &  158.69 ($A_B$=20$\pm$4\,mmag) \\
Additional amplitude, $A_{1,A}$($I$) [mmag]     &  39.5 ($A_B$=35$\pm$3\,mmag)    &  32.8                            &   33.2                         \\   % Prudil says for 16999 that A_A has Blazhko effect
$A_{1,D}$($I$)/$A_{1,A}$($I$)                   &      4.40                       &  3.48                            &   4.78                         \\ [0.1cm]

\multicolumn{4}{l}{\underline{\texttt{Gaia} (DR2) Survey:} Brown et al. (2018)}  \\ [0.04cm]
Identification number                           &   4107346185163417472           & 4059515539589312768              & 4059584606943065472   \\   % C7V5p828
$m_{\rm g}$\hskip 4pt\ [mag]                    &    17.154 $\pm$ 0.016           &  18.369 $\pm$ 0.013              &  18.886 $\pm$ 0.017     \\  %  C7V5p829
$m_{\rm bp}$ [mag]                              &    17.640 $\pm$ 0.060           &  18.940 $\pm$ 0.084              &  19.651 $\pm$ 0.085      \\
$m_{\rm rp}$\, [mag]                            &    16.259 $\pm$ 0.037           &  16.733 $\pm$ 0.035              &  17.298 $\pm$ 0.036                 \\
bp$\_$rp [mag]                                  &        1.381                    &    2.208                         &  2.354             \\
Parallax, $\pi$ [mas]                           & \,~[0.066 $\pm$ 0.118]          &   \dots                          &  [--1.1 $\pm$ 0.3]  \\   % C7V5p829
Proper motion (RA), $\mu_{\alpha}$\,[mas\,yr$^{-1}$]   &   --2.527 $\pm$ 0.212    &     \dots                        & ~\,[0.474 $\pm$ 0.575]     \\  %  C7V5p829
Proper motion (DEC), $\mu_{\delta}$\,[mas\,yr$^{-1}$]  &   --1.791 $\pm$ 0.142    &     \dots                        &  [--7.021 $\pm$ 0.402]   \\  %  C7V5p829

\bottomrule
\end{tabular}
}
\end{table*}

%  FIGURE 5  see C7p879 for c:\K2\pecRRd\I\PhasedLC-2021May17.ps --> .pdf --> pRRd_Fig5.pdf
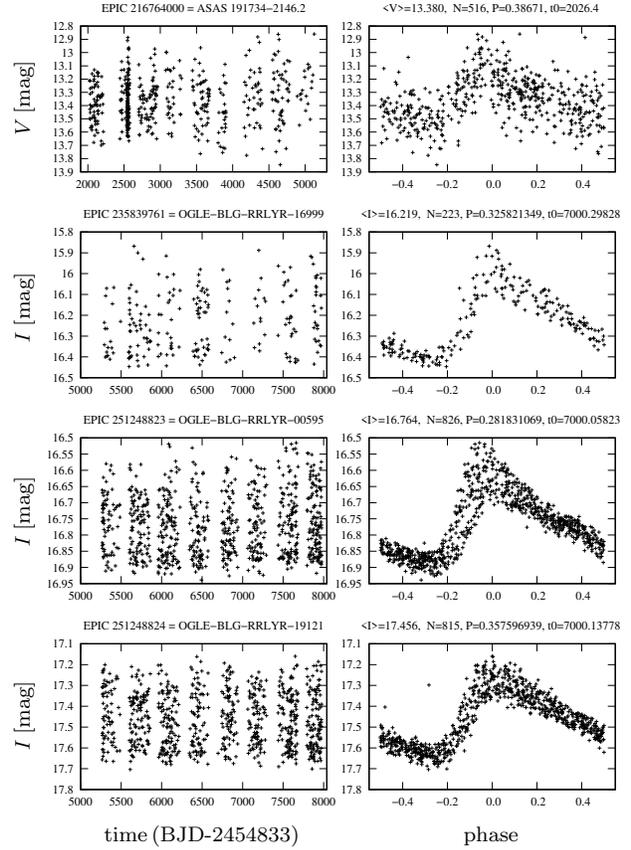
\begin{figure}
\begin{center}
\begin{overpic}[width=7.8cm]{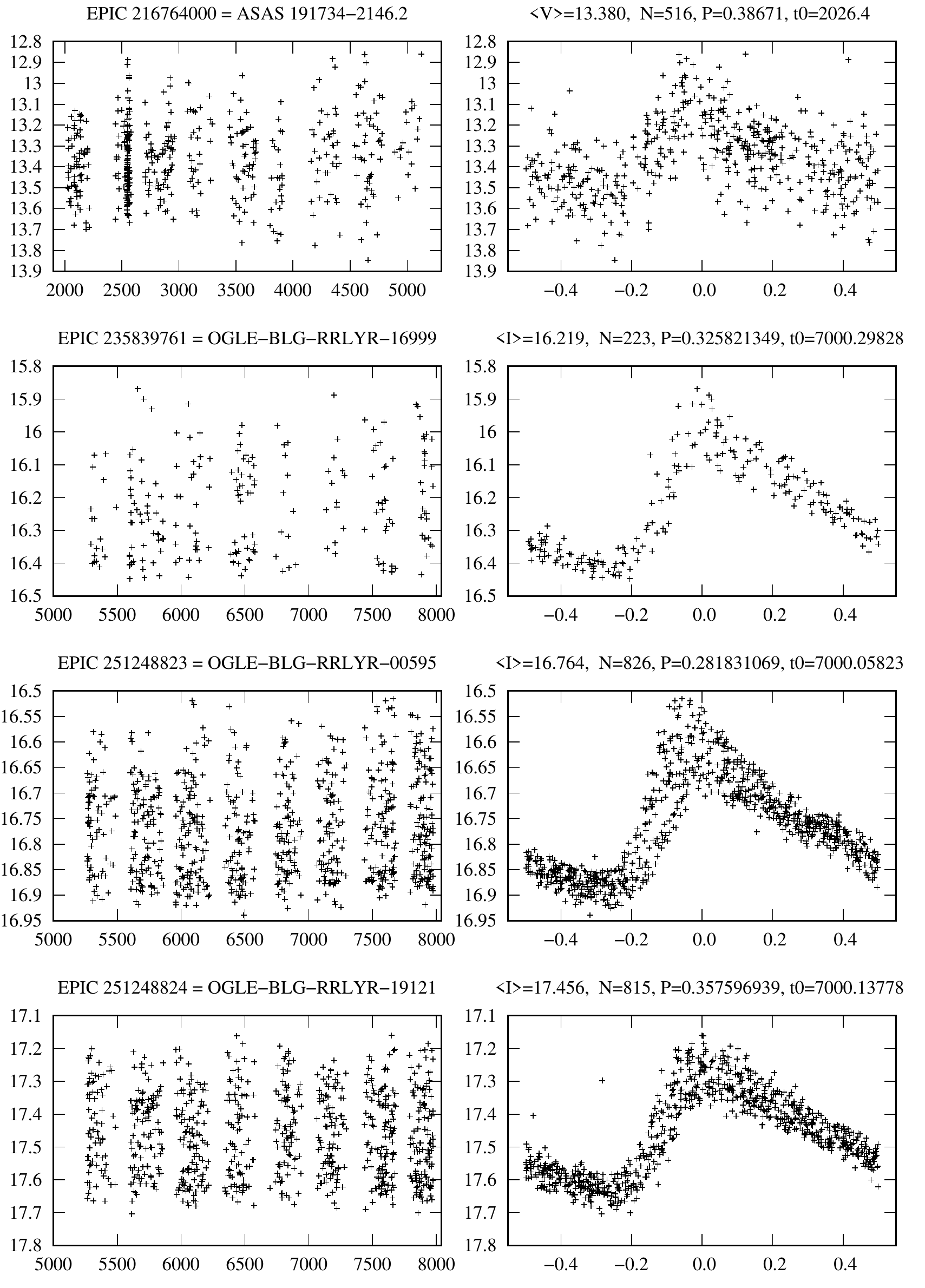} \put(7,-3){time\,(BJD-2454833)} \put(51,-3){phase}  \put(-4,8){\rotatebox{90}{\it I {\rm [mag]}}}  \put(-4,33){\rotatebox{90}{\it I {\rm [mag]}}}   \put(-4,58){\rotatebox{90}{\it I {\rm [mag]}}}   \put(-4,83){\rotatebox{90}{\it V {\rm [mag]}}}     \end{overpic}  %
\end{center}

\caption{Light curves reproduced from previous ground-based
photometry: (Top row) \texttt{ASAS} $V$-photometry for
EPIC\,216764000; (Bottom rows) \texttt{OGLE-IV} $I$-photometry
(2010-2017) for the three fainter Campaign\,11 stars. The left
panels show the photometry as a function of time, and the right
panels show phased light curves (all asymmetric).}

\end{figure}

% Subsection 2.3
\subsection{Other observations}

% the AAVSO Photometric All-Sky Survey\footnote{https://www.aavso.org/apass} (\texttt{APASS}),
%   https://exoplanetarchive.ipac.caltech.edu/docs/SuperWASPMission.html

Tables~1-2 summarize much of what was known prior to {\it K2}
about EPIC\,216764000 and the three fainter Campaign\,11 stars.
Apart from the Uitterdijk (1949) results, most of the information
comes from large-scale surveys: the Ecliptic Plane Input Catalog
(\texttt{EPIC})\footnote{https://archive.stsci.edu/k2/epic/search.php}
(Huber et al. 2016); the Optical Gravitational Lensing Experiment
(\texttt{OGLE})
surveys\footnote{http://ogledb.astrouw.edu.pl/ogle/OCVS} (see
Udalski et al. 2008, 2015; Soszy\'nski et al. 2014b, 2017b); the
All-Sky Automated Survey
(\texttt{ASAS})\footnote{www.astrouw.edu.pl/asas/} (Pojma\'nski
1997; Pojma\'nski \& Maciejewski 2005), and the \texttt{Gaia}
survey\footnote{https://www.cosmos.esa.int/web/gaia/home}.  Other
surveys considered here are the \texttt{2MASS}
survey\footnote{https://www.ipac.caltech.edu/2mass/}, the
\texttt{Pan-STARRS} survey\footnote{https://panstarrs.stsci.edu/},
the \texttt{WISE}
survey\footnote{https://www.nasa.gov/mission\_pages/WISE},  and the
\texttt{SuperWASP}
survey\footnote{https://en.wikipedia.org/wiki/Wide\_Angle\_Search\_for\_Planets}.
Particularly relevant are the periods and amplitudes reported in
P17, where the three Campaign\,11 stars were among the 42 RRd stars
identified as `peculiar' (triangles in Fig.\,1). Finding charts for
all four stars are available through the \texttt{Aladin Sky Atlas}
website, the most informative being the high-resolution colour
figures formed from \texttt{Pan-STARRS} $g,z$ images.

\iffalse
\footnote{A PDF file showing finding charts for the four `peculiar' RRd
stars observed by {\it K2} is included in the supplementary material.  The
\texttt{Aladin Sky Atlas} website is at
https://aladin.u-strasbg.fr/AladinLite.}.
\fi

%------------------------not used----------------------------------------------
\iffalse
In this section we discuss the photometric observations most relevant for
conducting a frequency analyses of the four stars.
    All three stars observed by {\it K2} in Campaign\,11 previously
had been observed by the \texttt{OGLE-IV} survey, with EPIC\,251248823 also having been
observed by the \texttt{OGLE-III} survey.  Since the \texttt{OGLE} data provide mean
$V,I$ magnitudes relevant for distance estimation  and for comparison with the
mean {\it Kepler} magnitudes, and colour information for placement in H-R
diagramsfor planning follow-up spectroscopy, EPIC\,216764000  had been observed by the All-Sky
Automated Survey, \texttt{ASAS}, and the three fainter stars had been extensively
observed by the \texttt{OGLE} Survey  All four stars were also observed by the
\texttt{GAIA} (DR2) Survey   While the \texttt{ASAS}
amd \texttt{OGLE} data do not have as high precision as the {\it K2}
photometry, or the advantage of consecutive observations, the photometry was
extensive and acquired over many years.
\fi
%------------------------------------------------------------------------------

% Section 2.3.1
\subsubsection{Previous photometry of {\rm EPIC\,216764000}}
% ASAS data copied to C:\K2\ASAS;  see C7V3p429-43, p474-475
% See C7p559 for ASAS P04
% P=0.3867075\,d is given on the latest \texttt{AAVSO} finding chart.

The variability of EPIC\,216764000 was discovered by Hendrik van Gent in the
1930's. Subsequently, 334 photographic plates taken by van Gent at the Union
Observatory (Johannesburg) between 1928 and 1937 were measured by Uitterdijk
(1949) who identified it as an RR~Lyrae star with pulsation period
0.3867145($\pm$13)\,d.  The light curve was shown to be asymmetric (risetime
0.20), with apparent brightness ranging from 13.4 to 14.3\,mag, and time (HJD)
of maximum light $t_0$=2426568.308($\pm$3).

More recently over 500 brightness measurements were made between 2001 and 2009
by the \texttt{ASAS-3} survey.  According to that survey, $P$=0.3867075\,d,
which is consistent with the pulsation period derived by Uitterdijk. No mention
was made of amplitude modulation or `additional' periods.  It also is not among
the RRd stars in the Galactic Plane identified by Szczygie{\l} \& Fabrycky
(2007). However, it was included among the 1455 Galactic RRab stars studied by
Szczygie{\l}, Pojma\'nski \& Pilecki (2009), whose paper is the source of most
of the \texttt{ASAS} properties in Table~1, including a derived photometric
metallicity, [Fe/H]=0.14$\pm$0.31\,dex. In Figure~5 (top panel) a phased light
curve reproduced from the \texttt{ASAS} $V$-photometry (assuming  $P$ and $t_0$
given in  the \texttt{ASAS} catalog) is shown. The asymmetry of the light
curve, hence the classification as an `ab-type' RR~Lyrae star, is consistent
with the findings of Uitterdijk. A reanalysis of the \texttt{ASAS} photometry
confirms the basic information given in the \texttt{ASAS} catalog.   The
pulsation period given there is 0.000040\,d shorter than the `dominant' period
derived from the later {\it K2} data, $P_D$=0.386748\,d (see below), suggesting
the possibility of an increasing dominant period. However, despite trying
different period search methods additional periodicity was not detectable in
the \texttt{ASAS} data.

% \subsubsection{ {\rm \texttt{SuperWASP}} photometry of {\rm EPIC\,216764000}}
% See C7p868-873,
% See C7p962-964 for PDM2 analysis of SuperWASP data.

EPIC\,216764000 was also observed by the \texttt{SuperWASP- South} cameras over
5600 times between 2006 May\,13 and 2008 May\,17 (see Butters et al. 2010).
From these data the dominant period was estimated to be $P$=0.38671815\,d
(Greer et al. 2017).  Unfortunately, significant gaps are present in the
time-series, and the uncertainties in the individual measurements are large, on
average $\pm$0.11\,mag.  Consequently, neither the `additional' period nor
amplitude variations were detected.  A reanalysis of the \texttt{SuperWASP}
data using \texttt{PDM2} found $P$=0.386721\,d (with $\theta_{\rm min}$=0.192)
but no detection of the additional period.

% Section 2.3.2
\subsubsection{Previous photometry of the Campaign\,11 stars}

The most extensive  photometric observations of the three fainter stars have
been made by the \texttt{OGLE} survey (see Table~2), mainly through $I$ filters
(for periods and light curve information), but also through $V$ filters (for
colour information).  All three stars were observed from 2010-2017 by the
\texttt{OGLE-IV} survey (Soszy\'nski et al. 2011, 2017b), and EPIC\,251248823
was also observed by the \texttt{OGLE-III} (2001-04) survey (Udalski et al.
2008). Despite having relatively short (dominant) periods, between 0.28\,d and
0.36\,d, and relatively low total amplitudes, 0.26--0.45\,mag, all three stars
were identified by the \texttt{OGLE} classifier as RRab stars, a classification
consistent with all three stars having asymmetric light curves when the
photometry is phased with the dominant period (see Fig.\,5).

% See C11p1213 for statistics on secondary periods
For EPIC\,251248824 (=\,\texttt{OGLE-BLG-RRLYR-19121}) a secondary period,
$P_A$=0.248191\,d, was detected by the \texttt{OGLE-IV} pipeline (see the
`\texttt{OGLE} Collection of Variable Stars' website at
http://ogledb.astrouw.edu.pl/$\sim$ogle/OCVS/).  This period agrees with the
`additional' period derived by P17 from their analysis of the same
$I$-photometry and confirmed by the {\it K2} data. In fact, eight of the 42
stars in the original pRRd sample are noted in the \texttt{OGLE} catalogue as
having secondary periods similar to those reported by P17.

% See C11Vol7
P17 also found that \texttt{OGLE-BLG-RRLYR-16999} and
\texttt{OGLE-BLG-RRLYR-19121} have modulated amplitudes.  For the former star
(=\,EPIC\,235839761) the {\it additional} mode has a modulation (Blazhko)
period $P_{B}$=507$\pm$11\,d, with $I$-amplitude 35$\pm$3\,mmag; and for the
latter (=\,EPIC\,251248824) the {\it dominant} mode has $P_{B}$=482$\pm$5\,d,
with $I$-amplitude 20$\pm$4\,mmag (see discussion in $\S$3.2.3 below).

% SECTION 3
\section{FREQUENCY ANALYSIS}

Detailed frequency analyses of the {\it K2} photometry were conducted for each
of the four pRRd stars using several different methods.  Initial estimates of
the frequencies were made using the \texttt{Period04} program (Lenz \& Breger
2005), the \texttt{SigSpec/Combine} programs (Reegen 2007), and the
\texttt{FNPEAKS} period-searching code (written by
Z.\,Ko{\l}aczkowski\footnote{Source available from
http://helas.astro.uni.wroc.pl/ \\ deliverables.php?lang=en\&active=fnpeaks},
see Moskalik \& Ko{\l}aczkowski 2009). These algorithms employ successive
prewhitenings of the data, which were applied manually in the case of
\texttt{Period04} and \texttt{FNPEAKS} and automatically for \texttt{SigSpec}.
At each step a multi-frequency model, which includes all the previously
identified frequencies, was fitted to the observations. The stopping criterion
for the \texttt{Period04} and \texttt{FNPEAKS} analyses was a signal-to-noise
ratio of four for the highest remaining peak in the periodogram (Breger et al.
1993). The \texttt{SigSpec} runs were stopped when the \texttt{Sig} index
reached the default value of five (Reegen 2011). Monte Carlo simulation
(\texttt{Period04}) was also used to estimate errors for the `dominant' and
`additional' pulsation frequencies. In addition to these parametric methods,
non-parametric period searches were conducted using the Phase Dispersion
Minimization (\texttt{PDM2}) method of Stellingwerf (1978, 2011).  All of these
approaches were useful in establishing the significant frequencies present in
the data, and in most cases produced consistent results.

Optimized estimates of the frequencies, amplitudes and phases were made using
our own non-linear least-squares (Levenberg--Marquardt) multi-frequency fitting
program, written in \texttt{SAS} and utilizing its \texttt{PROC NONLIN}
procedure. The two-frequency fitting formula was

%\begin{multline} m(t) = m_0 +
%    \sum_{i=1}^{N_D}                        A_{\rm i,D}        \sin \,
%    (i\omega_D \, [t-t_0]               +  \phi_{\rm i,D}) \,\,+ \\
%    \sum_{j=1}^{N_A }                       A_{\rm j,A}        \sin \,
%    (j\omega_A \, [t-t_0]               +  \phi_{\rm j,A}) \,+ \\
%    \sum_{i=1}^{N_D} \sum_{j=1}^{N_A}   [A^{+}_{\rm i,j}     \sin \,((i\omega_D
%    + j\omega_A) \, [t-t_0]  +  \phi^{+}_{\rm i,j}) \,+ \\ A^{-}_{\rm i,j}
%    \sin \, ((i\omega_D -  j\omega_A) \, [t-t_0]  +  \phi^{-}_{\rm i,j}) ],
%\end{multline}

\begin{multline}
m(t) = m_0 + \sum_{i=1}^{N_D} A_{\rm i,D}\sin\,(i\omega_D\,[t-t_0] + \phi_{\rm i,D})\,\,+ \\
      ~~~~~~~\sum_{j=1}^{N_A} A_{\rm j,A}\sin\,(j\omega_A\,[t-t_0] + \phi_{\rm j,A})\,\, + \\
     ~~~~~~~~\sum_{i=1}^{N_D}\sum_{j=1}^{N_A}\,[A^{+}_{\rm i,j}\sin\,((i\omega_D + j\omega_A)\,[t-t_0] + \phi^{+}_{\rm i,j})\,\,+ \\
                              A^{-}_{\rm i,j}\sin\,((i\omega_D - j\omega_A)\,[t-t_0] + \phi^{-}_{\rm i,j})],
\end{multline}

\vskip0.5truecm

\noindent where $m_0$ is the mean magnitude,  $\omega_D$ and
$\omega_A$ are the angular frequencies for the `dominant' and
`additional' modes, the $A$ and $\phi$ are the amplitudes and phases
of the various terms in the Fourier sums, up to ${N_D}$ terms for
the dominant mode and ${N_A}$ terms for the additional mode.  The
assumed ${N_D}$ and ${N_A}$ values were adjusted to include all
significant harmonic peaks in the amplitude spectra.

The derived amplitudes and phases were used to calculate the Fourier
decomposition parameters introduced by Simon \& Lee (1981):
amplitude ratio $R_{\rm i1}$ = $A_{i}$/$A_{1}$, and
epoch-independent phase difference $\phi_{\rm i1}$ = $\phi_{i}$ --
$i \phi_{1}$, where $i$ denotes the $i$th harmonic.  Superscripts
`s' or `c' on the $\phi_{\rm i1}$ parameter will be used to indicate
the use of either a sine or cosine series for the Fourier
summations.

% Section 3.1
\subsection{EPIC\,216764000 (=\,V1125\,S\lowercase{gr})}

% FIGURE 6
% See C7p586-7 for source files
% Prudil star See C7p247 for the creation of this figure;  see also C7p414-426
% See also C7p242;   C7p398 for EAP most recent analysis, which are same as Everest
% See C7V4p619 for 2019Apr1.pdf, later to be replaced Sept.2,2019 on C7pxxxx.
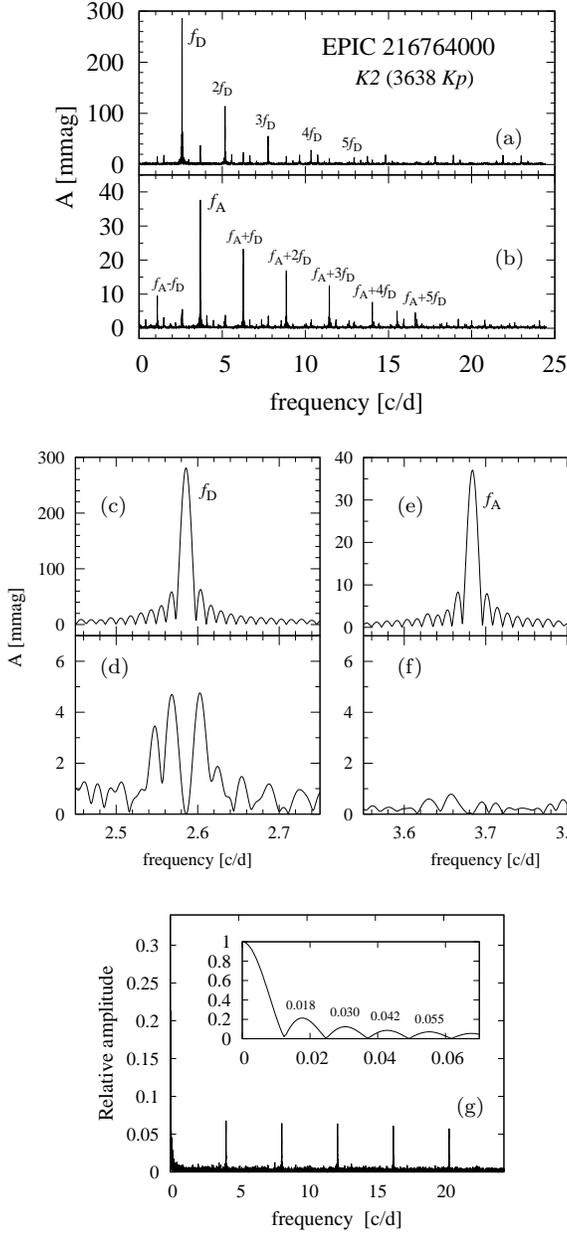
\begin{figure}
% \begin{overpic}[width=8.0cm]{216764000-p04spectrum-2019Mar15.pdf}  \put(60,85){(a)}  \put(60,67){(b)}  \put(60,45){(c)}  \put(59,24){(d)}  \end{overpic}
% \begin{overpic}[width=8.0cm]{216764000-p04spectrum-2019Apr8.pdf}  \put(75,78){(a)}  \put(75,57){(b)}  \put(75,29){(c)}  \end{overpic}
\hskip0.6truecm \begin{overpic}[width=7.0cm]{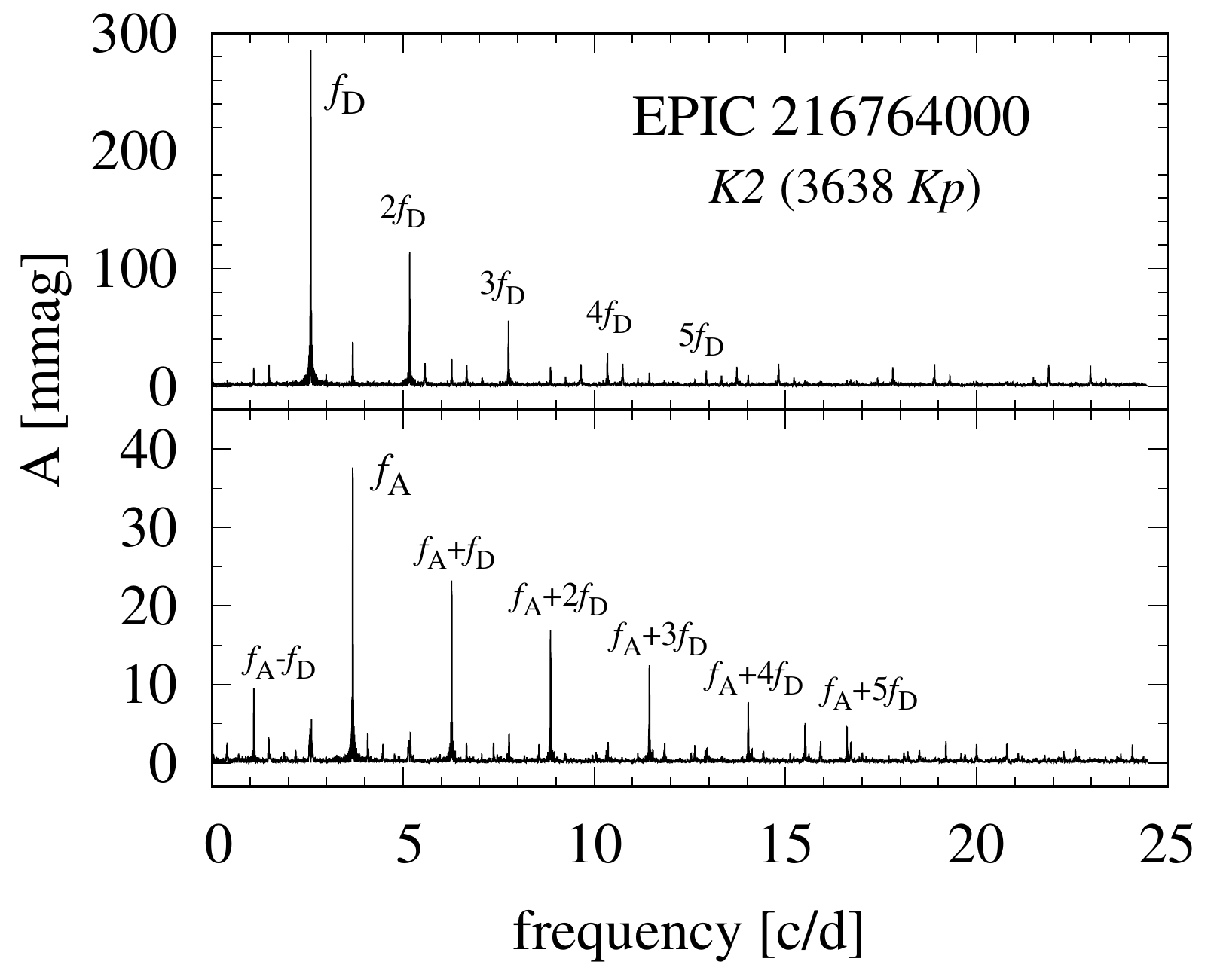}  \put(85,53){(a)}  \put(85,30){(b)}   \end{overpic}  % renamed from "216764000-p04spectrum-EAP-2021May19.pdf"  see C7p784b May19,21
\vskip0.3truecm
\begin{overpic}[width=4.25cm]{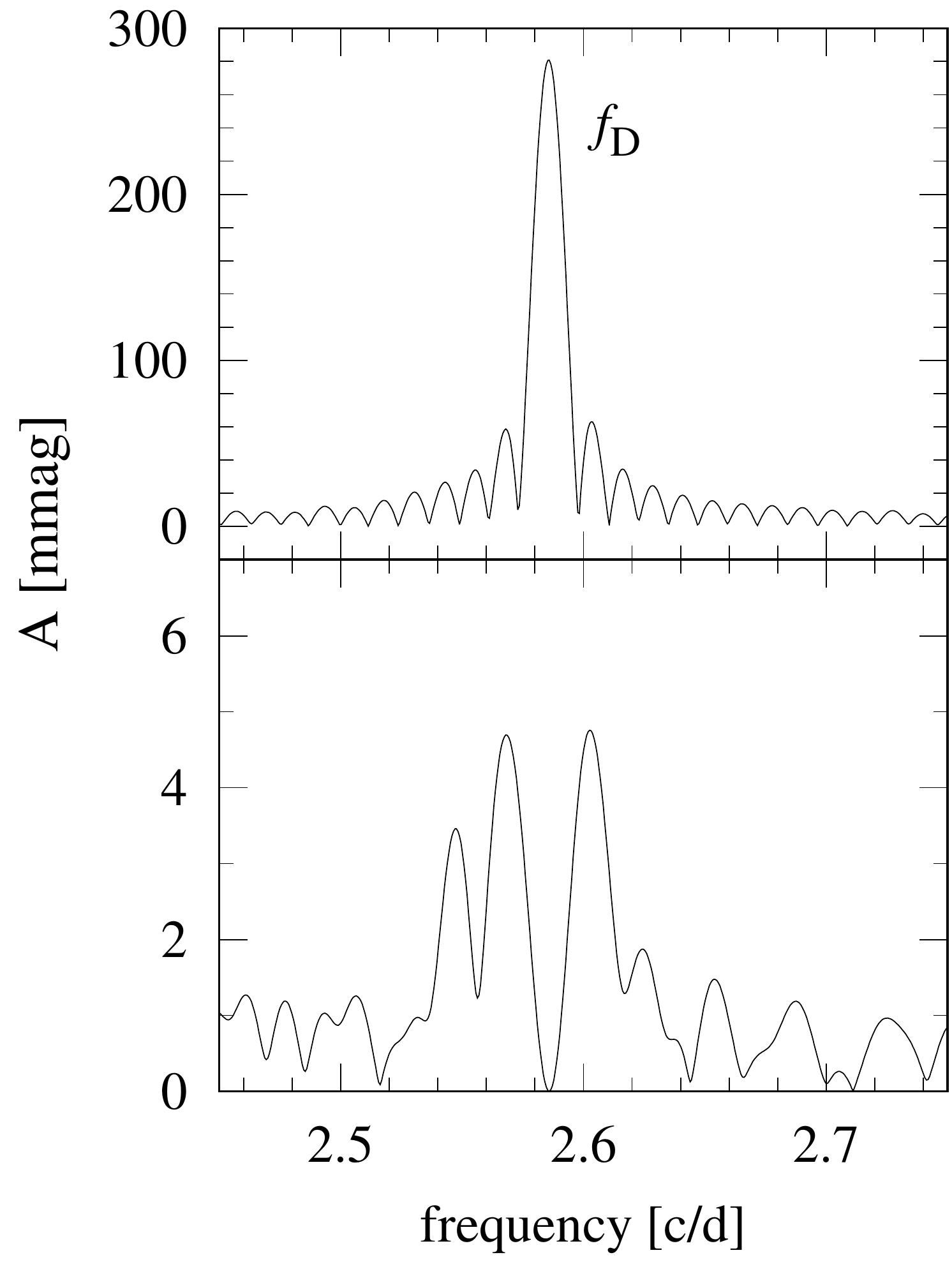}  \put(23,85){(c)}  \put(23,47){(d)}   \end{overpic}  % renamed from "2019Apr1.pdf" [C7p619,p643,p720]; p619 May19,2021
\begin{overpic}[width=3.73cm] {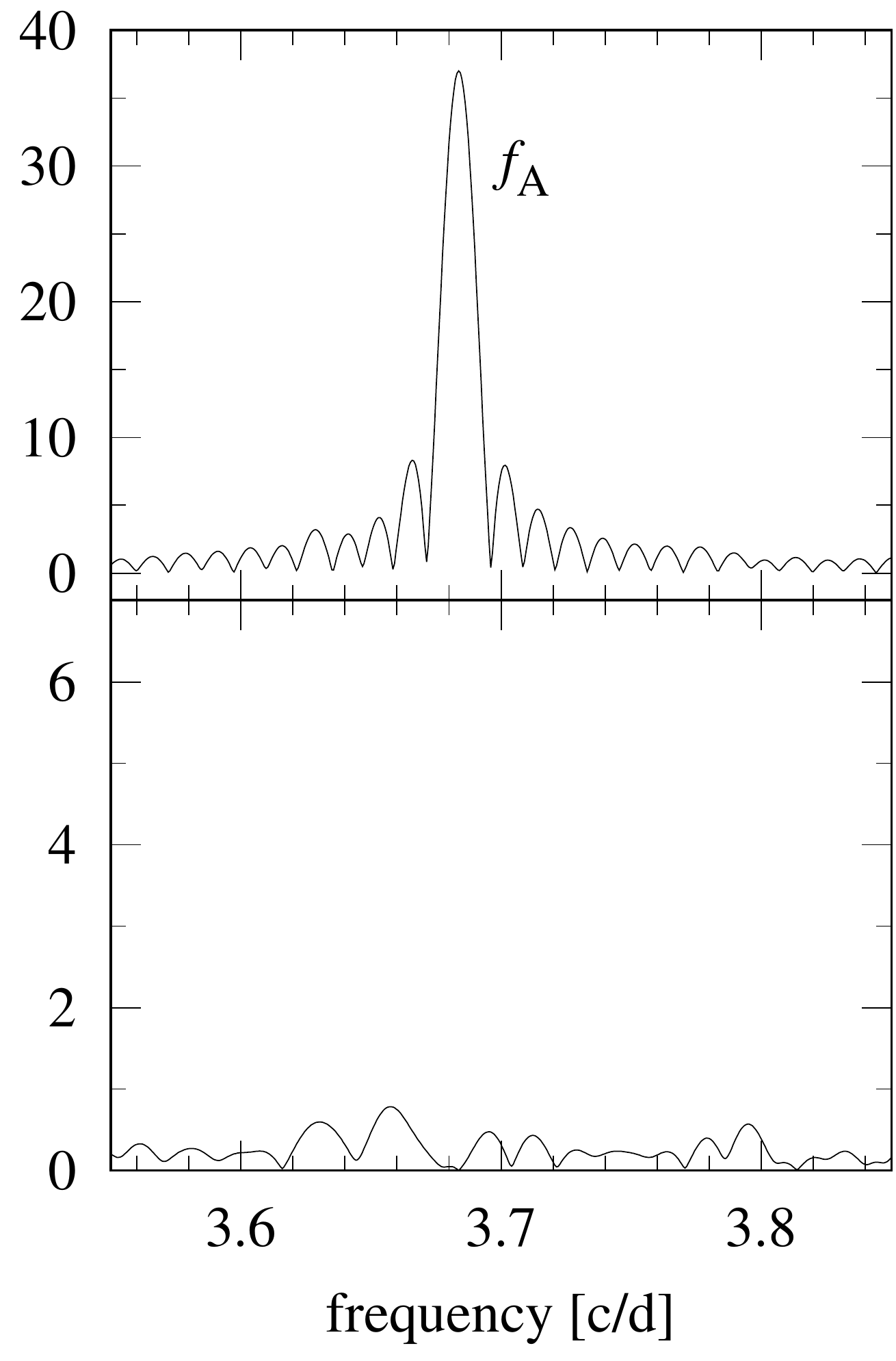}  \put(16,85){(e)}  \put(16,47){(f)}   \end{overpic}  % renamed from "2019Apr1add2.pdf" [C7]
\vskip0.5truecm
\hskip1.2truecm
\begin{overpic}[width=5.5cm]{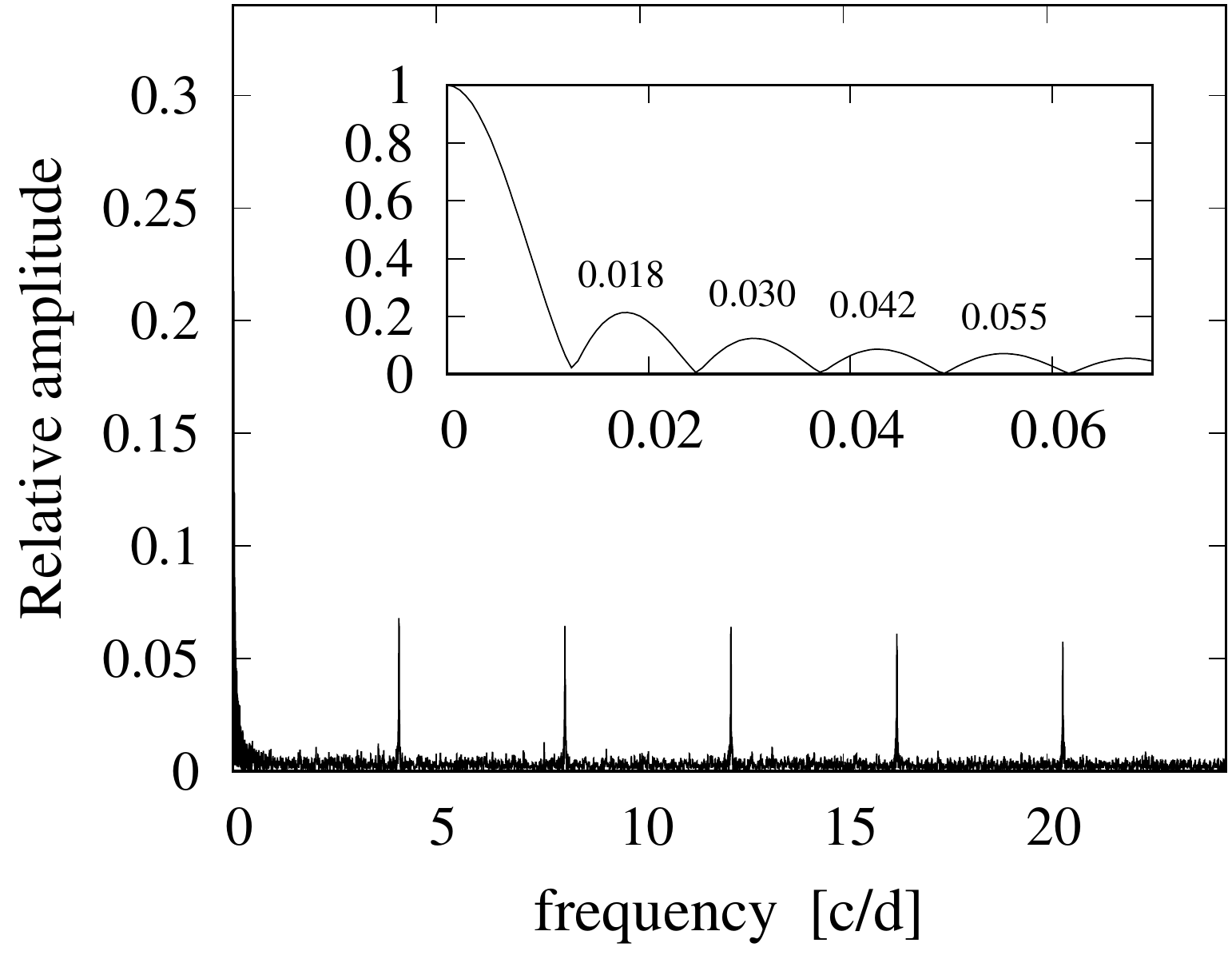}   \put(88,30){(g)}   \end{overpic}  % renamed from "WindowFn-2019May7b.pdf"
% \begin{overpic}[width=4.30cm]{2019Apr1.pdf}     \put(23,85){(a)}  \put(23,47){(b)}    \end{overpic}
% \begin{overpic}[width=3.8cm] {2019Apr1add.pdf}  \put(16,85){(c)}  \put(16,47){(d)}  \end{overpic}

\caption{Fourier amplitude spectra for EPIC\,216764000 ({\it K2}
photometry, \texttt{EAP} pipeline, \texttt{Period04}). (a) The
highest peak corresponds to the {\it dominant} frequency,
$f_D$=2.585663($\pm$39)\,d$^{-1}$, for which the first term in the Fourier
expansion has an amplitude 285.4$\pm$0.2\,mmag. The four most
powerful of the seven detected harmonics (see Table~3) have been
labelled.  (b) After prewhitening with $f_D$ and its harmonics the
highest remaining peak is that of the {\it additional} frequency,
$f_A$=3.683689($\pm$43)\,d$^{-1}$ (amplitude 37.8$\pm$0.2\,mmag).  The other
strong peaks are combination frequencies involving $f_A$ and $f_D$.
The harmonic at 2$f_A$ is visible but weak (amplitude 1.8\,mmag).
(c) Close-up of amplitude spectrum near $f_D$. (d) Close-up of
spectrum after prewhitening with $f_D$ and its harmonics. (e)
Close-up of amplitude spectrum in the vicinity of $f_A$. (f)
Amplitude spectrum after prewhitening with $f_A$ and its harmonics.
(g) Window function for the {\it K2} photometry.}

\label{fig:FT}
\end{figure}

%  TABLE 3
% For the frequencies see p.622: Mandy's SAS program run on Everest data; the ampl are a bit lower than the P04 values
% For the Sig values see C7pxxxxx
%  see C7p750 for PAM-EAP SAS output;  see C7p807 for improved ampl; see C7p1135-1137 for corrected ampl,phase
\begin{table}
{\fontsize{6}{7.2}\selectfont  %  <--- works and better; 2nd (skip) should be 1.2 times 1st (size)
\begin{threeparttable}

%{

\fontsize{6}{7.2}\selectfont  %  <--- works and better; 2nd (skip) should be 1.2 times 1st (size)

\caption{EPIC\,216764000 (=\,V1125\,Sgr): frequencies, amplitudes
and phases derived from the {\it K2} photometry. The independent
frequencies, $f_D$ and $f_A$, have been underlined, and the
uncertainties in the amplitudes and phases are random errors (i.e.,
do not include systematic errors). The quantity \texttt{Sig} is the
significance of the peak in the amplitude spectrum (see Reegen 2007,
2011). The mean magnitude, $<${\it Kp}$>$=13.7412$\pm$0.0001\,mag,
agrees well with the \texttt{EPIC} mean. The assumed time of maximum
light (for the phases) was $t_0$=BJD\,2454833.0000+2468.9335.}

%}

\begin{tabular}{lrcrrr}
\toprule
Label & \multicolumn{1}{c}{Frequency}    & Period & \multicolumn{1}{c}{Ampl.}     & \multicolumn{1}{c}{Phase} & \texttt{Sig} \\
      & \multicolumn{1}{c}{[d$^{-1}$]}   & {[d]}  & \multicolumn{1}{c}{[mmag]}    & \multicolumn{1}{c}{[rad]} &              \\
      &                                  &        & \multicolumn{1}{c}{$\pm$0.20} &                           &              \\
\midrule
    \multicolumn{6}{c}{(a) {\it Dominant} frequency, $f_D$\,(=\,f1), and its harmonics}                \\ [0.04cm]
  ~\,f1          & \underline{2.58566} & 0.386748 & 285.40         & --2.270$\pm$0.001 &  690          \\
    2f1          &            5.17132  & 0.193374 & 113.60         & --1.797$\pm$0.002 &  562          \\
    3f1          &            7.75698  & 0.128916 &  55.06         & --1.249$\pm$0.004 &  385          \\
    4f1          &           10.34264  & 0.096687 &  26.87         & --0.804$\pm$0.008 &  269          \\
    5f1          &           12.92830  & 0.077350 &  13.20         & --0.131$\pm$0.016 &  201          \\
    6f1          &           15.51398  & 0.064458 &   5.47         &   0.572$\pm$0.037 &  114          \\
    7f1          &           18.09964  & 0.055250 &   1.65         &   1.132$\pm$0.120 &   33          \\
    8f1          &           20.68528  & 0.048344 &   0.49         &   1.449$\pm$0.403 &   13          \\ [0.1cm]
%   9f1          &           23.27094  & 0.042972 &   0.10         & --2.391$\pm$1.945 &    5          \\ [0.1cm]

\multicolumn{6}{c}{(b) {\it Additional} frequency, $f_A$\,(=\,f2), and its harmonics}                  \\ [0.04cm]
  ~\,f2          & \underline{3.68369} & 0.271468 &  37.81         & --1.236$\pm$0.008 &  330          \\
    2f2          &            7.36738  & 0.135734 &   1.81         &   0.382$\pm$0.111 & $<$5          \\  [0.1cm]
%   3f2          &           11.05107  & 0.090489 &   0.1:\;\;\!\! & --2.710$\pm$1.953 & $<$5          \\  [0.1cm]

\multicolumn{6}{c}{(c) Linear combination frequencies}                                                 \\ [0.04cm]
  ~\,f1+f2       &            6.26935  & 0.159506 &  23.23         & --1.142$\pm$0.011 &  294          \\
    2f1+f2       &            8.85501  & 0.112930 &  17.24         & --0.818$\pm$0.013 &  244          \\
    3f1+f2       &           11.44067  & 0.087407 &  12.48         & --0.333$\pm$0.017 &  234          \\
    4f1+f2       &           14.02633  & 0.071294 &   7.77         &   0.159$\pm$0.026 &  163          \\
    5f1+f2       &           16.61199  & 0.060197 &   4.59         &   0.701$\pm$0.044 &   92          \\
    6f1+f2       &           19.19765  & 0.052090 &   2.62         &   1.291$\pm$0.076 &   58          \\
    7f1+f2       &           21.78333  & 0.045907 &   1.23         &   1.920$\pm$0.162 &   27          \\
    8f1+f2       &           24.36897  & 0.041036 &   0.51         &   2.541$\pm$0.389 &   13          \\ [0.1cm]
%   9f1+f2       &           26.95463  & 0.037099 &   0.13         & --2.836$\pm$1.556 & $<$5          \\ [0.1cm]

    2f1--f2      &            1.48763  & 0.672210 &   2.88         & --0.764$\pm$0.069 &   61          \\
    3f1--f2$^a$  &            4.07273  & 0.245502 &   2.7:\;\;\!\! & --0.598$\pm$0.053 &   58:\;\!\!\! \\
    4f1--f2      &            6.65895  & 0.150174 &   2.48         &   0.287$\pm$0.080 &   55          \\
    5f1--f2      &            9.24463  & 0.108171 &   1.23         &   1.694$\pm$0.162 &   27          \\
    6f1--f2      &           11.83029  & 0.084529 &   0.67         &   1.541$\pm$0.295 &   14          \\
    7f1--f2      &           14.41595  & 0.069368 &   0.61         &   1.728$\pm$0.323 &   14          \\
    8f1--f2      &           17.00162  & 0.058818 &   0.56         &   2.453$\pm$0.356 &   13          \\ [0.1cm]
%   9f1--f2      &           19.58725  & 0.051054 &   0.28         &   3.140$\pm$0.712 &    6          \\ [0.1cm]

  ~\,f2--f1      &            1.09803  & 0.910725 &   9.72         &   3.088$\pm$0.021 &  205\;\!\!    \\
    2f2--f1      &            4.78171  & 0.209130 &   0.57         &   1.013$\pm$0.349 &   13          \\ [0.1cm]
%   2f2--2f1     &            2.19606  & 0.455361 &   0.24         &   1.588$\pm$0.826 &    6          \\ [0.1cm]
%   3f2--f1      &            8.46541  & 0.118128 &   0.09         &   0.485$\pm$2.313 & $<$5          \\ [0.1cm]

  ~\,f1+2f2      &            9.95304  & 0.100472 &   0.97         & --0.060$\pm$0.205 &   25          \\
    2f1+2f2      &           12.53870  & 0.079753 &   1.15         &   0.581$\pm$0.173 &   28          \\
    3f1+2f2      &           15.12436  & 0.066119 &   0.93         &   1.058$\pm$0.214 &   24          \\
    4f1+2f2      &           17.71002  & 0.056465 &   0.55         &   1.582$\pm$0.363 &   14          \\
    5f1+2f2      &           20.29568  & 0.049272 &   0.37         &   1.771$\pm$0.534 &   12          \\ [0.1cm]
%   6f1+2f2      &           22.88136  & 0.043704 &   0.25         &   2.284$\pm$0.798 &    8          \\
%   7f1+2f2      &           25.46700  & 0.039267 &   0.19         &   2.961$\pm$1.065 & $<$5          \\ [0.1cm]
%   5f1--2f2     &            5.56094  & 0.179826 &   0.17         & --1.729$\pm$1.138 &    5          \\ [0.1cm]
%   6f1--2f2$^b$ &            8.14660  & 0.122751 &   0.2:\;\;\!\! &   1.787$\pm$1.121 &    7:\;\!\!\! \\ [0.1cm]

% \multicolumn{6}{c}{(d) Thruster frequency and its harmonics}     \\ [0.04cm]
%   1\,f\_th$^a$ &            4.074586 & 0.245424 &   3.3:         &   1.981           &   89:\;\!\!\! \\
%   2\,f\_th$^b$ &            8.157279 & 0.122590 &   0.4          &   2.435           &   12          \\
%   3\,f\_th     &           12.235136 & 0.081732 &   0.2          &   1.102           &    6          \\
%   4\,f\_th     &           16.314436 & 0.061295 &   0.2          &   3.940           &    5          \\
%   5\,f\_th$^c$ &           20.499572 & 0.048782 &   0.2          &   5.625           &    6          \\
%   6\,f\_th$^d$ &           24.599301 & 0.040652 &   0.3          &   2.182           &    9          \\
\bottomrule
\end{tabular}
\begin{tablenotes}

{\fontsize{6}{7.2}\selectfont  %  <--- works and better; 2nd (skip) should be 1.2 times 1st (size
\item Notes:  (a) The frequency  3f1--f2 = 4.073\,d$^{-1}$ coincides with the {\it Kepler}/{\it K2} thruster-firing
    frequency, $f_{\rm th}$=4.075\,d$^{-1}$, making measurement of amplitudes and phases of both frequencies highly unreliable.}
%   (b) The frequency  6f1--2f2 = 8.147\,d$^{-1}$ is close to 2\,$f_{\rm th}$=8.157\,d$^{-1}$.  }
%   (c) the thruster-frequency harmonic 5\,$f_{\rm th}$ is close to the combination frequency 5f1+2f2;  and (d) the
%   thruster-frequency harmonic 6$f_{\rm th}$ is close to the combination frequency 8\,f1+f2.   }

\end{tablenotes}
\end{threeparttable}
}
\end{table}

% maybe incorporate into table?
\iffalse
\footnote{The uncertainties in Table~4 are lower
limits, having been derived using the Fourier multi-frequency \texttt{SAS} program with
photometry from the \texttt{EAP} pipeline (improved with additional
detrending).   They do not include systematic errors or errors arising from
autocorrelation of the {\it K2} data. Note the obvious strong correlation
between the amplitudes of the peaks and the \texttt{Sig} statistic defined by
Reegen (2007, 2011).}
\fi

% FIGURE 7  See C7p658-659,972,667(May21,2021) for source;  PAM-EAP-OLr pipeline
\begin{figure}
\begin{center}
\begin{overpic}[width=8.1cm]{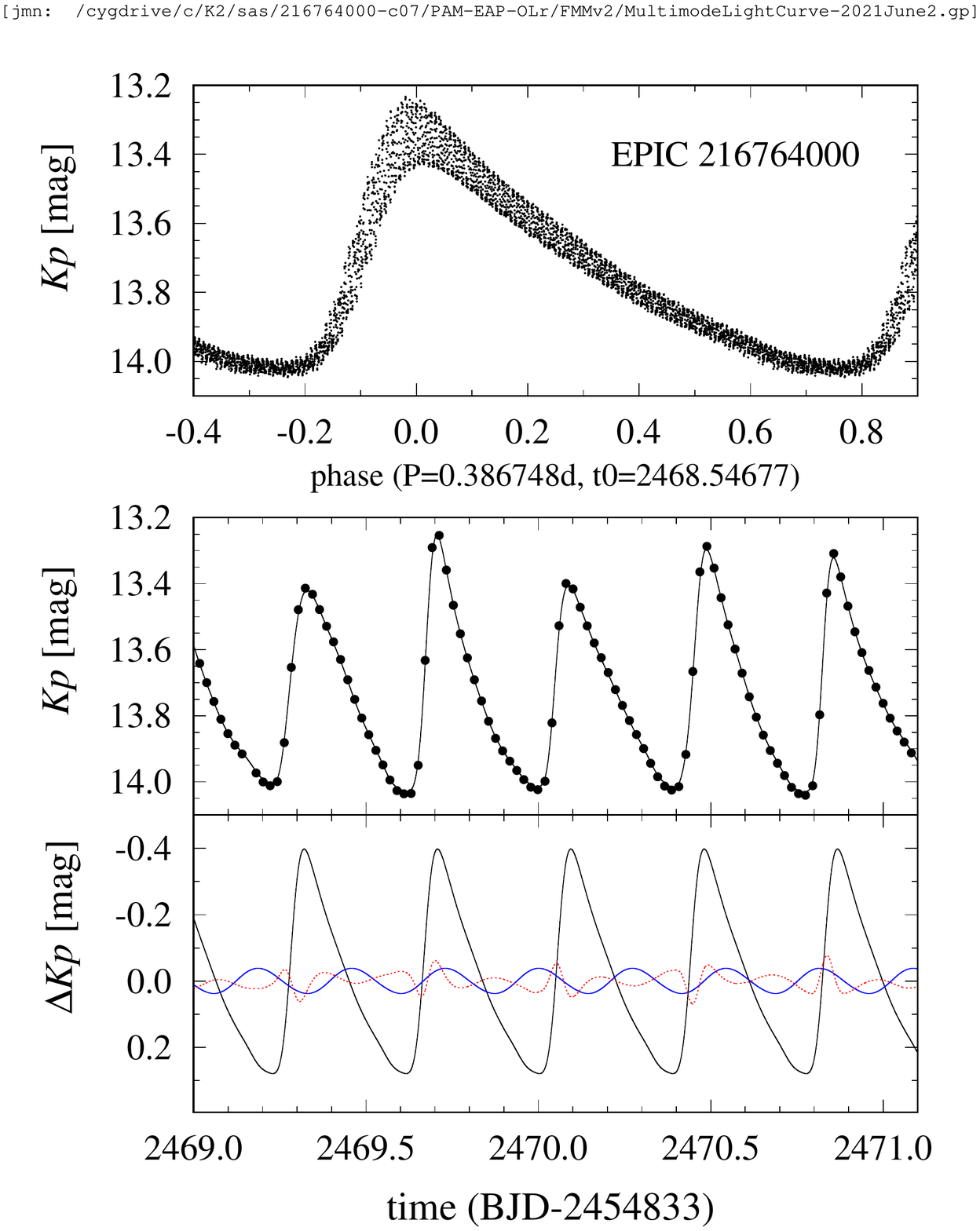} \put(16,92){(a)} \put(16,55){(b)} \put(16,30){(c)}   \end{overpic}  % renamed from "216764000-c07-mmLC-2020June27.pdf"
% \begin{overpic}[width=8.4cm]{201585823-c01-mmLC.pdf}             \put(17,65){(c)}  \put(17,17){(d)}    \end{overpic}
\end{center}
\vskip-0.4truecm

\caption{Observed and fitted light curves for EPIC\,216764000.  (a) Light curve
phased with the dominant period (the scatter is due to the unremoved additional
oscillation and combination frequencies); (b) Close-up comparison of the fitted
and observed {\it K2} light curves ($\sim$19 points per cycle); (c) Component
light curves derived from non-linear least-squares fitting of the two-frequency
model. The lower-amplitude nearly-sinusoidal (blue) curve is for the additional
mode with period $P_A$=0.271468\,d, and the dotted (red) curve is the
contribution from the combination frequencies.}

\label{fig:FT}
\end{figure}

% Section 3.1.1
\subsubsection{Frequencies, Amplitudes and Phases}

% see C2p361 for Pawel results:
% PDM2:  See C7p899
% gLS:   and $f_A$=3.68378\,d$^{-1}$ with $P(\omega)$=0.28.
% PDM2:  and $f_A$=3.68362\,d$^{-1}$ at $\theta$(min)=0.58.
% For S/N see K2C19

\iffalse
\footnote{A \texttt{gLS} analysis of \texttt{PDCsap} photometry found $f_D$ =
2.58569\,d$^{-1}$ with $P(\omega)$=0.77.  A \texttt{PDM2} analysis of
\texttt{Everest} data found $f_D$=2.58567 d$^{-1}$ with $\theta$(min)=0.027
(extremely low). In general, analysis of the \texttt{Everest} data gave
similar frequencies but slightly smaller amplitudes than analysis of the
\texttt{PDCsap} and \texttt{EAP} data.}
\fi

Fourier amplitude spectra for EPIC\,216764000 are shown in Figure~6. The 32
most significant frequencies present in the {\it K2} data are summarized in
Table~3, which also contains the derived Fourier amplitudes and phases. The
highest peak in the top panel corresponds to  the {\it dominant} frequency,
$f_D$=2.585663($\pm$3)\,d$^{-1}$, for which the signal-to-noise ratio (S/N) is
$\sim$120 and the amplitude of the first term of the Fourier sum  is
$A_{1,D}$=285.4$\pm$0.2\,mmag.   The corresponding period,
$P_D$=0.386748($\pm$1)\,d, is slightly longer than the periods determined by
the earlier surveys but is consistent with other estimates based on analysis of
the {\it K2} data processed through other pipelines. A total of seven high
harmonics were detected, four of which are labelled in Fig.\,6(a), indicating
asymmetry in the shape of the underlying light curve.

% In general, analysis of the \texttt{Everest} data gave similar frequencies but
% slightly smaller amplitudes than analysis of the \texttt{PDCsap} data:
% $f_D$=2.585675\,d$^{-1}$, $f_A$=3.68372\,d$^{-1}$, $A_{1,D}$=280\,mmag,
% $A_{1,A}$=37.1\,mmag.  These comparisons give some idea of the systematic
% errors.

% numbers from Pawel e-mail C2p361;  see p.10 of Pawel's comments on the paper
After prewhitening with $f_D$ and its harmonics,  the {\it additional}
frequency at $f_A$=3.68369($\pm$3)\,d$^{-1}$ ($P_A$=0.271468($\pm$2)\,d,
S/N=15.7) is clearly identified in Fig.\,6(b). The main effect of this
low-amplitude ($A_{1,A}$=37.8$\pm$0.2\,mmag) oscillation is to beat with the
dominant oscillation, leading to variations of the total amplitude (see
Fig.\,3). Note that the harmonics at 2$f_D$ and 3$f_D$ have more power than the
additional-mode pulsation.  The only detectable harmonic of $f_A$ was the first
one at 2$f_A$=7.36738\,d$^{-1}$ (amplitude 1.8$\pm$0.2\,mmag). The peak at the
next harmonic, at 3$f_A$=11.05\,d$^{-1}$, is hidden in the noise and cannot be
detected.  Also prominent in Fig.\,6(b) are many frequencies that are linear
combinations of $f_A$ and $f_D$.  Their presence argues strongly that both
frequencies  originate in EPIC\,216764000 and do not result from blending with
another star.  The {\it K2} thruster-firing frequency near $f_{\rm
th}$=4.08\,d$^{-1}$ (amplitude $\sim$3\,mmag) is present but coincides with the
combination frequency at $3f_D$--$f_A$ = 4.07\,d$^{-1}$, making measurement of
amplitudes and phases of both frequencies highly unreliable.

Figs.\,6(c)-(f) show close-up views of the amplitude spectra in the vicinity of
$f_D$ and $f_A$.  Fig.\,6(d) reveals residual power after prewhitening with
$f_D$ and its harmonics. In general, amplitude and frequency modulations are
revealed by sidelobes separated from the main peak by a frequency difference
($\Delta$f) equal to the reciprocal of the Blazhko period. The sidelobes may be
placed symmetrically on the opposite sides of the main peak (equidistant
triplets), on one side only (doublets), or occur in multiplets. However in this
case the residual peaks with amplitudes $\sim$5\,mmag, and the next set of
sidelobes with amplitudes at $\sim$3 and 2\,mmag  are not well resolved from
$f_D$. They correspond to the wings of the broadened $f_D$ left after imperfect
prewhitening, i.e., they are related to non-stationarity of $f_D$.  On the
other hand, Fig.\,6(f) shows no significant signal left after prewhitening with
$f_A$ and its harmonic(s), thus there is no evidence in the Fourier transform
for non-stationarity of $f_A$.  Time variations of the amplitudes and
frequencies of both $f_D$ and $f_A$ are investigated further in $\S$3.1.3
below.

% \footnote{Our best \texttt{FNPEAK} solution included 46 fitting
% frequencies and required an additional 19 frequencies to remove
% residual power caused by unstable components requiring more than one
% sine function to be prewhitened.}

Fig.\,6(g) shows the window function for the {\it K2} photometry,
for frequencies out to the long-cadence Nyquist frequency at
24.5\,d$^{-1}$.  The constant amplitude spikes at 4.0746, 8.1573,
12.2351, 16.3144 and 20.4996\,d$^{-1}$ are frequencies associated
with the {\it K2} telescope thruster firings every six hours. The
inset shows the central peak, with measured FWHM=0.0148\,d$^{-1}$
(only slightly larger than the Rayleigh criterion
$f_R$=0.0123\,d$^{-1}$), and its close sidelobes (labelled with
frequency offsets).

% see C7p890-891 for Rij,phi_i1 values (with errors)

% Section 3.1.2
\subsubsection{Light curves and Fourier parameters}

Observed and fitted light curves for EPIC\,216764000 are plotted in
Figure~7. The top panel shows the observed {\it Kp} light
curve phased with the period 0.386748\,d, confirming the asymmetric
shape seen by Uitterdijk (1949) and by the \texttt{ASAS} survey (top
panel of Fig.\,5), and significantly improving upon the definition
of the shape.  Most of the light variation occurs near maximum
brightness. There are no obvious {\it humps} on the rise to maximum
light, {\it bumps} near minimum light (see Christy 1966;
Gillett \& Crowe 1988; Chadid et al. 2014), or other such features.
The cycle-to-cycle light curve shapes predicted by the two-frequency
model (Eqn.\,1) are compared with the observed photometry for the
first couple of days of observations in Fig.\,7(b). From this figure
(and Fig.\,3) it is evident that the total amplitude (i.e.,
min-to-max), $A_{\rm tot}$, and to a lesser extent the risetime, RT
(see Sandage 2004), change from cycle to cycle. In general the
scatter about the fitted light curve is small, $\lesssim$10\,mmag.
%  See C7p739 for P04 analysis that gives residual 0.0033\,mag

% \subsubsection{Fourier parameters}

% see C11p1169-1172 for Prudil et al. Fig. 3 (Atot,R21,R31,PHI21,PHI31 versus Period graphs)

% Prudil et al. abstract:   "10 stars from our sample exhibit equidistant peaks
% in the frequency spectrum, which suggests the Blazhko-type modulation of the
% main pulsation frequency and/or additional periodicity. The Fourier
% coefficients R21 and R31 are some of the lowest among fundamental-mode RR~Lyrae
% stars, but among the highest for the first-overtone pulsators. For the phase
% Fourier coefficients phi21 and phi31, our stars lie between RRab and RRc stars. The
% stars discussed were compared with radial linear pulsation models. Their
% position in the Petersen diagram cannot be reproduced by assuming that two
% radial modes are excited and their physical parameters are like those
% characteristic of RR~Lyrae stars."

% If the light curve is stationary then single numbers suffice for each
% parameter; but when there are significant variations over time, either
% frequency modulation (FM) or amplitude modulation (AM), time-dependent
% functions are needed to describe the cycle-to-cycle changes (see, for example,
% Figs.\,2 and 3 of Nemec et al. 2013).

Fourier decomposition parameters, mean risetimes and total amplitudes for
EPIC\,216764000 are given in column (2) of Table~4, where the quantities for
the dominant and additional pulsation modes are identified by the subscripts
`D' and `A'.  The pulsation period of the dominant mode, $P_D$=0.387\,d, and
the asymmetric light curve (characterized by the short risetime,
RT$_D$=0.235$\pm$0.001), are both consistent with fundamental-mode pulsation in
a metal-rich RR~Lyrae star (see fig.\,4 of Nemec et al.  2013).  The short
period for the additional mode, $P_A$=0.271\,d, and the long risetime,
RT$_A$=0.472$\pm$0.001 (which is only slightly smaller than 0.500 for a pure
sinusoid), are both consistent with first-overtone pulsation. The ratio of the
total amplitudes, $A_{{\rm tot},D}$/$A_{{\rm tot},A}$=8.89$\pm$0.13, is
significantly larger than the ratio derived using the first term of the Fourier
decomposition, $A_{1,D}$/$A_{1,A}$=7.55$\pm$0.09 (both based on {\it Kp}
magnitudes).

%  TABLE 4 -  see C7p890 for hand calculations;see C7p832 for confirmation
%  For OGLE values see pp.xxxxxx
\begin{table}
{
\fontsize{6}{7.2}\selectfont  %  <--- works and better; 2nd (skip) should be 1.2 times 1st (size)
\fontsize{5}{6.0}\selectfont  %  <--- works and better; 2nd (skip) should be 1.2 times 1st (size)
% \begin{threeparttable}
%\fontsize{6}{7.2}\selectfont  %  <--- works and better; 2nd (skip) should be 1.2 times 1st (size)

\caption{Fourier decomposition parameters, risetimes and total
amplitudes for both modes of the four pRRd stars observed by {\it
K2} and by the \texttt{OGLE} survey.  The amplitude units are
[mmag]. Note: $\phi^c_{\rm 21}$ = $\phi^s_{\rm 21}$ + $\pi$/2, and
$\phi^c_{\rm 31}$ = $\phi^s_{\rm 31}$ -- $\pi$.  }

% For ease of comparison with the parameters derived from the {\it K2}
% photometry both sine-series (superscript `s') and cosine-series (superscript
% `c') phase parameters are given.

\begin{tabular}{lcccc}
\toprule
\multicolumn{1}{l}{\texttt{EPIC} Id:} &  \multicolumn{1}{c}{216764000} &   \multicolumn{1}{c}{235839761} &   \multicolumn{1}{c}{251248823} &   \multicolumn{1}{c}{251248824}  \\
\multicolumn{1}{l}{{\it Kp}\,[mag]:}  &        13.74                   &                  17.59          &               18.83             &             19.84                \\
\multicolumn{1}{l}{\texttt{OGLE} Id:} &    --                          &              \texttt{16999}     &           \texttt{00595}        &          \texttt{19121}          \\
\multicolumn{1}{c}{(1)}     &  \multicolumn{1}{c}{(2)}         &   \multicolumn{1}{c}{(3)}       &   \multicolumn{1}{c}{(4)}       &   \multicolumn{1}{c}{(5)}        \\
\midrule
%                            216764000             235839761        251248823         251248824
%                           C7p890,939,1134-9     C11p962,1116      C11p992,1263         C11V6pxxxx
    \multicolumn{5}{c}{(a) {\it K2}-photometry, {\it Kp}-passband (2016) }    \\ [0.05cm]
$f_D$\,[d$^{-1}$]         &  2.585663($\pm$3)  & 3.069166($\pm$4)& 3.548224($\pm$3)& 2.796445($\pm$2)\\
$P_D$\,[d]                &  0.386748($\pm$1)  & 0.325821($\pm$1)& 0.281831($\pm$1)& 0.357597($\pm$1)\\  % see C11p1116 for 1st, p1109 for 2nd;  p1112 for 3rd,
RT$_D$                    &  0.235$\pm$0.001   & 0.279$\pm$0.001 & 0.289$\pm$0.001 & 0.277$\pm$0.001 \\   % see C7p939-940; C11p1050-1052
$A_{\rm tot,D}$           & \!\!\!\!\!\!676$\pm$1 & \!\!\!\!\!\!627$\pm$3 & \!\!\!\!\!\!548$\pm$5 & \!\!\!\!\!\!771$\pm$4       \\      % C7p939 and 959;  C11p1050-52
$A_{1,D}$                 & \!\!\!\!\!\!285.40$\pm$0.20 & \!\!\!\!\!\!284.3$\pm$1.5 & \!\!\!\!\!\!237.7$\pm$2.6 & \!\!\!\!\!\!345.7$\pm$2.2   \\
$R_{\rm 21,D}$            &  0.398$\pm$0.001   & 0.350$\pm$0.008 & 0.370$\pm$0.013 & 0.324$\pm$0.008 \\
$R_{\rm 31,D}$            &  0.193$\pm$0.001   & 0.145$\pm$0.006 & 0.169$\pm$0.012 & 0.121$\pm$0.007 \\
$R_{\rm 41,D}$            &  0.094$\pm$0.003   & 0.064$\pm$0.006 & 0.083$\pm$0.011 & 0.050$\pm$0.007 \\
$R_{\rm 51,D}$            &  0.045$\pm$0.002   & 0.021$\pm$0.006 & 0.032$\pm$0.011 & 0.014$\pm$0.006 \\
$R_{\rm 61,D}$            &  0.018$\pm$0.001   & 0.013$\pm$0.006 & 0.023$\pm$0.011 &   \dots         \\
% $R_{\rm 71,D}$          &  0.005$\pm$0.001   &  \dots          &   \dots         &  \dots          \\
$\phi^s_{\rm 21,D}$       &  2.743$\pm$0.002   & 2.81$\pm$0.03   &  2.57$\pm$0.05  &  3.04$\pm$0.03  \\
$\phi^s_{\rm 31,D}$       &  5.561$\pm$0.004   & 5.62$\pm$0.06   &  5.20$\pm$0.10  &  5.93$\pm$0.06  \\
$\phi^s_{\rm 41,D}$       &  1.994$\pm$0.001   & 2.21$\pm$0.11   &  1.74$\pm$0.18  &  2.64$\pm$0.15  \\
$\phi^s_{\rm 51,D}$       &  4.937$\pm$0.020   & 5.13$\pm$0.29   &  4.21$\pm$0.39  &  5.86$\pm$0.49  \\ [0.2cm]
% $\phi^s_{\rm 61,D}$     &  1.63$\pm$0.04     & 1.51$\pm$0.48   &  0.62$\pm$0.53  &   \dots         \\ [0.1cm]
% $\phi^s_{\rm 71,D}$     &  4.46$\pm$0.13     &  \dots          &  \dots          &   \dots         \\
%                             C7p979,1137            C11V6p962
%                                                    C11p1606
% \multicolumn{5}{c}{(b) `Additional' mode ({\it K2}, {\it Kp}-passband, 2016) }  \\ [0.05cm]
$f_A$\,[d$^{-1}$]         &  3.68368($\pm$3)   & 4.32216($\pm$13)&  5.00709         & 4.02916         \\
$P_A$\,[d]                &  0.271468($\pm$2)  & 0.231366($\pm$7)&  0.199717        & 0.248191        \\
RT$_A$                    &  0.472$\pm$0.001   & 0.436$\pm$0.002 &  0.421$\pm$0.002 & 0.439$\pm$0.002 \\
$A_{\rm tot,A}$           & \!\!\!76$\pm$2 & \!\!\!\!\!\!225$\pm$3 & \!\!\!\!\!\!147$\pm$5 & \!\!\!\!\!\!198$\pm$4     \\
$A_{1,A}$                 & \!\!\!37.81$\pm$0.20 & \!\!\!\!\!\!114.1$\pm$1.5 & \!\!\!69.2$\pm$2.6 & \!\!\!97.4$\pm$2.2   \\
$R_{\rm 21,A}$            &  0.048$\pm$0.005   & 0.12$\pm$0.02   &  0.15$\pm$0.04   & 0.07$\pm$0.02   \\
$\phi^s_{\rm 21,A}$       &  2.85$\pm$0.11     & 2.71$\pm$0.35   &  3.05$\pm$0.32   &  2.28$\pm$0.37  \\
% $R_{\rm 31,A}$          &     \dots          &   \dots         &    \dots         &   \dots         \\
% $\phi^s_{\rm 31,A}$     &  1.0$\pm$2.0       &   5$\pm$4       &  2.6$\pm$2.6     &   \dots         \\
% $R_{\rm 41,A}$          &    \dots           &    \dots        &    \dots         &   \dots         \\
% $\phi^s_{\rm 41,A}$     &    \dots           &   4$\pm$6       &  4.6$\pm$1.3     &  \dots          \\
\\
%                                                 C11p702,1678           C11p698           C11p700
%                                                   16999            00595            19121
\multicolumn{5}{c}{(b) \texttt{OGLE-IV}-photometry, $I$-passband (2010-2017) }  \\ [0.05cm]
% $f_D$\,[d$^{-1}$]         &    \dots          &  3.0691662      & 3.5482250       & 2.7964446       \\  % calculated from the period given on C11p1211
% $P_D$\,[d]                &   \dots           &  0.32582139     & 0.28183106      & 0.35759693      \\  % from C11p1211
% $\epsilon$($P_D$)         &   \dots           &  0.00000042     & 0.00000019      & 0.00000028      \\  % from C11p1211
$A_{\rm tot,D}$             &  \dots            &      446        &    265          &   360           \\  % from C11p1211, from OGLE on-line data directory
%$\Delta$$A_{\rm tot,D}$    &  \dots            &      --181      &    --283       &   --411         \\  % Above value minus value on C11p1211;  avg=-292 mmag
% Atot(I) 2010-2013         &                   &      466        &                 &                 \\  %  see C11p1154, on-line catalog, 2010-13
%$P_D$\,[d]                 &  \dots            &  0.32582139     & 0.28183106      & 0.35759693      \\
%$R_{\rm 21}$($I$)2010-2013 &  \dots            &    0.327        &   0.482         &  0.360          \\  % see C11p1154
$R_{\rm 21,D}$              &  \dots            & 0.39$\pm$0.03   & 0.39$\pm$0.01   &  0.34$\pm$0.02  \\  % see C11p1211, C11p1407;  C11p1290; C11p1175
%$\Delta$$R_{\rm 21,D}$     &  \dots            & 0.04$\pm$0.04   & 0.02$\pm$0.03  &  0.00$\pm$0.03  \\  %  avg of absolute values = 0.022;  OGLE - K2
%$\phi^c_{\rm 21}$($I$)2010-2013  &  \dots        &    4.520        &   4.552         &  4.607          \\ % see C11p1154
%$\phi^s_{\rm 21}$($I$)     &  \dots            &    6.091        &   6.122         &  6.178          \\
%$\phi^c_{\rm 21,D}$        &  \dots            &    4.553        &   4.450         &  4.598          \\ %  see C11p1211
$\phi^s_{\rm 21,D}$         &  \dots            &  2.99$\pm$0.10  &  2.91$\pm$0.03  &  3.01$\pm$0.06  \\
%$\Delta$$\phi^s_{\rm 21,D}$ &  \dots            &    0.17        &   0.31          &  --0.01        \\ %  avg of absolute values =
%$R_{\rm 31}$($I$)2010-2013 &  \dots            &    0.197        &   0.258         &  0.158          \\ % see C11p1154
$R_{\rm 31,D}$              &  \dots            & 0.19$\pm$0.03   & 0.18$\pm$0.01   &  0.14$\pm$0.02  \\ % see C11p1211; C11p1407; 11p1636
%$\Delta$$R_{\rm 31,D}$      &  \dots            & 0.04$\pm$0.04   & 0.03$\pm$0.03  &  0.01$\pm$0.03  \\
% $\phi^c_{\rm 31}$         &  \dots            &    2.571        &   2.663         &  2.851          \\
% $\phi^s_{\rm 31}$         &  \dots            &    5.713        &   5.805         &  5.993          \\
%$\phi^c_{\rm 31,D}$        &  \dots            &    2.631        &   2.346         &  2.876          \\ % see C11p1211
$\phi^s_{\rm 31,D}$         &  \dots            &  5.92$\pm$0.22  &   5.55$\pm$0.06 &  5.92$\pm$0.15  \\
%$\Delta$$\phi^s_{\rm 31,D}$ &  \dots           &    0.15         &   0.29          &  0.09           \\ [0.1cm]
%
$R_{\rm 21,A}$              &  \dots            &   0.13$\pm$0.10 & 0.13$\pm$0.02   &  0.04$\pm$0.06  \\  % see C11p1635,
%$\Delta$$R_{\rm 21,A}$      &  \dots           &   0.01$\pm$0.12 & 0.x2$\pm$0.03   &  x.00$\pm$0.03  \\  %
$\phi^s_{\rm 21,A}$         &  \dots            &   2.33$\pm$0.88 & 3.14$\pm$0.20   &  1.9$\pm$1.4          \\  [0.1cm] %
%$\Delta$$\phi^s_{\rm 21,A}$ &  \dots           & --0.38$\pm$1.23 &   x.31          &   x.01          \\  %  avg of absolute values =

%                                                 C11p1678            C11p698,1679       C11p700,1680
%                                                   16999               00595               19121
\multicolumn{5}{c}{(c) Differences (in the sense $I$--{\it Kp}) }  \\ [0.05cm]
$\Delta$$A_{\rm tot,D}$     &  \dots            &      --181      &    --283         &     --411        \\  % Above value minus value on C11p1211;  avg=-292 mmag
$\Delta$$R_{\rm 21,D}$      &  \dots            &   0.04$\pm$0.03 &   0.02$\pm$0.02  &   0.02$\pm$0.02  \\  %  avg of absolute values = 0.022;  OGLE - K2
$\Delta$$\phi^s_{\rm 21,D}$ &  \dots            &   0.18$\pm$0.10 &   0.34$\pm$0.06  & \!\!\!--0.03$\pm$0.07  \\  %  avg of absolute values =
$\Delta$$R_{\rm 31,D}$      &  \dots            &   0.04$\pm$0.03 &   0.01$\pm$0.02  &   0.02$\pm$0.02  \\  %
$\Delta$$\phi^s_{\rm 31,D}$ &  \dots            &   0.30$\pm$0.23 &   0.35$\pm$0.12  & \!\!\!--0.01$\pm$0.16  \\  %
$\Delta$$R_{\rm 21,A}$      &  \dots            &   0.01$\pm$0.10 & \!\!\!--0.02$\pm$0.04 & \!\!\!--0.03$\pm$0.06  \\  %
$\Delta$$\phi^s_{\rm 21,A}$ &  \dots            & \!\!\!--0.4$\pm$0.9 &   0.09$\pm$0.38 & \!\!\!--0.4$\pm$1.4   \\  %  avg of absolute values =

\bottomrule
\end{tabular}
% \begin{tablenotes}
% {\fontsize{6}{7.2}\selectfont  %  <--- works and better; 2nd (skip) should be 1.2 times 1st (size
% \item Notes: see text for definitions of $R_{\rm i1}$ and $\phi^s_{\rm i1}$, and for transformations
% between Fourier sine (s) and cosine (c) series.   }
% \end{tablenotes}
% \end{threeparttable}
}
\end{table}

% are similar for the four stars, with mean risetimes RT($D$)=0.270$\pm$0.010 and
% RT($A$)=0.442$\pm$0.009, and mean total amplitudes $A_{\rm
% tot}$($D$)=660$\pm$40\,mmag and $A_{\rm tot}$($A$)=160$\pm$30\,mmag, the latter
% ranging from 76 to 225\,mmag.  Other averages for the dominant mode include:
% $R_{\rm 21}$($D$)=0.366$\pm$0.013, $R_{\rm 31}$($D$)=0.157$\pm$0.013,
% $\phi_{\rm 21}$($D$)=2.79$\pm$0.09, and $\phi_{\rm 31}$($D$)=5.58$\pm$0.13.

% For the additional modes the mean values are:  $R_{\rm
% 21}$($A$)=0.098$\pm$0.021, $R_{\rm 31}$($A$)=0.015$\pm$0.004 and $\phi^s_{\rm
% 21}$($A$)=2.73$\pm$0.14; the mean $\phi^s_{\rm 31}$($A$) is too uncertain to
% quote.

Disentangled light curves for the two modes of EPIC\,216764000 are plotted in
the top row of Figure~8. The standard deviations of the residuals for the two
modes are 8.0 and 2.5\,mmag, respectively.  The light curve for the dominant
mode (left panel) was produced by subtracting from the time-series data all the
Fourier components (Eqn.1, Table~3) related to the additional mode (n$f_A$) and
all the components corresponding to the frequency combinations
(m$f_D$$\pm$n$f_A$), leaving only components related to the fundamental mode,
which were then phased with the period of the dominant mode. Because of
non-stationarity of both modes (see Fig.\,10), the subtraction of unwanted
Fourier components was performed with the time-dependent prewhitening method
(Moskalik et al. 2015). The light curve of the additional mode (right panel)
was produced in a similar way: all components related to the dominant mode
(m$f_D$) and all components corresponding to the frequency combinations
(m$f_D$$\pm$n$f_A$) were subtracted, and the results phased with the additional
period.

% FIGURE 8 - Original from Pawel  See C11p698-702 for Web page information.  See
% C11V5p711+ for how Pawel made the light curves
% See C7V5p973 for top graph;  see C7V5p979
\begin{figure}
\begin{center}
\begin{overpic}[width=8.4cm]{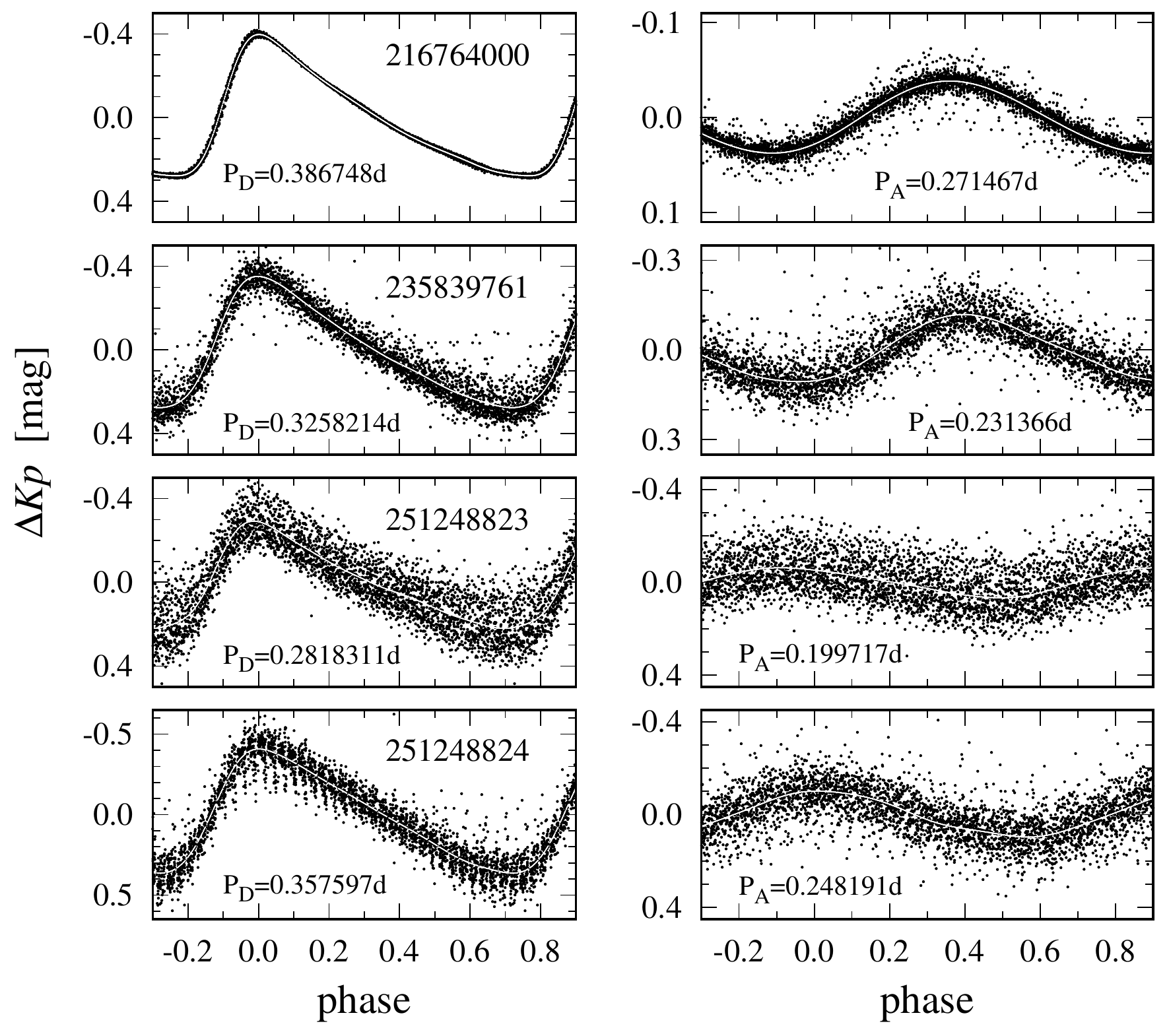}   \end{overpic}  % renamed from "PhasedLC-2019Sept1.pdf" C7V5p579
% \begin{overpic}[width=8.0cm]{2019Apr18_f222.pdf} \end{overpic}
\end{center}

\caption{Phased light curves for EPIC\,216764000 and the three fainter pRRd
stars, revealing the shapes of the dominant and additional modes. Note the
different vertical scales for the two modes and for the different stars. The
dominant and additional pulsation periods ($P_D$ and $P_A$) are given on each
panel.  The adopted zero-points (for the phases) were $t_0$ = 2454833 +
2468.93357 = 2457301.9335 for EPIC\,216764000, and the \texttt{OGLE-IV} $t_0$
values for the three Campaign\,11 stars (see Table~2). }

\label{fig:PawelLCs}
\end{figure}

% Section 3.1.3
\subsubsection{Amplitude and Frequency Variations}

Despite the {\it K2} observations of EPIC\,216764000 spanning only 82.6\,d, a
search was made for possible frequency (or phase) and amplitude variations.  In
Figure~9 the residual magnitudes (observed minus fitted) are plotted, where the
fitted magnitude values were calculated by substituting into Eqn.~1 the
parameter estimates given in Table~3. The residuals are clearly  not normally
distributed but appear to be of two types: low amplitude (from --10 to
+10\,mmag), and higher amplitude (from --40 to +40\,mmag), the latter starting
out negative, turning positive between $t$=2501 and 2531, and then switching
back to negative. If the up-down variations are periodic then the period is
$\sim$60\,d. The large-amplitude spikes are well resolved (with 3-4
observations per spike) and  periodic with a frequency close to $f_D$.  The
spikes are reminiscent of fig.\,2 of Kov\'acs (2018b) which shows a similar
pattern when the assumed period was purposely set to a value slightly larger
than the actual value.

Fig.\,9 reveals that the dominant pulsation period is unstable and that $P_D$
varies back and forth, which is equivalent to saying that the pulsation phase
$\phi_{1,D}$ varies back and forth (see Fig.\,10). Because the ascending branch
of the light curve (see top left panel of Fig.\,8) is much steeper than the
descending branch, any phase difference between observed pulsation and the
ephemeris (i.e., a horizontal shift) will correspond to a vertical difference
which is much larger on the ascending than the descending branch. If the
observed maximum happens earlier than the predicted maximum then we will
measure large positive residuals (on the ascending branch) and small negative
residuals (on the descending branch). It will be the other way around if the
observed maximum comes later than the predicted one. This also explains why the
spikes repeat with a period close to $P_D$.

The fact that the low-amplitude residuals tend to be negative when the
high-amplitude residuals are positive is simply a consequence of the asymmetric
light curve. The ascending branch of EPIC\,216764000 lasts $\sim$0.2$P_D$, and
the descending branch lasts $\sim$0.8$P_D$.  Thus, the ascending branch is
roughly four times steeper than the descending branch. Consequently we can
expect that the high-amplitude residuals (coming from the ascending branch)
should be roughly four times larger than the low-amplitude residuals. And this
is exactly what is seen.

% FIGURE 9
% See C7V4p749, C7V5pp785;  EAP pipeline; label fixed May19,2021
\begin{figure}
\begin{center}
\begin{overpic}[width=8cm]{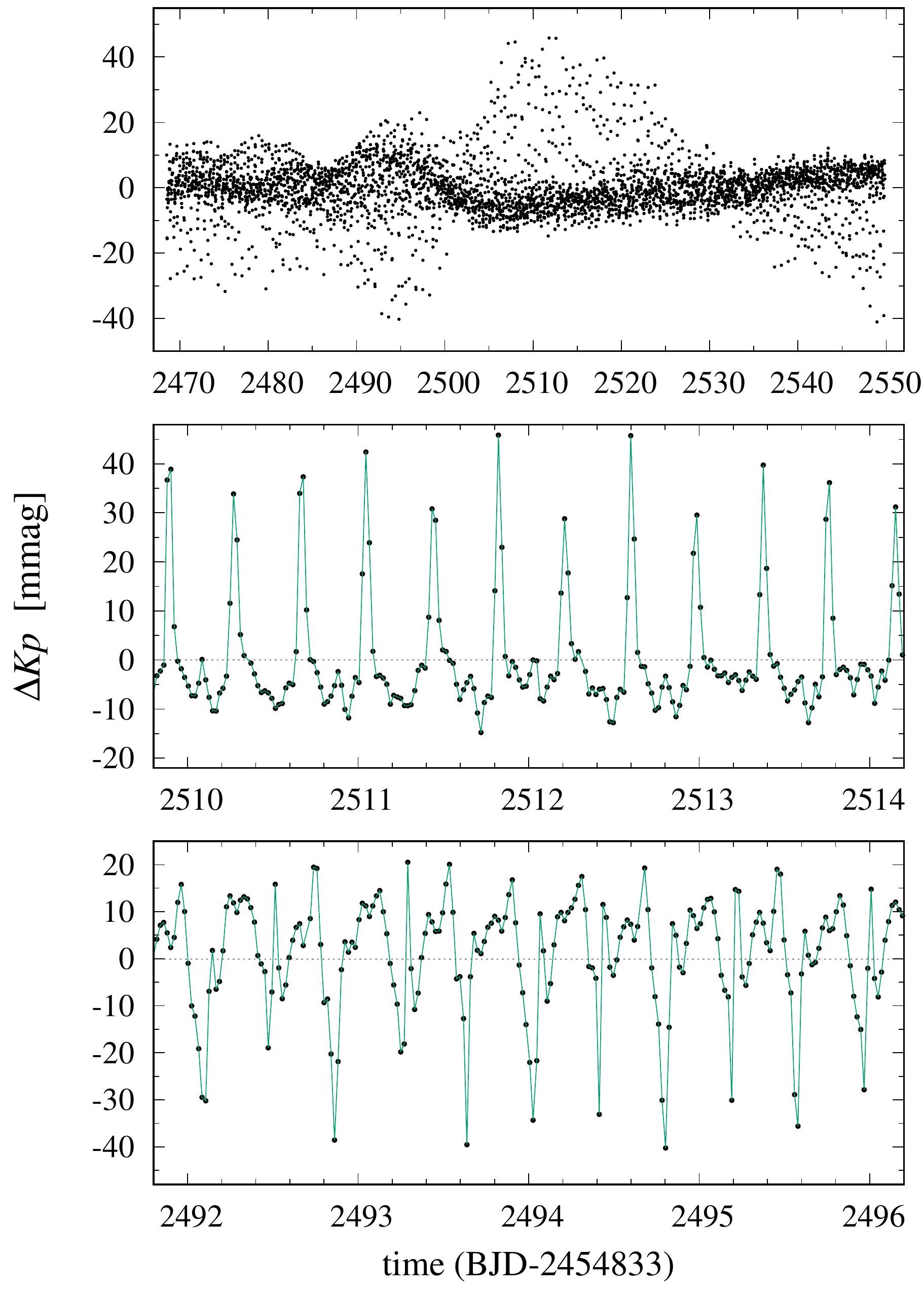} \put(13,94){(a)} \put(13,64){(b)} \put(13,32){(c)} \end{overpic} % renamed from "resids-PAM-EAP-2freq.pdf"
\end{center}
\vskip0.2cm

\caption{EPIC\,216764000: residuals after prewhitening with
$f_D$=1/0.386748\,d$^{-1}$, $f_A$=1/0.271468\,d$^{-1}$, their harmonics and
linear combination frequencies. (a) Residuals over the entire 82.6-d interval
of the {\it K2} data. (b) Close-up of a 4-d interval when the large-amplitude
residuals are positive. (c) Close-up of a 4-d interval when the large-amplitude
residuals are negative. The large-amplitude spikes are resolved (with 3-5
observations per spike) and have a period close to $P_D$.  }

\label{fig:216764000_residuals}
\end{figure}

% FIGURE 10
% For top 4 panels see C7V5p952
% See C7V5p953-961 for source of PDM2 analysis of Everest data (bottom two plots)';  labels fixed May19,2021  C7p1185-86
\begin{figure}
\begin{center}
\begin{overpic}[width=7.5cm]{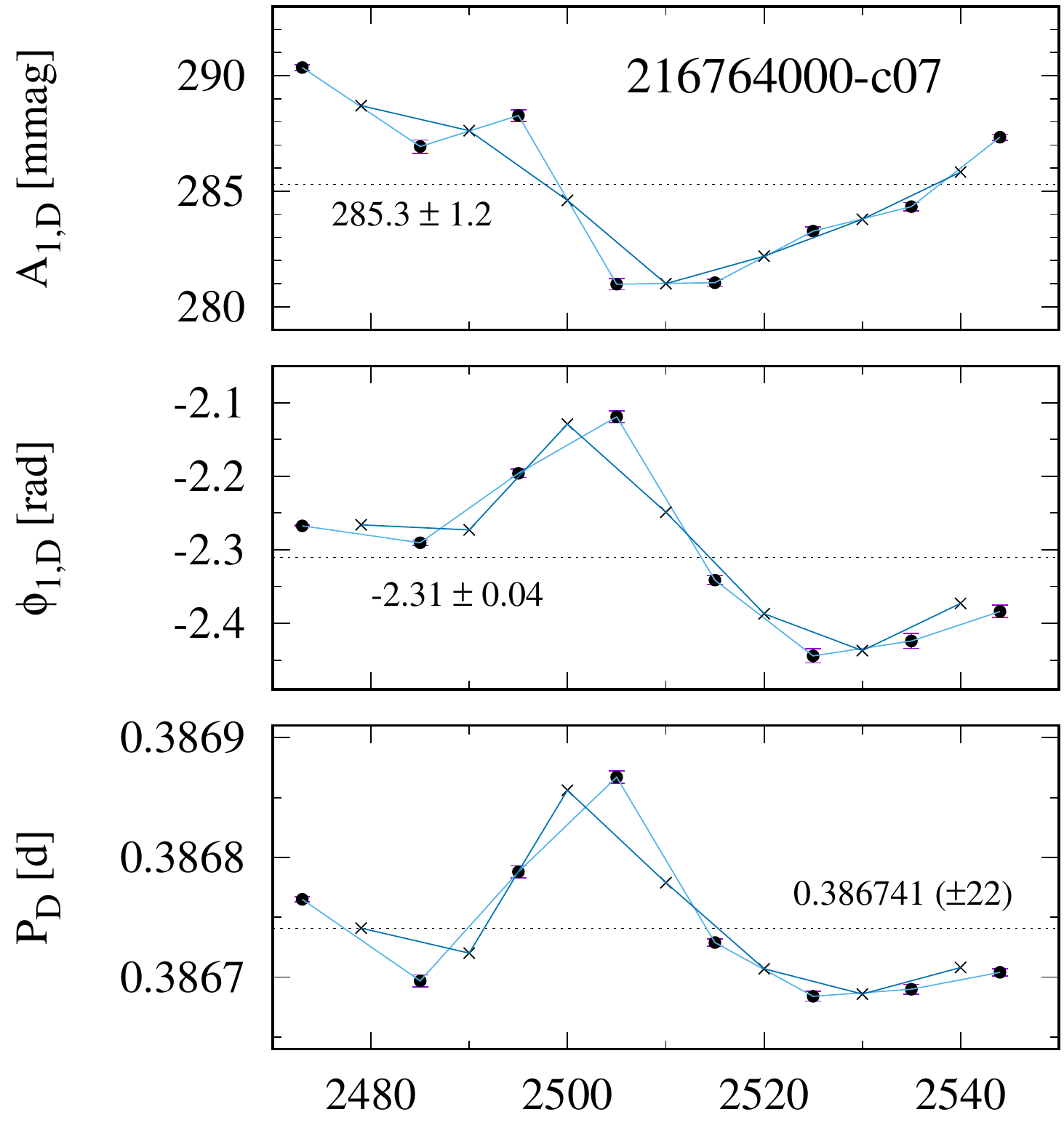}   \end{overpic}  % renamed from "216764000-10dayBins-2020July7_D.pdf"
\begin{overpic}[width=7.5cm]{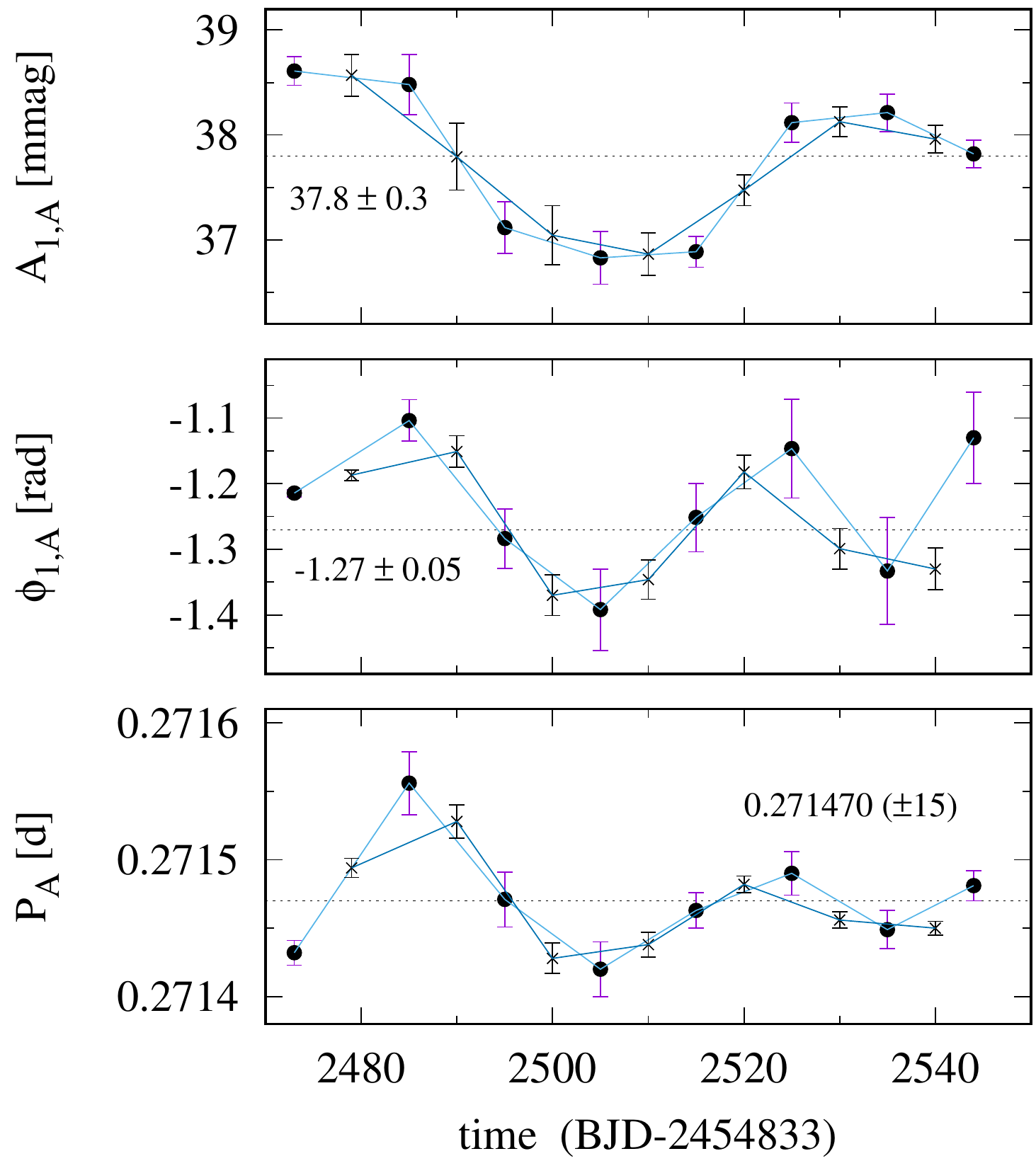}   \end{overpic}  % renamed from "216764000-10dayBins-2020July7_A.pdf"
\end{center}

\caption{Time variations of the periods and Fourier first-term
amplitudes and phases for the dominant (D) and additional (A)
pulsation modes for EPIC\,216764000.  Clearly, both modes exhibit
amplitude, phase and period variations.}

\label{fig:TimeVariations}
\end{figure}

% see C7p1155-1166 for FMMv2 solutions
% see C7p959 for PDM2 results

To investigate further the variations of the residual magnitudes seen in
Fig.\,9 a time-dependent analysis of the {\it K2} photometry was performed, the
results of which are graphically summarized in Figure~10. The data were
subdivided into bins and for each bin Fourier first-term amplitudes ($A_1$) and
phases ($\phi_1$) were calculated for the dominant and additional frequencies
(eight harmonics for the dominant mode and three harmonics for the additional
mode) and for the linear combination frequencies. The analysis was performed
with our Fourier least-squares (Levenberg--Marquardt) multi-frequency fitting
program, taking as starting values the $P_D$ and $P_A$ values derived for the
entire data set (see Table~3). Owing to the high precision (mmag), relatively
short cadence (30\,min), long baseline (81.8\,d) and large number of {\it K2}
brightness measurements (N$\sim$3700), it was also possible to estimate {\it
directly} the time variations of $P_D$ and $P_A$.

In Fig.\,10 the upper three panels show the time variations for the dominant
mode, and the lower three panels show the variations for the additional mode.
In each panel two sets of points are plotted: (a) the solid dots were derived
by dividing the photometry into eight non-overlapping 10-day-wide bins and
calculating the $A_1$, $\phi_1$ and $P$ values for each bin, the (labelled)
horizontal line indicating the mean value of the eight bins.  Each bin
contained $\sim$480 points, and the standard deviations of the residuals of the
light curve fits ranged from 4.5 to 2.0\,mmag.  For both modes the mean  $P$,
$A$ and $\phi$ values agree well with the values derived from the combined
81.8\,d data set. The $P_D$ value for each bin also was derived using the
non-parametric phase dispersion method, specifically, \texttt{PDM2}
(Stellingwerf 1978, 2011), and the periods were found to be very similar to the
values derived using our multi-frequency program; (b) the crosses were derived
using seven overlapping 20-day-wide bins with 10-day offsets. Clearly both
modes show amplitude, phase and period variations, with the ranges of the
variations exceeding the uncertainties of the individual measurements (as
indicated by the error bars), indicating that the variations are almost
certainly real and are the cause of the residuals shown in Fig.\,9.

% See C17p741-751 for Chadid et al. (2010) paper on V1127 Aql.
As expected the pulsation periods and phases are related for both
modes. For the dominant mode the periods and phases are {\it
anticorrelated} with the amplitudes, the smallest $A_{\rm 1,D}$
values occuring shortly after the longest $P_D$ and greatest
$\phi_{\rm 1,D}$ values. For the additional mode all three
quantities are {\it correlated} for the first 40 days, the smallest
$A_{\rm 1,A}$ values occurring for the shortest $P_A$ periods, but
then the correlation seems to break down for the last 20 days. An
anticorrelation of the dominant period and amplitude was also seen
for V1127\,Aql (see fig.\,12 of Chadid et al. 2010), a Blazhko star
that we believe to be another pRRd star (see $\S$4.1 below).

We cannot exclude that the observed variations of the pulsation modes in
EPIC\,216764000 might be periodic and that the star is undergoing a slow
Blazhko modulation (like V1127\,Aql). However, because the {\it K2} time-base
is only 82\,d, our data are insufficient to make such claims. What can be
concluded is that the data collected so far are consistent with non-repetitive
(irregular) phase and amplitude variability.

\subsubsection{Long-term changes of the dominant period?}

The four main photometric data sets for EPIC\,216764000 (Johannesburg plates
measured by Uitterdijk, \texttt{ASAS-3}, \texttt{SuperWASP} and {\it K2} -- see
Table~1) were acquired over more than 80 years, which usually is sufficient for
deriving a period change rate, d$P$/d$t$.  Unfortunately, the $\sim$65-yr
gap between the Johannesburg photographic plates (1928-1937) measured by
Uitterdijk and the \texttt{ASAS-3} (2001-2009) and \texttt{SuperWASP}
(2006-2008) photometry makes keeping track of pulsation cycles almost
impossible.  When only the \texttt{ASAS-3}, \texttt{SuperWASP} and {\it K2}
observations  are considered, the dominant period  appears to be increasing,
the estimated periods for the three epochs being 0.386708\,d, 0.386718\,d and
0.386748\,d, respectively.  Adopting these periods, and their respective epochs
of maximum light (see Table~1), and assuming a linearly increasing period gives
an estimated d$P$/d$t$ = 7.75$\times$10$^{-9}$\,d\,d$^{-1}$ =
2.83\,d\,Myr$^{-1}$.  Disappointingly, this period change rate gives a
predicted period for the 1930's that is considerably smaller than the period
derived by Uitterdijk (which is mid-way between the \texttt{ASAS-3} and
\texttt{SuperWASP} values). One possible explanation for this high rate of
period change, which is three orders of magnitude too fast to be of
evolutionary origin, is that the period may have irregular variations such as
were observed by Le\,Borgne et al. (2007) for many RRab stars (see their
fig.\,4).  Or possibly the period change rate is non-linear. Almost
certainly, the period is not constant.

% Section 3.2
\subsection{Campaign\,11 Stars}

The three pRRd stars observed during Campaign\,11 have mean {\it Kp}
magnitudes ranging from 17.59 to 19.84\,mag (see $\S$2.2) and are
considerably fainter than EPIC\,216764000.  As noted above, all
three stars were extensively observed by the \texttt{OGLE-IV} survey
and are among the 42 `peculiar' RRd stars discussed by P17.
Because the results of the frequency analyses of the {\it K2} data
are similar for the three stars they will be discussed as a group.

% Section 3.2.1
\subsubsection{Frequencies, Amplitudes and Phases}

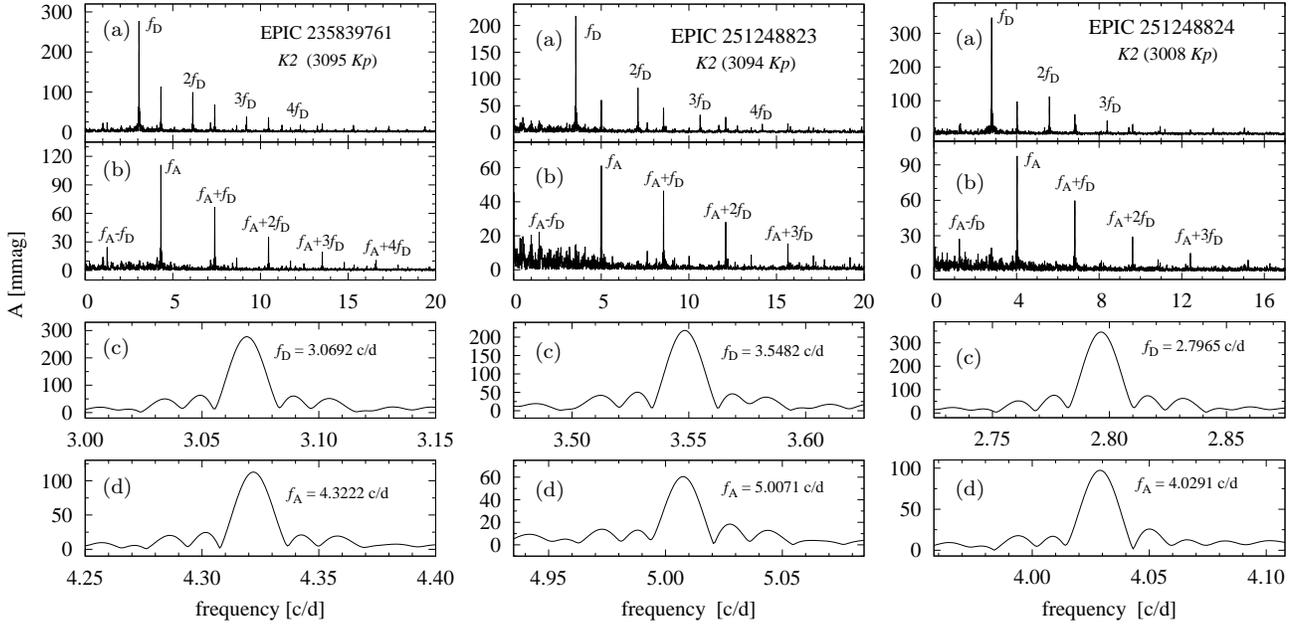
\begin{figure*}
\begin{center}
% see C11p748 for source files
\begin{overpic}[width=6.11cm]{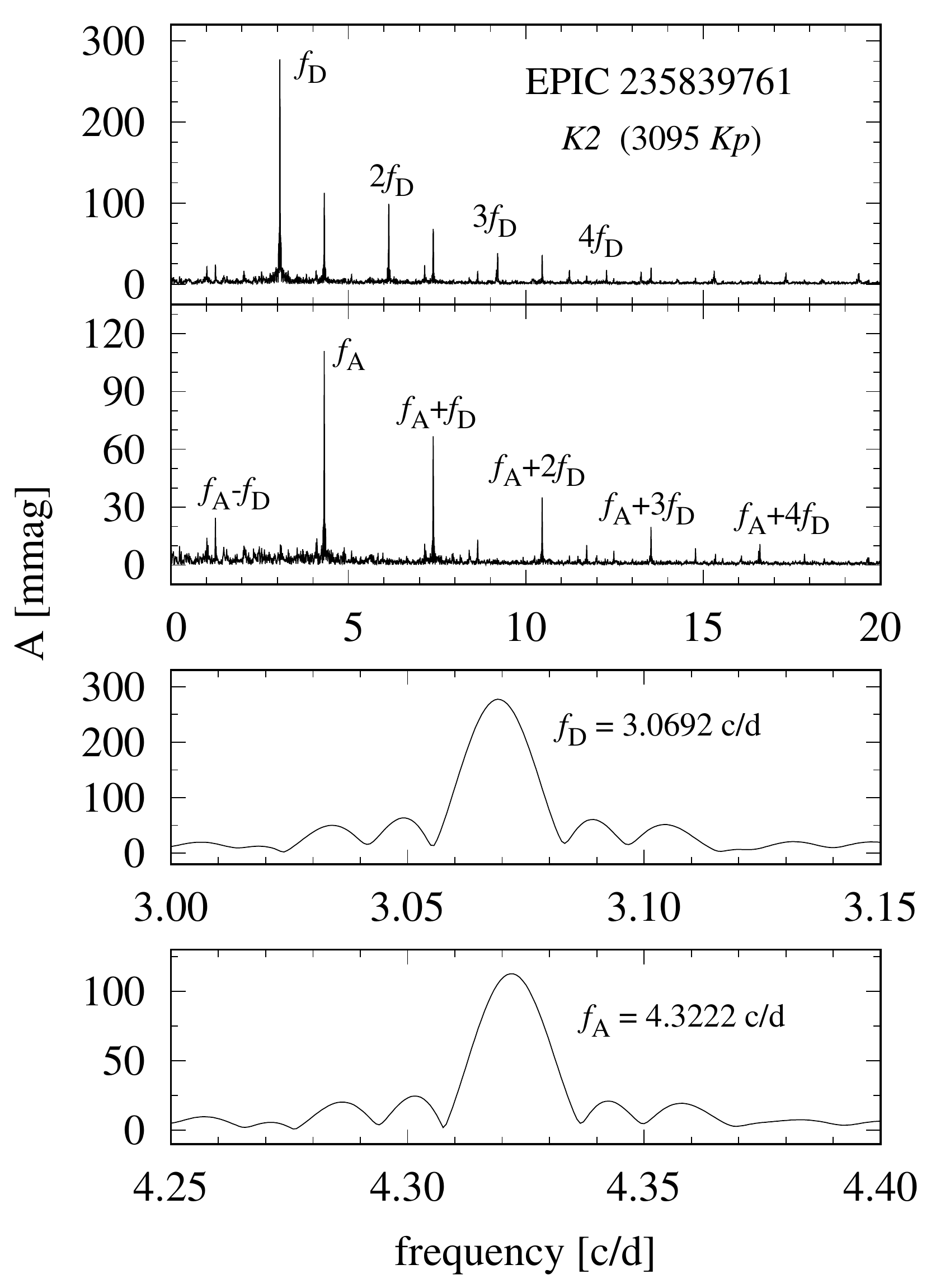}  \put(16,93){(a)}  \put(16,71){(b)}  \put(16,43){(c)}  \put(16,21){(d)}   \end{overpic}  % renamed from "235839761-p04-PDCsap-AmplSpec-2019Mar17b.pdf" [C11p756; C19p687]
\begin{overpic}[width=5.5cm]{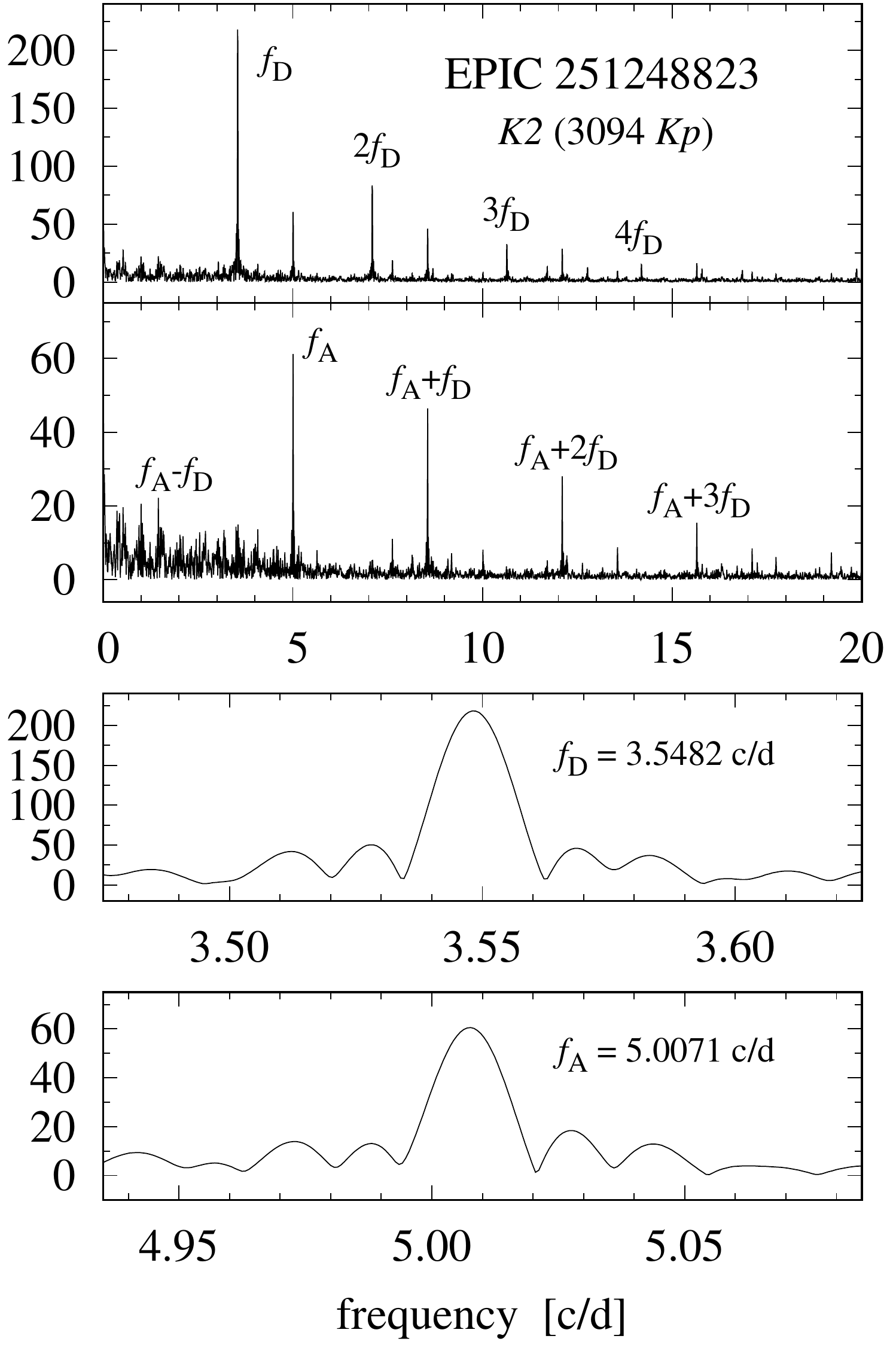}  \put(11,93){(a)}  \put(11,71){(b)}  \put(11,43){(c)}  \put(11,21){(d)}   \end{overpic}  % renamed from "251248823-p04-AmplSpec-2019Mar15.pdf" [C19p688]
\begin{overpic}[width=5.40cm]{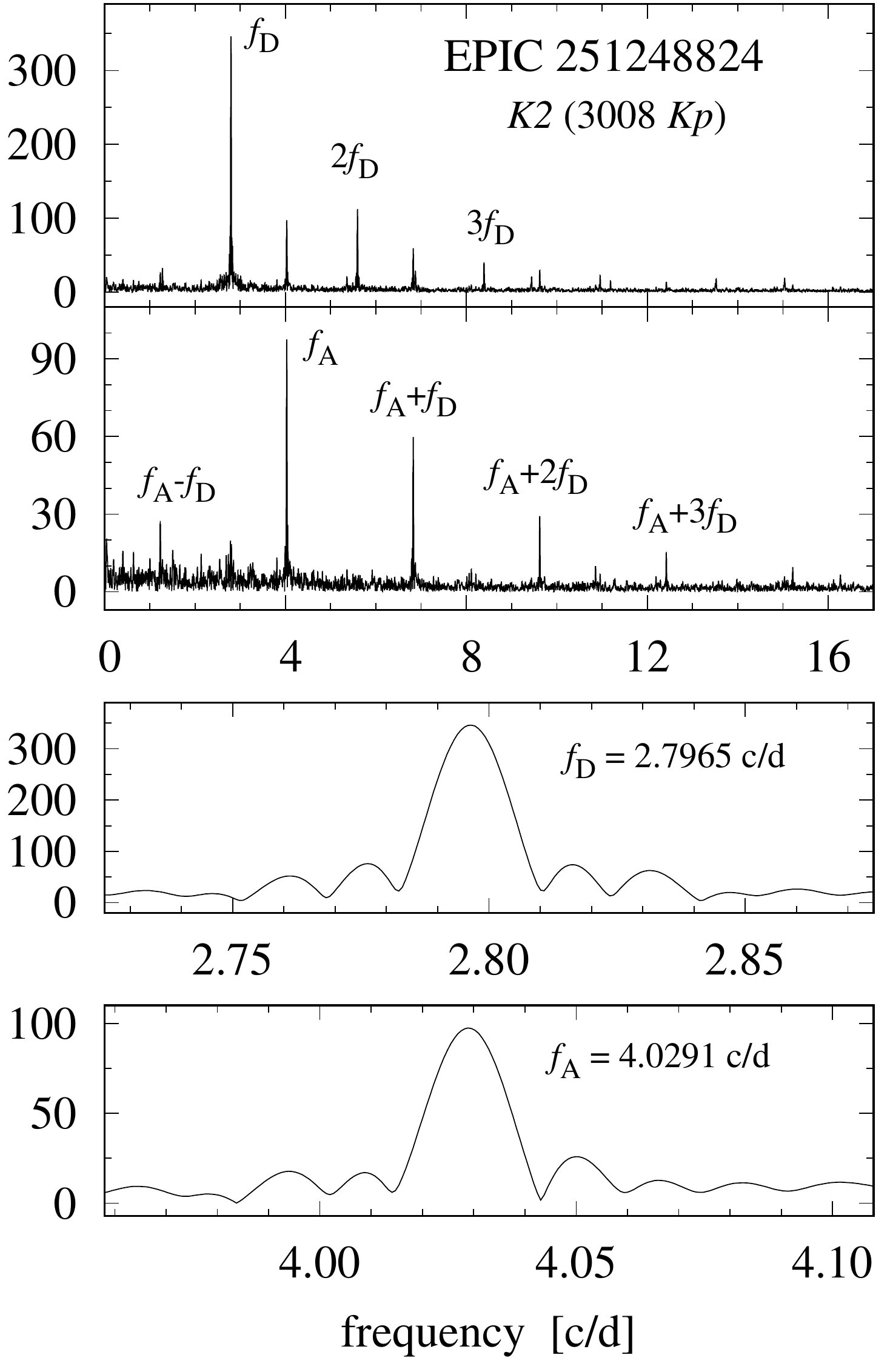}  \put(11,93){(a)}  \put(11,70){(b)}  \put(11,42){(c)}  \put(11,21){(d)}   \end{overpic}  % renamed from "251248824-p04-ampspec-2019Oct28.pdf" [C19p689]
% see C11p754 for source of next picture
% \begin{overpic}[width=7.8cm]{235839761-p04-OGLE-AmplSpec-2019Mar17.pdf}  \put(12,93){(e)}  \put(12,71){(f)}  \put(12,43){(g)}  \put(12,21){(h)}  \end{overpic}
\end{center}

\caption{Fourier amplitude spectra for the three Campaign\,11 pRRd stars, where
the spectra for each star were derived using more than 3000 \texttt{PDCsap}
measurements acquired by {\it K2} over 74 days in 2016.  The top two panels for
each star show: (a) the {\it dominant} peaks at frequency $f_D$ and their
harmonics (a few of which are labelled); and (b) after prewhitening with $f_D$
and its harmonics, the {\it additional} frequency peak at $f_A$ and prominent
peaks corresponding to linear combinations of $f_A$ and $f_D$ (several of which
are labelled) -- the harmonics at 2$f_A$ are barely visible. See Fig.6g for
window function.  The bottom panels (c,\,d) show close-ups of the $f_D$ and
$f_A$ peaks.}

\label{Fig11}
\end{figure*}

% new FIGURE 12 (after deleting O-C figure)
% See C7p586-7,592;   C11p756 for K2, C11p762 for OGLE for source files
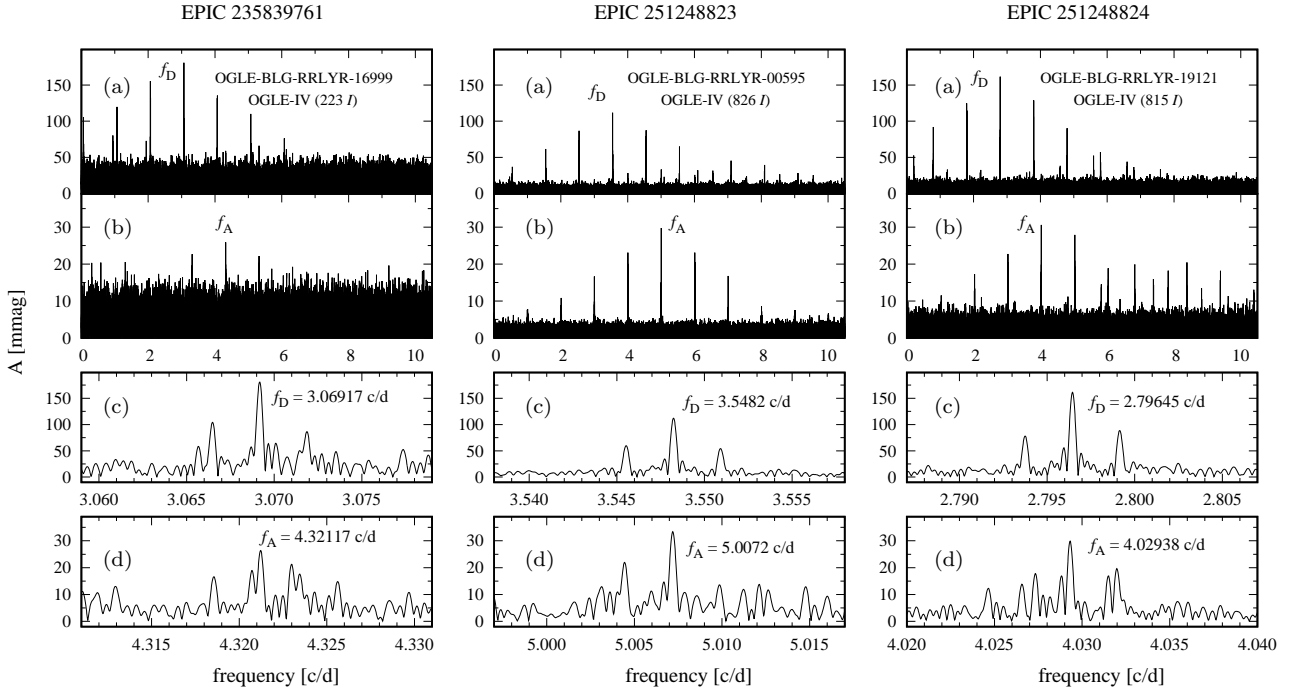
\begin{figure*}
\begin{center}
% see C11p748 for source files
% \begin{overpic}[width=8.2cm]{235839761-p04-PDCsap-AmplSpec-2019Mar17.pdf}   \put(16,93){(a)}  \put(16,71){(b)}  \put(16,43){(c)}  \put(16,21){(d)}  \end{overpic}
% see C11p754 for source of next picture
\begin{overpic}[width=17cm]{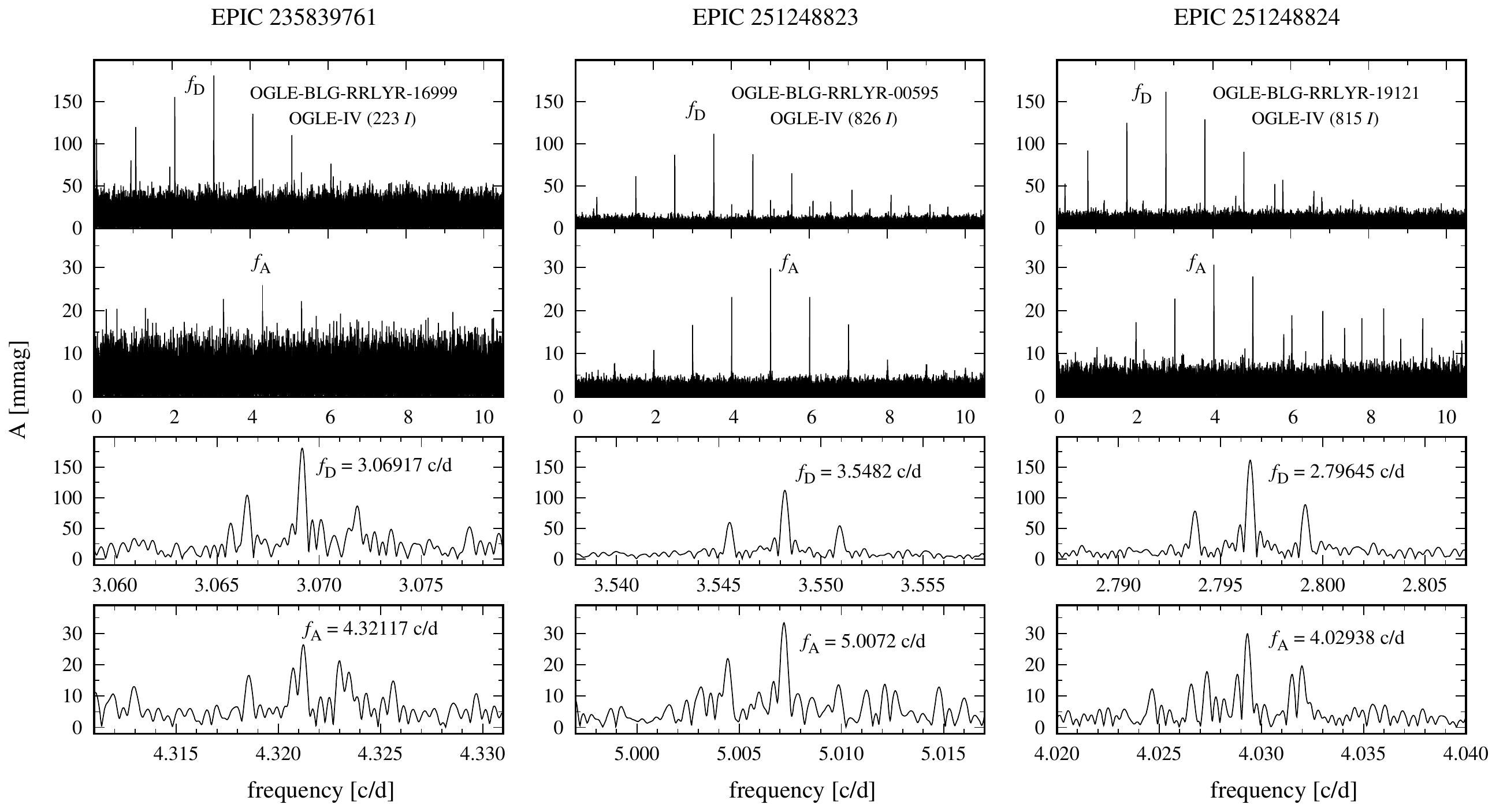} \put(8,47){(a)}  \put(8,36){(b)}  \put(8,22){(c)}  \put(8,10){(d)}   \put(41,47){(a)}  \put(41,36){(b)}  \put(41,22){(c)}  \put(41,10){(d)}   \put(73,47){(a)}  \put(73,36){(b)}  \put(73,22){(c)}  \put(73,10){(d)}  \end{overpic}   % see K2XC19pxxx
%\begin{overpic}[width=5.2cm]{pRRd_Fig12a.pdf} \put(12,93){(a)}  \put(12,69){(b)}  \put(12,40){(c)}  \put(12,17){(d)} \put(24,103){EPIC\,235839761}  \put(-8,40){\rotatebox{90}{A [mmag]}}   \end{overpic} %renamed "235839761-p04-OGLE-AmplSpec-2019Mar17b.pdf" C11p762,C19p691
%\begin{overpic}[width=5.1cm]{pRRd_Fig12b.pdf} \put(12,93){(a)}  \put(12,69){(b)}  \put(12,40){(c)}  \put(12,17){(d)} \put(24,102){EPIC\,251248823} \put(25,-4){frequency [c/d]}  \put(-45,-4){frequency [c/d]}  \put(100,-4){frequency [c/d]} \end{overpic} %renamed "251248823-AmpSpec-2019Mar20.pdf" C19p692
%\begin{overpic}[width=5.6cm]{pRRd_Fig12c.pdf} \put(10,92){(a)}  \put(10,73){(b)}  \put(10,52){(c)}  \put(10,30){(d)} \put(10,12){(e)} \put(22,101){EPIC\,251248824} \end{overpic} % renamed "251248824-AmpSpec-2019Mar23.pdf" C19p693
\end{center}
\vskip0.1cm
\caption{Fourier amplitude spectra derived from ground-based \texttt{OGLE-IV}
$I$-band photometry (acquired over 7.3 years, from 2010 to 2017). The upper
panels for each star reveal the `dominant' and `additional' frequencies, where
the extra peaks are mainly one-day aliasing peaks.  The lower panels show
close-ups in the vicinity of $f_D$ and $f_A$, prominently showing one-year
aliasing peaks at $\Delta$f = $\pm$0.0027\,d$^{-1}$.  The double peaks in the
bottom panel for EPIC\,235839761 (left) reveal the 507$\pm$11\,d modulation of
the additional mode, while for EPIC\,251248824 (right) there is no evidence for
the 482$\pm$5\,d modulation of the dominant mode claimed by P17; instead one
sees evidence for a 481\,d modulation of the additional mode.}

\label{fig:FT}
\end{figure*}

%  TABLE 5 -  see C11,Vol.5,p.771 for ampl, but NOT phases;  where from?
\begin{table}
{\fontsize{6}{7.2}\selectfont  %  <--- works and better; 2nd (skip) should be 1.2 times 1st (size)
\begin{threeparttable}
\fontsize{6}{7.2}\selectfont  %  <--- works and better; 2nd (skip) should be 1.2 times 1st (size)

\caption{EPIC\,235839761 (=\,\texttt{OGLE-BLG-RRLYR-16999}): Fourier
frequencies, amplitudes and phases derived from the {\it K2}
photometry (C112 \texttt{PDCsap}, range 47.7\,d, N=2039), assuming
the time of maximum light from the \texttt{OGLE-IV} study
(Soszy\'nski et al. 2017b): $t_0$=BJD\,2454833.00000+2167.29828 (=UT 2014
December\;8, 19:09:31). The fitted mean magnitude was $<${\it
Kp}$>$= 17.596$\pm$0.001\,mag, with $\sigma$(residuals)=49\,mmag.
}

\begin{tabular}{lrcrrr}
\toprule
Label & \multicolumn{1}{c}{Frequency}    & Period & \multicolumn{1}{c}{Ampl.}    & \multicolumn{1}{c}{Phase}     & \texttt{Sig} \\
      & \multicolumn{1}{c}{[d$^{-1}$]}   & {[d]}  & \multicolumn{1}{c}{[mmag]}   & \multicolumn{1}{c}{[rad]} &              \\
      &                                  &        & \multicolumn{1}{c}{$\pm$1.5} &                               &              \\
\midrule
\multicolumn{6}{c}{(a) {\it Dominant} frequency, $f_D$\,(=\,f1), and its harmonics}       \\
  ~\,f1      & \underline{3.06917} & 0.325821 & 284.3 & --2.188$\pm$0.005 &     452       \\   % See p.937, p.1059, p.1093, p1255
    2f1      &            6.13834  & 0.162911 &  99.1 & --1.546$\pm$0.016 &      98       \\
    3f1      &            9.20751  & 0.108607 &  39.1 & --0.925$\pm$0.040 &      89       \\
    4f1      &           12.27668  & 0.081455 &  17.4 & --0.240$\pm$0.088 &      25       \\
    5f1      &           15.34585  & 0.065164 &   5.7 &   0.532$\pm$0.272 &       5       \\
    6f1      &           18.41502  & 0.054303 &   3.5 &   0.880$\pm$0.444 &    $<$5       \\
    7f1      &           21.48419  & 0.046546 &   1.9 &   2.070$\pm$0.826 &    $<$5       \\ [0.1cm]
%   8f1      &           24.55333  & 0.040728 &   1.0 &   2.425$\pm$1.578 &    $<$5       \\ [0.1cm]

\multicolumn{6}{c}{(b) {\it Additional} frequency, $f_A$\,(=\,f2), and its harmonics}     \\
  ~\,f2      & \underline{4.32216} & 0.231366 & 114.1 &   2.036$\pm$0.014 &     222       \\   % see p.1093 for improved ampl., p1255
    2f2      &            8.64431  & 0.115683 &  13.2 &   0.512$\pm$0.117 &      16       \\
    3f2      &           12.96647  & 0.077122 &   1.3 & --1.405$\pm$1.176 &    $<$5       \\ [0.1cm]
%   4f2      &           17.28862  & 0.057842 &   1.3 & --0.019$\pm$1.189 &    $<$5       \\
%   5f2      &           21.61078  & 0.046273 &   0.6 &   1.171$\pm$2.807 &    $<$5       \\ [0.1cm]

\multicolumn{6}{c}{(c) Linear combination frequencies }                                   \\
  ~\,f1+f2   &            7.39133  & 0.135294 &  66.0 &   2.271$\pm$0.023 &     189       \\   % see p.938. p1256
    2f1+f2   &           10.46050  & 0.095598 &  35.0 &   2.831$\pm$0.044 &      85       \\
    3f1+f2   &           13.52967  & 0.073912 &  18.6 & --2.636$\pm$0.082 &      33       \\
    4f1+f2   &           16.59884  & 0.060245 &  10.8 & --2.013$\pm$0.143 &      13       \\
    5f1+f2   &           19.66801  & 0.050844 &   4.4 & --1.135$\pm$0.347 & $\sim$5       \\
    6f1+f2   &           22.73718  & 0.043981 &   1.7 & --0.772$\pm$0.930 &    $<$5       \\ [0.1cm]
%   7f1+f2   &           25.80635  & 0.038750 &   0.8 &   0.393$\pm$1.730 &    $<$5       \\ [0.1cm]

    2f1--f2  &            1.81618  & 0.550605 &   7.2 &   2.292$\pm$0.214 &       6       \\
    3f1--f2  &            4.88535  & 0.204693 &   6.8 &   2.868$\pm$0.227 &       7       \\
    4f1--f2  &            7.95452  & 0.125715 &   5.6 & --2.741$\pm$0.277 &       5       \\
    5f1--f2  &           11.02369  & 0.090714 &   2.7 & --1.767$\pm$0.569 &    $<$5       \\
    6f1--f2  &           14.09286  & 0.070958 &   2.0 & --1.176$\pm$0.763 &    $<$5       \\ [0.1cm]
%   7f1--f2  &           17.16203  & 0.058268 &   0.7 &   1.166$\pm$2.227 &    $<$5       \\ [0.1cm]

  ~\,f2--f1  &            1.25299  & 0.798093 &  21.4 & --2.894$\pm$0.072 &      47       \\
    2f2--f1  &            5.57514  & 0.179368 &   3.8 &   1.052$\pm$0.406 & $\sim$5       \\
%   3f2--f1  &            9.89730  & 0.101038 &   1.3 & --0.388$\pm$1.192 &    $<$5       \\ [0.1cm]
    2f2--2f1 &            2.50597  & 0.399047 &   6.7 &   3.033$\pm$0.230 &    $<$5       \\ [0.1cm]

  ~\,f1+2f2  &           11.71348  & 0.085372 &  10.1 &   0.646$\pm$0.154 &      11       \\
    2f1+2f2  &           14.78265  & 0.067647 &   8.7 &   1.053$\pm$0.177 &      10       \\
    3f1+2f2  &           17.85182  & 0.056017 &   7.2 &   2.033$\pm$0.213 &       6       \\
    4f1+2f2  &           20.92099  & 0.047799 &   3.1 &   2.775$\pm$0.490 &    $<$5\;\!\! \\
    5f1+2f2  &           23.99016  & 0.041684 &   3.0 & --2.693$\pm$0.504 &    $<$5\;\!\! \\ [0.1cm]
%   6f1+2f2  &           27.05933  & 0.036956 &   1.0 & --2.003$\pm$1.535 &    $<$5       \\ [0.1cm]

\bottomrule
\end{tabular}
% \begin{tablenotes}
% {\fontsize{6}{7.2}\selectfont  %  <--- works and better; 2nd (skip) should be 1.2 times 1st (size
% \item Notes:   }
% \end{tablenotes}
\end{threeparttable}
}
\end{table}

%  TABLE 6 -  see C11,Vol.5,p.771 for ampl, but NOT phases;  where from?  C11p1253
%  (C11-2 only, N=2039, PDCsap with additional custom detrending and normalization to Kp=18.826$\pm$0.002; $\sigma$(residuals)=90\,mmag).

\begin{table}
{\fontsize{6}{7.2}\selectfont  %  <--- works and better; 2nd (skip) should be 1.2 times 1st (size)
\begin{threeparttable}
\fontsize{6}{7.2}\selectfont  %  <--- works and better; 2nd (skip) should be 1.2 times 1st (size)

\caption{EPIC\,251248823\,(=\,\texttt{OGLE-BLG-RRLYR-00595}):\,Fourier
frequencies, amplitudes and phases derived from the {\it K2}
photometry (C112 \texttt{PDCsap} with additional detrending and
normalization to $<${\it Kp}$>$=18.826$\pm$0.002\,mag, N=2039,
$\sigma$(residuals)=90\,mmag).  For the phases  we adopted
$t_0$=BJD\,2454833.00000+2167.05823 (=UT 2014 December\;8, 13:23:51), the
value from the latest \texttt{OGLE-IV} study (Soszy\'nski et al.
2017b).}

\begin{tabular}{lrcrrr}
\toprule
Label & \multicolumn{1}{c}{Frequency}    & Period & \multicolumn{1}{c}{Ampl.}    & \multicolumn{1}{c}{Phase}     & \texttt{Sig} \\
      & \multicolumn{1}{c}{[d$^{-1}$]}   & {[d]}  & \multicolumn{1}{c}{[mmag]}   & \multicolumn{1}{c}{[rad]} &              \\
      &                                  &        & \multicolumn{1}{c}{$\pm$2.6} &                               &              \\
\midrule
\multicolumn{6}{c}{(a) {\it Dominant} frequency, $f_D$\,(=\,f1), and its harmonics} \\
  ~\,f1     & \underline{3.54822} & 0.281831 & 237.7 & --2.064$\pm$0.011 &  324  \\   % See p.992
    2f1     &            7.09645  & 0.140916 &  87.9 & --1.563$\pm$0.030 &  110  \\   % see p.1001-2 for Sig values
    3f1     &           10.64467  & 0.093944 &  40.1 & --0.989$\pm$0.065 &   43  \\
    4f1     &           14.19290  & 0.070458 &  19.8 & --0.231$\pm$0.132 &   13  \\
    5f1     &           17.74112  & 0.056366 &   7.7 &   0.177$\pm$0.339 &    4  \\
    6f1     &           21.28935  & 0.046972 &   5.6 &   0.807$\pm$0.468 & $<$4  \\ [0.1cm]
%   7f1     &           24.83757  & 0.040262 &   2.2 &   2.957$\pm$1.186 & $<$4  \\
%   8f1     &           28.38580  & 0.035229 &   0.5 & --2.914$\pm$5.076 & $<$4  \\ [0.1cm]

\multicolumn{6}{c}{(b) {\it Additional} frequency, $f_A$\,(=\,f2), and its harmonics} \\
  ~\,f2     & \underline{5.00709} & 0.199717 &  69.2 & --1.211$\pm$0.038 &   98  \\
    2f2     &           10.01417  & 0.099859 &  10.7 &   0.632$\pm$0.245 &    5  \\ [0.1cm]
%   3f2     &           15.02126  & 0.066572 &   1.1 & --1.007$\pm$2.470 & $<$5  \\ [0.1cm]
%   4f2     &           20.02834  & 0.049929 &   2.3 & --0.228$\pm$1.125 & $<$5  \\
%   5f2     &           25.03543  & 0.039943 &   1.3 & --1.580$\pm$2.032 & $<$5  \\ [0.1cm]

\multicolumn{6}{c}{(c) Linear combination frequencies }                          \\
  ~\,f1+f2  &            8.55531  & 0.116886 &  49.0 & --0.698$\pm$0.053 &   55  \\   % see p.992,1014
    2f1+f2  &           12.10353  & 0.082621 &  32.3 & --0.238$\pm$0.080 &   29  \\
    3f1+f2  &           15.65176  & 0.063891 &  16.3 &   0.550$\pm$0.160 &   10  \\
    4f1+f2  &           19.19998  & 0.052083 &  11.3 &   0.987$\pm$0.231 &    7  \\
    5f1+f2  &           22.74821  & 0.043960 &   7.7 &   2.066$\pm$0.333 & $<$5  \\
    6f1+f2  &           26.29643  & 0.038028 &   2.1 &   3.022$\pm$1.276 & $<$5  \\ [0.1cm]
%   7f1+f2  &           29.84466  & 0.033507 &   2.4 & --2.495$\pm$1.071 & $<$5  \\ [0.1cm]

  ~\,f2--f1 &            1.45886  & 0.685467 &  26.2 & --0.593$\pm$0.100 &   24  \\
%   2f2--f1 &            6.46595  & 0.154656 &   2.8 &   1.421$\pm$0.918 & $<$5  \\
%   3f2--f1 &           11.47303  & 0.087161 &   1.1 &   3.099$\pm$2.408 & $<$5  \\ [0.1cm]
    2f1--f2 &            2.08936  & 0.478615 &   5.6 & --0.881$\pm$0.462 & $<$5  \\
    3f1--f2 &            5.63759  & 0.177381 &  11.8 & --0.525$\pm$0.222 &    7  \\
    4f1--f2 &            9.18581  & 0.108863 &   8.4 &   1.021$\pm$0.313 &    6  \\
%   5f1--f2 &           12.73403  & 0.078530 &   2.9 &   0.814$\pm$0.908 & $<$5  \\
    6f1--f2 &           16.28226  & 0.061417 &   3.8 &   2.043$\pm$0.690 & $<$5  \\ [0.1cm]
%   7f1--f2 &           19.83049  & 0.050427 &   1.1 & --2.488$\pm$2.311 & $<$5  \\ [0.1cm]

  ~\,f1+2f2 &           13.56239  & 0.073733 &   9.3 &   0.841$\pm$0.278 &    7  \\
    2f1+2f2 &           17.11062  & 0.058443 &   7.2 &   1.886$\pm$0.360 &    7  \\
    3f1+2f2 &           20.65884  & 0.048405 &   5.8 &   2.119$\pm$0.449 & $<$5  \\
    4f1+2f2 &           24.20707  & 0.041310 &   5.2 &   2.712$\pm$0.501 & $<$5  \\ [0.1cm]
%   5f1+2f2 &           27.75529  & 0.036029 &   3.0 & --2.441$\pm$0.879 & $<$5  \\
%   6f1+2f2 &           31.30352  & 0.031945 &   0.7 & --1.534$\pm$3.599 & $<$5  \\ [0.1cm]

%           &            0.9478   &          &  20.7 &                   &   10  \\
%           &            0.4276   &          &  20.6 &                   &   10  \\
%           &            0.4024   &          &  27.9 &                   &   11  \\
%  3f1--2f2 &            0.63050  & 1.586034 &  10.1 & --2.651$\pm$0.260 & \dots \\   %  see p1016
%   6f1+5f2 &           46.32477  & 0.021587 &  10.0 &   0.785$\pm$0.259 & \dots \\
%  3f2--4f1 &            0.82836  & 1.207209 &   9.1 & --2.490$\pm$0.284 & \dots \\
%   7f1+4f2 &           44.86591  & 0.022289 &   8.8 & --2.492$\pm$0.302 & \dots \\
%   7f1+5f2 &           49.87300  & 0.020051 &   6.8 & --0.076$\pm$0.378 & \dots \\
%  5f1--3f2 &            2.71987  & 0.367665 &   5.9 &   0.685$\pm$0.441 & \dots \\
%  2f2--2f1 &            2.91772  & 0.342733 &   5.8 &   2.948$\pm$0.447 & \dots \\
%  6f1--4f2 &            1.26101  & 0.793017 &   5.5 & --2.447$\pm$0.472 & \dots \\ [0.1cm]
\bottomrule
\end{tabular}
% \begin{tablenotes}
% {\fontsize{6}{7.2}\selectfont  %  <--- works and better; 2nd (skip) should be 1.2 times 1st (size
% \item Notes:   }
% \end{tablenotes}
\end{threeparttable}
}
\end{table}

%  TABLE 7 -  see C11,Vol.5,p.771 for ampl, but NOT phases;  where from?
\begin{table}
{\fontsize{6}{7.2}\selectfont  %  <--- works and better; 2nd (skip) should be 1.2 times 1st (size)
\begin{threeparttable}
\fontsize{6}{7.2}\selectfont  %  <--- works and better; 2nd (skip) should be 1.2 times 1st (size)

\caption{EPIC\,251248824 (=\,\texttt{OGLE-BLG-RRLYR-19121}):
significant frequencies derived from {\it K2} photometry (C111+C112,
\texttt{PDCsap}, mean magnitude $<${\it
Kp}$>$=19.844$\pm$0.002\,mag, N=3087, $\sigma$(residuals)=85\,mmag).
For the phases the assumed time of maximum light was
$t_0$=BJD\,2454833.00000+2167.13778 (=UT 2014 December\;8,
15:18:24), the value from the latest \texttt{OGLE-IV} study.}

\begin{tabular}{lrcrrr}
\toprule
Label & \multicolumn{1}{c}{Frequency}    & Period & \multicolumn{1}{c}{Ampl.}    & \multicolumn{1}{c}{Phase}     & \texttt{Sig} \\
      & \multicolumn{1}{c}{[d$^{-1}$]}   & {[d]}  & \multicolumn{1}{c}{[mmag]}   & \multicolumn{1}{c}{[rad]} &              \\
      &                                  &        & \multicolumn{1}{c}{$\pm$2.2} &                               &              \\
\midrule
\multicolumn{6}{c}{(a) {\it Dominant} frequency, $f_D$\,(=\,f1), and its harmonics} \\
  ~\,f1      & \underline{2.79644} & 0.357597 & 345.7 & --2.250$\pm$0.006 &  491 \\   % See p.1040, 1103
    2f1      &            5.59289  & 0.178798 & 112.0 & --1.459$\pm$0.019 &  190 \\   % see p.1041 for Sig values
    3f1      &            8.38933  & 0.119199 &  41.9 & --0.820$\pm$0.051 &   63 \\
    4f1      &           11.18578  & 0.089399 &  17.3 & --0.081$\pm$0.125 &   14 \\
    5f1      &           13.98222  & 0.071519 &   4.7 &   0.893$\pm$0.458 & $<$5 \\ [0.1cm]
%   6f1      &           16.77867  & 0.059599 &   1.5 &   0.344$\pm$1.464 & $<$5 \\
%   7f1      &           19.57511  & 0.051085 &   0.8 &   0.450$\pm$2.616 & $<$5 \\ [0.1cm]

\multicolumn{6}{c}{(b) {\it Additional} frequency, $f_A$\,(=\,f2), and its harmonics} \\
  ~\,f2      & \underline{4.02916} & 0.248191 &  97.4 & --1.756$\pm$0.022 &  190 \\   % p1103
    2f2      &            8.05831  & 0.124096 &   6.7 & --1.234$\pm$0.323 & $<$5 \\
    3f2      &           12.08746  & 0.082730 &   2.6 &   2.256$\pm$0.836 & $<$5 \\
    4f2      &           16.11662  & 0.062048 &   3.5 & --0.357$\pm$0.611 & $<$5 \\ [0.1cm]

\multicolumn{6}{c}{(c) Linear combination frequencies}                           \\
  ~\,f1+f2   &            6.82560  & 0.146507 &  59.0 & --1.276$\pm$0.037 &  108 \\   % see p.1040
    2f1+f2   &            9.62204  & 0.103928 &  29.3 & --0.703$\pm$0.074 &   35 \\
    3f1+f2   &           12.41849  & 0.080525 &  14.6 & --0.106$\pm$0.147 &   11 \\
    4f1+f2   &           15.21493  & 0.065725 &   9.0 &   0.596$\pm$0.240 &    5 \\
    5f1+f2   &           18.01138  & 0.055521 &   3.4 &   2.151$\pm$0.634 & $<$5 \\ [0.1cm]
%   6f1+f2   &           20.80782  & 0.048059 &   1.7 &   2.267$\pm$1.244 & $<$5 \\ [0.1cm]
%   7f1+f2   &           23.60427  & 0.042365 &   0.8 & --3.078$\pm$2.667 & $<$5 \\ [0.1cm]

    2f1--f2  &            1.56373  & 0.639495 &  10.0 & --0.469$\pm$0.215 &    7 \\
    3f1--f2  &            4.36018  & 0.229348 &   3.5 &   1.231$\pm$0.616 & $<$5 \\
    4f1--f2  &            7.15662  & 0.139731 &   3.0 &   1.273$\pm$0.722 & $<$5 \\
    5f1--f2  &            9.95307  & 0.100472 &   3.7 &   1.132$\pm$0.578 & $<$5 \\ [0.1cm]
%   6f1--f2  &           12.74951  & 0.078434 &   1.0 &   3.136$\pm$2.121 & $<$5 \\
%   7f1--f2  &           15.54596  & 0.064325 &   1.3 & --0.609$\pm$1.645 & $<$5 \\ [0.1cm]

  ~\,f2--f1  &            1.23271  & 0.811221 &  27.0 &   0.128$\pm$0.080 &   31 \\
    2f2--f1  &            5.26187  & 0.190047 &   5.8 &   2.731$\pm$0.372 & $<$5 \\ [0.1cm]
%   3f2--f1  &            9.29102  & 0.107631 &   1.3 & --0.860$\pm$1.677 & $<$5 \\ [0.1cm]

  ~\,f1+2f2  &           10.85475  & 0.092126 &   9.7 & --0.384$\pm$0.223 &    5 \\
    2f1+2f2  &           13.65120  & 0.073254 &   3.9 &   0.713$\pm$0.555 & $<$5 \\
    3f1+2f2  &           16.44764  & 0.060799 &   2.9 &   1.559$\pm$0.746 & $<$5 \\ [0.1cm]
%   4f1+2f2  &           19.24409  & 0.051964 &   0.5 & --0.688$\pm$4.082 & $<$5 \\
%   5f1+2f2  &           22.04053  & 0.045371 &   3.2 &   1.688$\pm$0.678 & $<$5 \\ [0.1cm]

    4f2--5f1 &            2.13440  & 0.468516 &  13.4 & --0.008$\pm$0.161 &   10 \\
    3f1--2f2 &            0.33102  & 3.020930 &   9.0 &   0.053$\pm$0.241 & $<$5 \\
    3f2--4f1 &            0.90169  & 1.109033 &   7.9 & --1.477$\pm$0.272 &    5 \\
    2f2--2f1 &            2.46542  & 0.405610 &   6.0 &   1.827$\pm$0.357 & $<$5 \\
    6f1--4f2 &            0.66205  & 1.510465 &   5.3 & --1.990$\pm$0.410 & $<$5 \\
    3f2--3f1 &            3.69813  & 0.270407 &   5.0 & --2.803$\pm$0.435 & $<$5 \\
    5f1--2f2 &            5.92391  & 0.168807 &   4.5 & --0.596$\pm$0.480 & $<$5 \\
    7f1--4f2 &            3.45849  & 0.289143 &   3.5 & --0.549$\pm$0.622 & $<$5 \\
    4f1--2f2 &            3.12747  & 0.319747 &   3.4 & --0.341$\pm$0.641 & $<$5 \\
    5f1--3f2 &            1.89476  & 0.527772 &   3.1 &   0.569$\pm$0.704 & $<$5 \\ [0.1cm]
% ~\,f1+3f2  &           14.88391  & 0.067187 &   2.6 & --1.281$\pm$0.829 & $<$5 \\
%   6f1--3f2 &            4.69120  & 0.213165 &   2.6 & --1.585$\pm$0.837 & $<$5 \\
%   7f1--3f2 &            7.48765  & 0.133553 &   2.4 &   2.353$\pm$0.907 & $<$5 \\
%   6f1--2f2 &            8.72036  & 0.114674 &   2.2 & --0.401$\pm$0.989 & $<$5 \\
\bottomrule
\end{tabular}
% \begin{tablenotes}
% {\fontsize{6}{7.2}\selectfont  %  <--- works and better; 2nd (skip) should be 1.2 times 1st (size
% \item Notes:   }
% \end{tablenotes}
\end{threeparttable}
}
\end{table}

% Frequency analysis of the K2 data
% See C11p1260 for 235839761 (C112 only because of C111 offset) - agrees with LBs Fourier.py analysis;

Fourier amplitude spectra derived from the {\it K2} photometry are
plotted in Figure~11. The spectra are  similar for all three
stars and resemble closely that of EPIC\,216764000.  The highest
peaks in the top panels of Fig.\,11 reveal clearly the {\it
dominant} frequencies, $f_D$, at 3.0692, 3.5482 and
2.7964\,d$^{-1}$, respectively.  The Fourier first-term amplitudes
of the dominant peak, $A_{\rm 1,D}$({\it Kp}) = 284.3$\pm$1.5,
237.7$\pm$2.6 and 345.7$\pm$2.2\,mmag  (with S/N = 77, 61 and 66),
are comparable  to the corresponding amplitude for EPIC\,216764000
(=285.4$\pm$0.2\,mmag).  Several of the observable harmonics of the
dominant frequencies can be seen in the top panels and have been
labelled; their presence is consistent with the stars having
asymmetric light curves when phased with the dominant period, as
shown in Fig.\,8. Amplitude spectra after prewhitening with the
dominant frequencies and their harmonics are plotted in the second
row of Fig.\,11. The highest peak in each panel (with S/N = 31, 17
and 20) reveals the presence of the respective {\it additional}
frequency, $f_A$, at 4.3222, 5.0071 and 4.0292\,d$^{-1}$.  The
amplitudes, $A_{\rm 1,A}$({\it Kp}) = 114.1$\pm$1.5, 69.2$\pm$2.6
and 97.4$\pm$2.2\,mmag, are all higher than the corresponding
amplitude for EPIC\,216764000 (=37.8$\pm$0.2\,mmag). Also
prominent are many peaks at combination frequencies involving $f_A$
and $f_D$, several of which have been labelled.  The bottom two rows
of Fig.\,11 show close-ups in the vicinity of the $f_D$ and $f_A$
peaks. The average values of the measured FWHM widths of the
dominant and additional peaks, 0.016\,d$^{-1}$ and 0.017\,d$^{-1}$,
are only slightly larger than the reciprocal of the {\it K2}
Campaign\,11 observation time interval, 1/74.2\,d = 0.013\,d$^{-1}$.

Frequencies, amplitudes and phases derived from the {\it K2}
Campaign\,11 data are summarized in Tables~5-7.  The number of
fitted frequencies for EPIC\,235839761, 251248823 and 251248824
(=\,29, 23, 33 respectively) are based on the C112 data only and are
a subset of all the fitted frequencies; they differ somewhat from
the number of frequencies that were used by the \texttt{FNPEAK}
analyses (C111+C112 datasets).

In Figure~12 amplitude spectra  based on the \texttt{OGLE-IV} $I$-passband
photometry for the three Campaign\,11 stars are plotted.   The top panels show
peaks corresponding to the $f_D$ and $f_A$ frequencies, both of which are
flanked by numerous one-day aliases. Prominent in the close-ups shown in the
bottom two panels are the one-year alias frequencies at
$\Delta$$f_D\pm0.0027$\,d$^{-1}$.   As expected, given that the
\texttt{OGLE-IV} data have a 7.5-year time-base compared with only 74.2 days
for the {\it K2} data, the $f_D$ and $f_A$ peaks are much narrower (FWHM widths
$\sim$0.0004\,d$^{-1}$) than those seen in Fig.~11 for the {\it K2} data.

The modulation of EPIC\,235839761 is revealed in the bottom left panel of
Fig.\,12 by the significant extra side-peak near 4.32305\,d$^{-1}$, which
corresponds to a modulation period 538\,d.  On the other hand, the modulation
of the dominant mode of EPIC\,251248824 is not confirmed (see bottom right
panels of Fig.\,12). It appears that there is modulation but that it is of the
additional mode, with the same period and amplitude as claimed for the dominant
mode.  Panel\,(c) shows the `dominant' peak at $f_D$=2.79645\,d$^{-1}$
($A_D$=157.7\,mmag) and its one-year alias peaks at
$f_D$$\pm$0.00274\,d$^{-1}$.  Panel\,(d) shows the `additional' peak at
$f_A$=4.02938\,d$^{-1}$ (amplitude 30\,mmag), its one-year alias peaks (again
offset by $\pm$0.00274\,d$^{-1}$), and extra interior side-peaks at
$f_A$$\pm$0.00208\,d$^{-1}$ (amplitude 18\,mmag), the difference corresponding
to $P_B$=481\,d.

% FIGURE 13
\begin{figure*}
\begin{center}
\begin{overpic}[width=6.18cm] {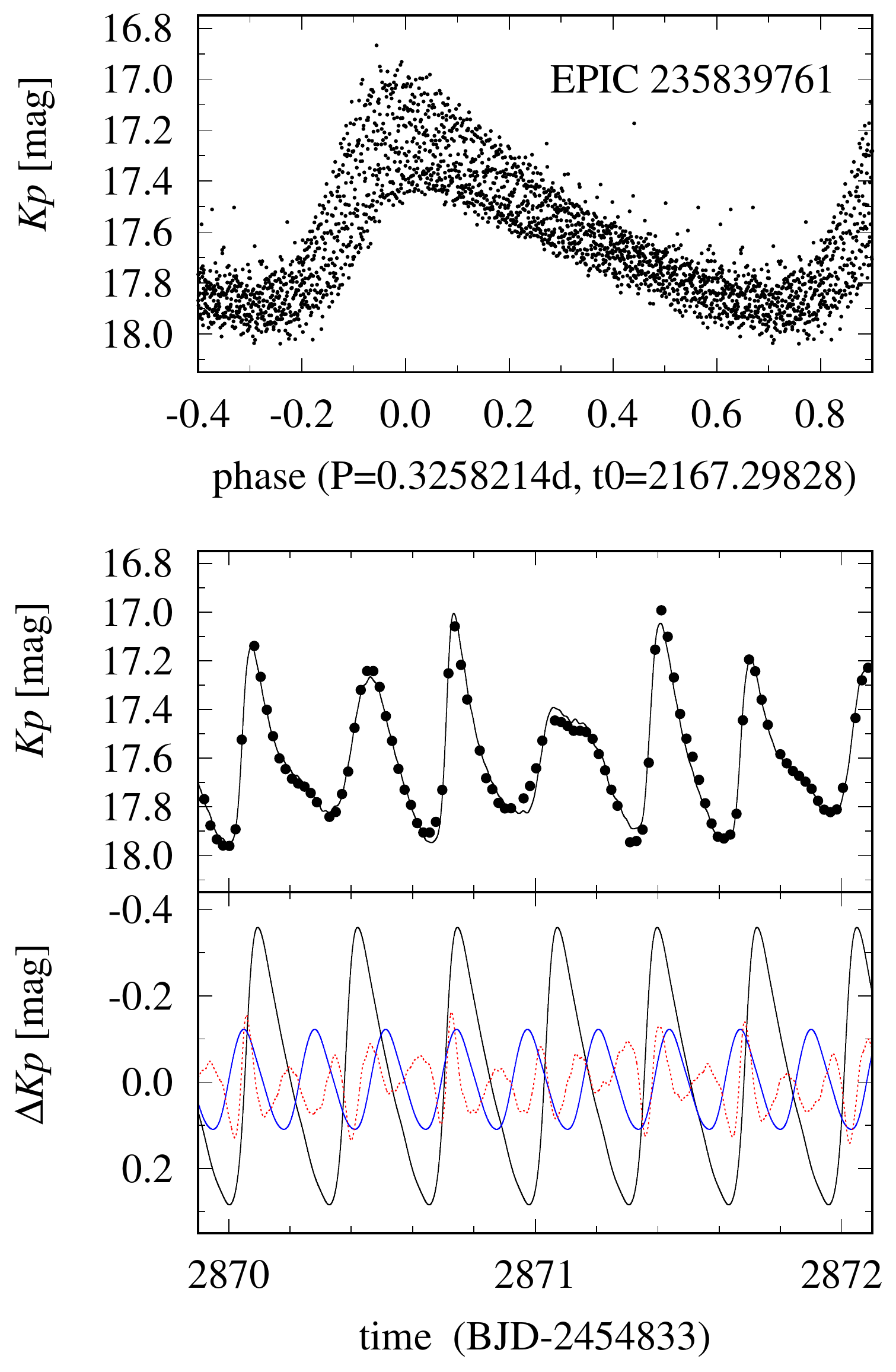} \end{overpic} % C11p1610 for source;  renamed from "235839761-c11-mmLC-2021May18.pdf"
\begin{overpic}[width=5.50cm]{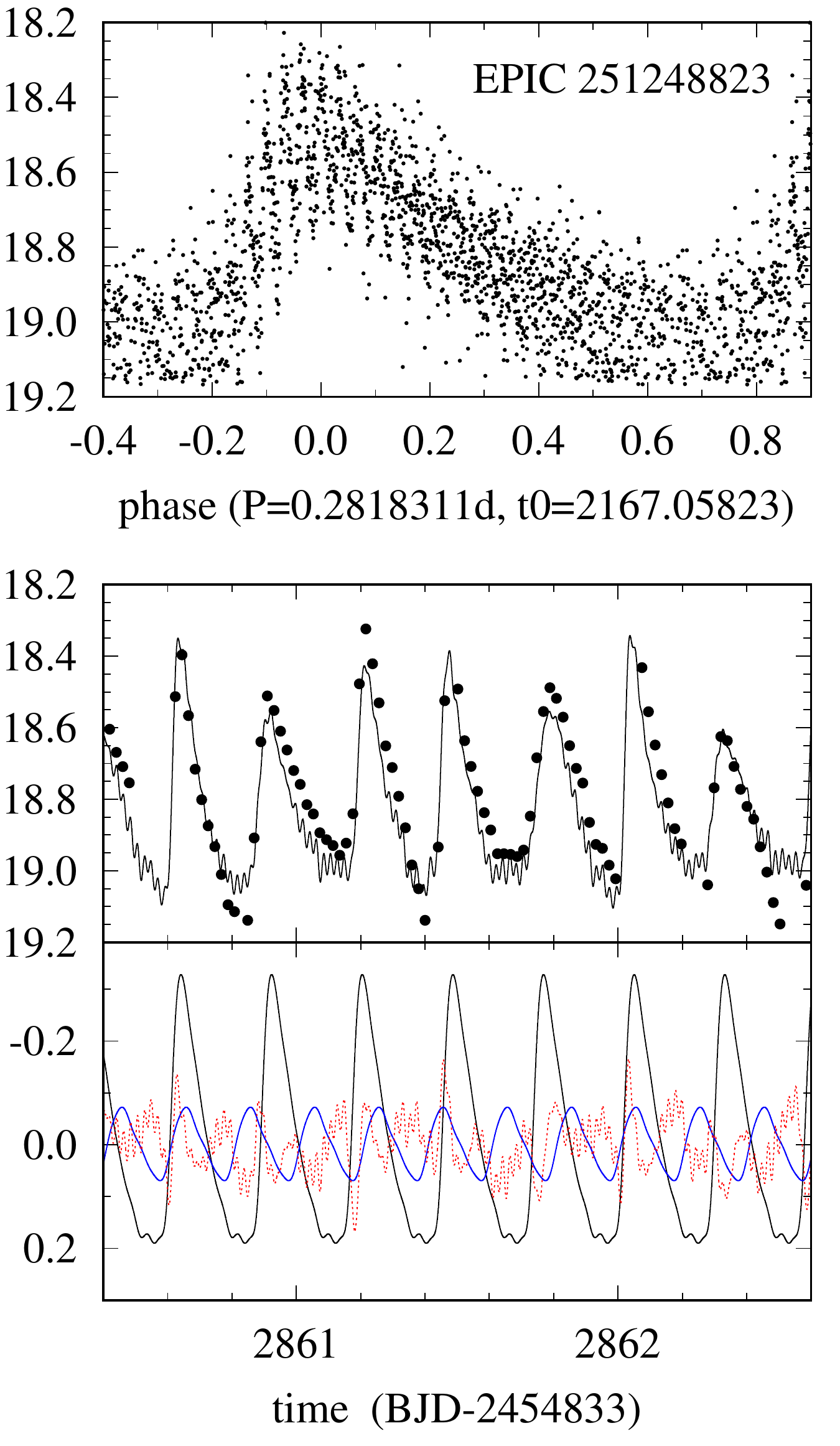} \end{overpic} % C11p1615 for source;  renamed from "251248823-c11-mmLC-2021May18.pdf"  c112only
\begin{overpic}[width=5.45cm]{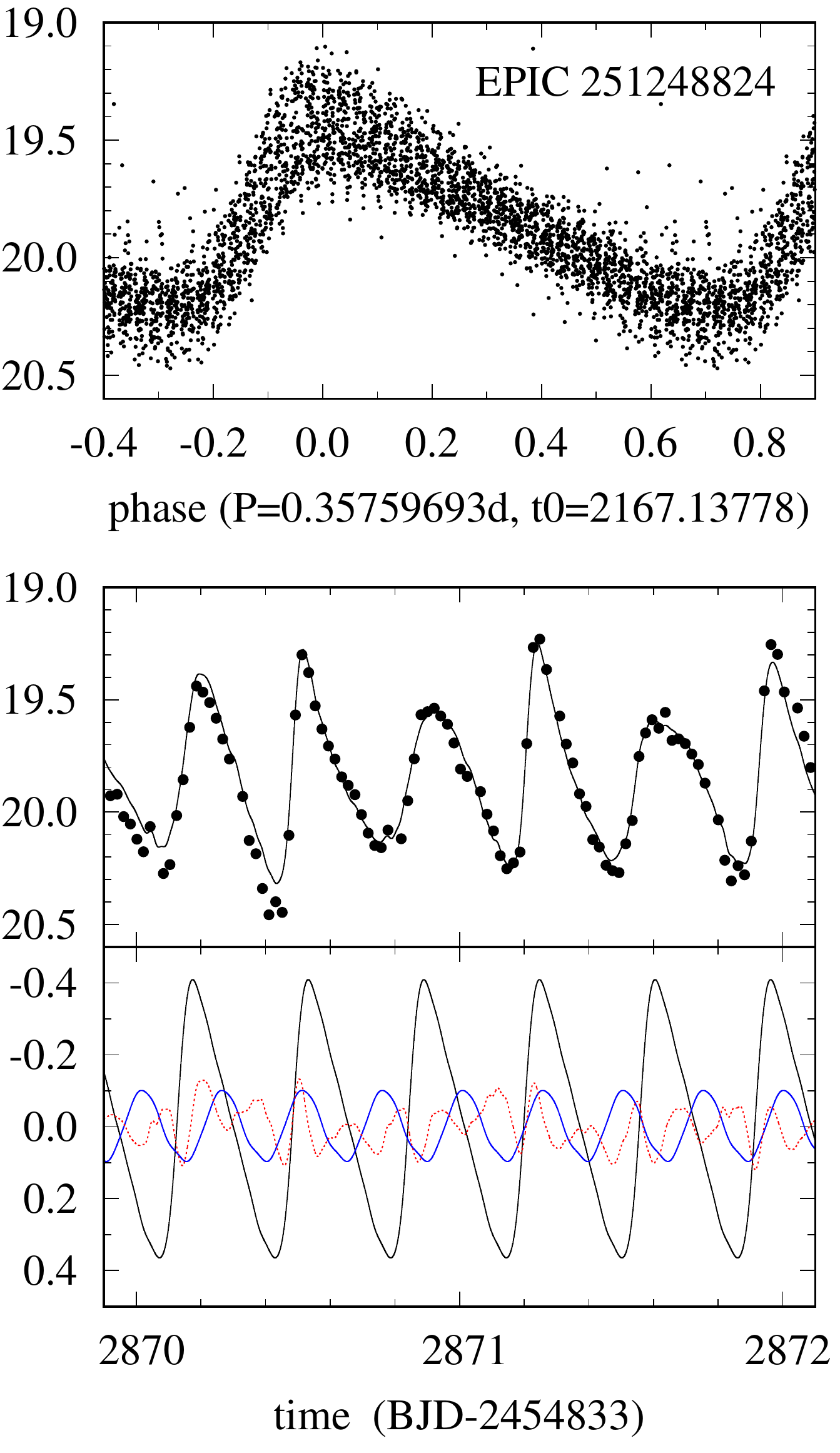} \end{overpic} % C11p1621 for source;  renamed from "251248824-c11-mmLC-2020June28-2870-73.pdf"
\end{center}
\vskip-0.4truecm

\caption{Light curves for the three pRRd stars observed
during Campaign\,11. (Top row) Photometry phased with the
\texttt{OGLE-IV} dominant periods and times of maximum light;
for EPIC\,251248823 only the C112 data are plotted. (Middle row)
Observed and fitted light curves (2.2\,d segments). (Bottom row)
Component light curves derived from the fitted two-frequency model.
As was the case for EPIC\,216764000 (see Fig.\,7) the amplitudes of
the combination frequencies (dotted red curves) are comparable in
strength to those of the additional modes (blue curves).}

\label{fig:FT}
\end{figure*}

% Section 3.2.2
\subsubsection{Light Curves and Fourier parameters}
% see C11p1211 for mean differences table

Light curves derived from the {\it K2} photometry are plotted in Figure~13. The
three panels for each star are as described for EPIC\,216764000  (see $\S$3.1.2
and Fig.\,7). The top row shows light curves phased with the \texttt{OGLE-IV}
dominant period and times of maximum light. Use of the latest high-precision
\texttt{OGLE-IV} periods and $t_0$-values served as a useful check on the {\it
K2} periods, and showed that the light curves constructed with the {\it K2}
photometry are not offset in time or phase from the \texttt{OGLE-IV} light
curves. For all three stars the influence of the combination frequencies on the
overall light curve is seen to be comparable to that of the additional mode.

Fourier decomposition parameters, total amplitudes, and risetimes for the
Campaign\,11 pRRd stars, all derived from the {\it Kp} photometry, are given in
columns (3)-(5) of Table~4.  For comparison purposes  \texttt{OGLE-IV} total
amplitudes and Fourier parameters derived from the $I$-passband data, and the
differences, $\Delta$, between the {\it K2} {\it Kp} and \texttt{OGLE-IV} $I$
values, are also recorded in the Table~4.  The dominant-mode total amplitudes
are similar for all four stars, but the additional-mode total amplitude is
smallest for EPIC\,216764000.  As expected the {\it Kp} total amplitudes are
larger than the $I$-passband total amplitudes, and the mean differences between
the {\it K2} and \texttt{OGLE-IV} $R_{\rm 21}$ and $R_{\rm 31}$ ratios are
practically negligible, 0.03$\pm$0.01 and 0.02$\pm$0.01.  The mean differences
between the phase parameters are also small, 0.16$\pm$0.11 for
$\Delta$$\phi_{\rm 21}$ and 0.21$\pm$0.11 for $\Delta$$\phi_{\rm 31}$.

Table~4 also reveals that the Fourier parameters and other descriptors for all
four stars are quite similar.   Inspection of the amplitude ratios ($R_{\rm
21}$, $R_{\rm 31}$, $R_{\rm 41}$, \dots) reveals an exponential fall-off, at
least for the first four terms. A similar fall-off was seen for the harmonic
components of the RR~Lyrae star V1127\,Aql observed by \texttt{CoRoT} (see
Chadid et al. 2010, and $\S$4.1 below). One also sees close agreement between
the parameter values derived from the {\it Kp} and \texttt{OGLE}
$I$-photometry.

% FIGURE 14  See C11p1089-1094 for 235839761;  C11p1095-1100 for 251248823;     C11p1103-->1251 for 251248824
\begin{figure*}
\begin{center}
\begin{overpic}[width=5.64cm]{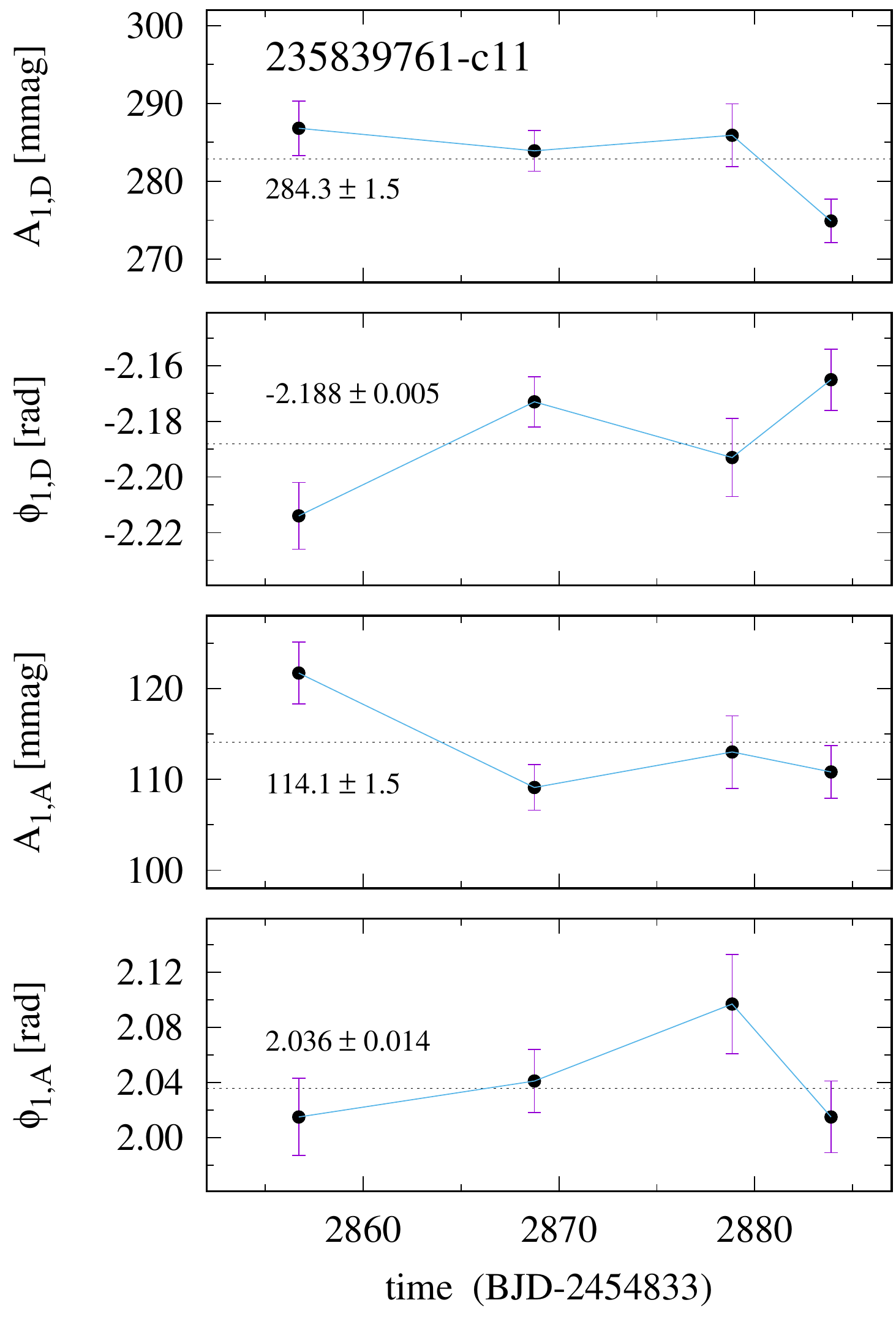}    \end{overpic}  % renamed from "235839761-10dayBins-2019Nov11.pdf"
\begin{overpic}[width=5.3cm] {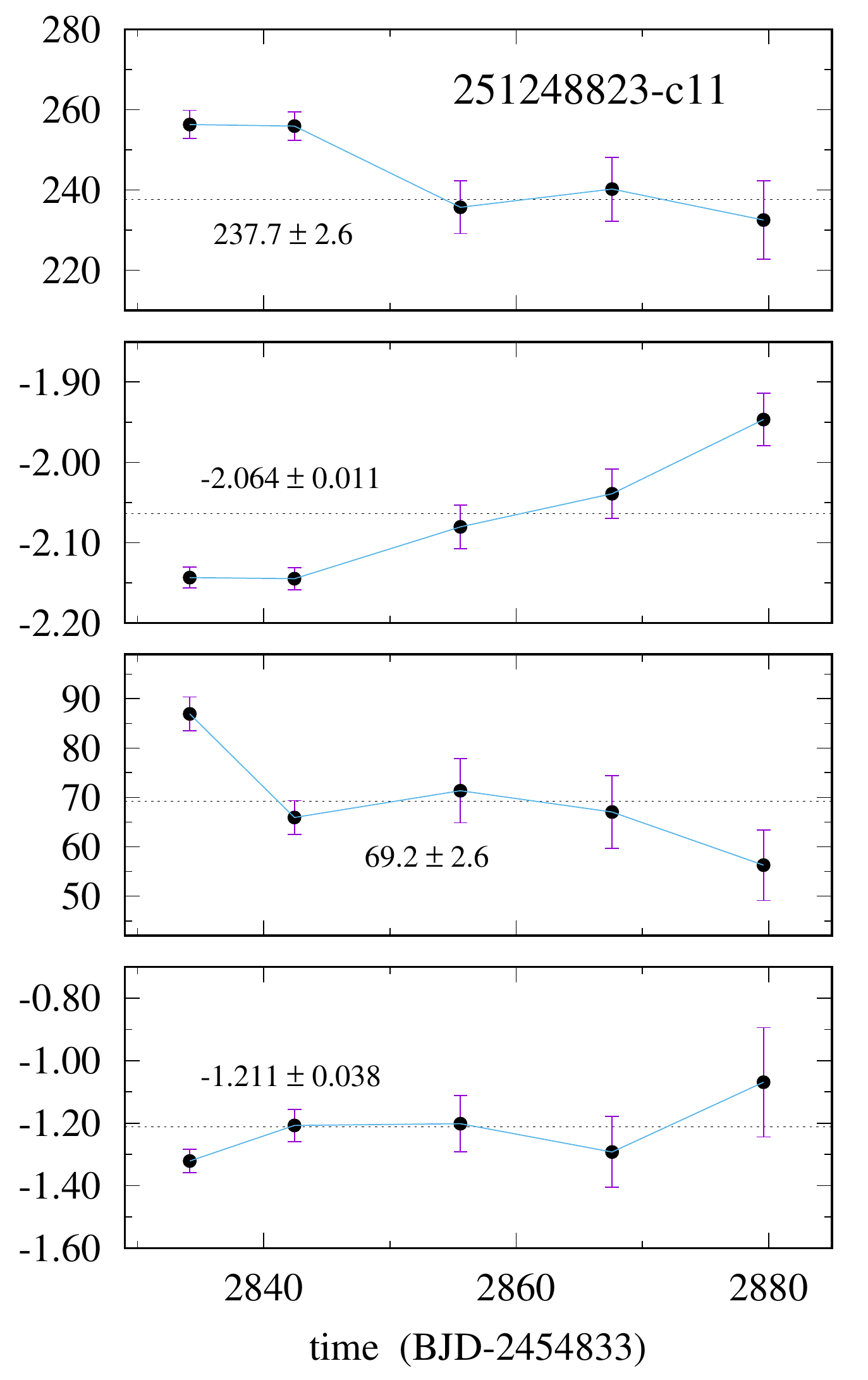}    \end{overpic}    % was ...2019Aug2;  see p1095-1100,1252; renamed from "251248823-12dayBins-2019Nov10.pdf"
\begin{overpic}[width=5.4cm] {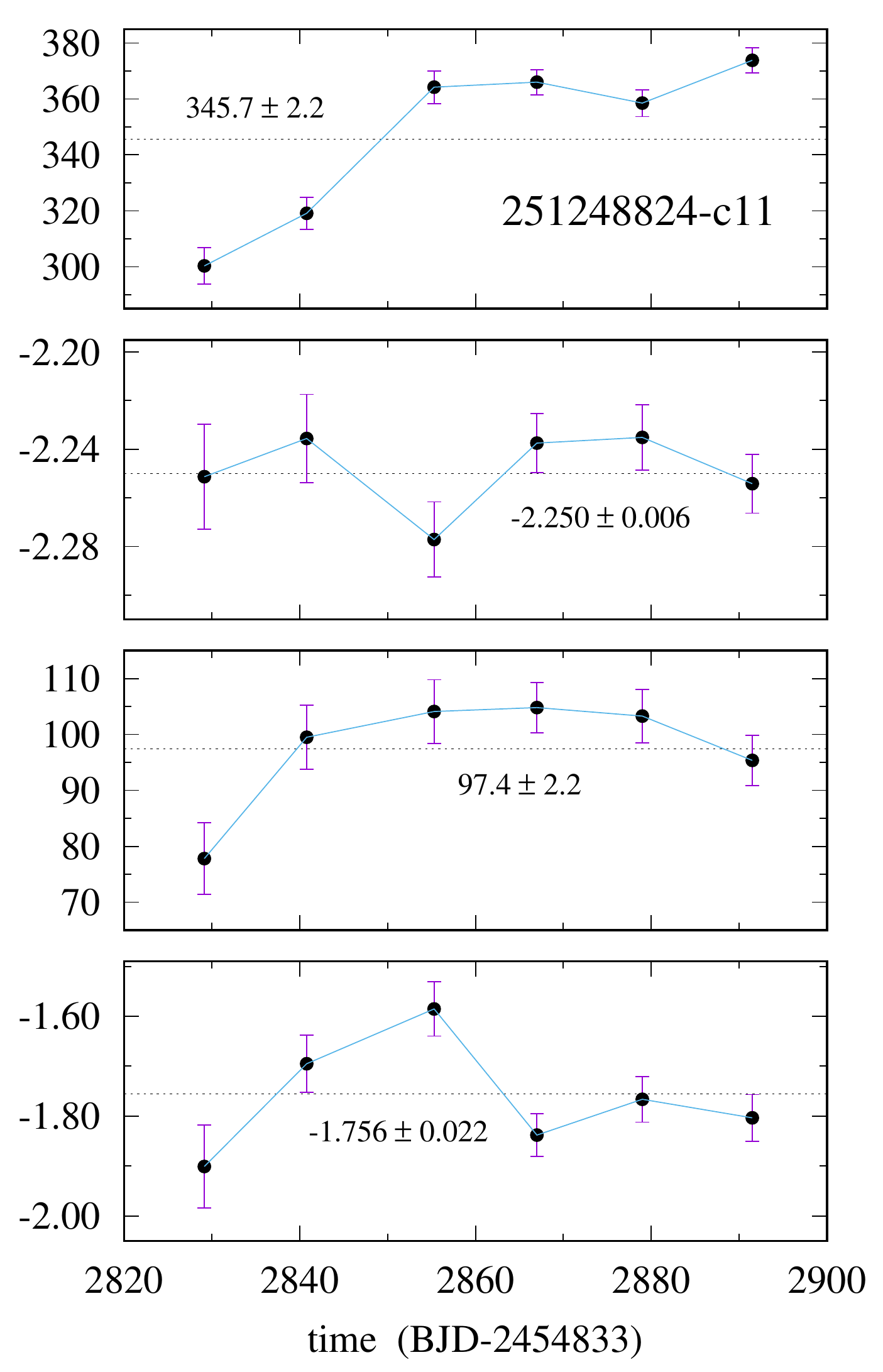}    \end{overpic}    % was ...2019Aug3.pdf; see p.1103,1251;  renamed from "251248824-12dayBins-2021Feb6.pdf"
\end{center}
\vskip-0.4truecm

\caption{Time variations for the three Campaign\,11 pRRd stars,
assuming the same periods and times of maximum light used for
Tables~5-7. For each star the top two panels show the variations of
the Fourier amplitudes and phases of the dominant mode, and the
bottom two panels show the variations for the additional mode.
Horizontal lines are drawn at the average values (see labels).   }

\label{fig:Fig14}
\end{figure*}

% Section 3.2.3
\subsubsection{Amplitude and Phase modulations}

% see C11p1180-1181 for SAS analysis 223 I-data (2010-2017)
As noted previously P17 found, based on \texttt{OGLE} $I$ photometry, that
EPIC\,235839761 and EPIC\,251248824 have modulated amplitudes (see last
paragraph of $\S$2.3.2, and $\S$3.2.1).   Unfortunately, the 74.2\,d
observation interval of the {\it K2} Campaign\,11 observations is insufficient
to confirm such long-period modulations, but the data do permit the
investigation of possible shorter-time-scale variations in the same way as
described above for EPIC\,216764000 (see Fig.\,10), i.e., by binning the {\it
K2} data and analyzing the photometry in each bin separately. The resulting
time-variation graphs are shown in Figure~14. Several possible trends can be
seen;  however, because the Campaign\,11 data were divided into two
subcampaigns having different detector sensitivities we could not be confident
that the trends are real.   Nevertheless, and despite the three stars being
significantly fainter than EPIC\,216764000, we are encouraged that there is
considerable internal consistency of the amplitudes and phases.

%  TABLE 8 -  see C11,Vol.5,p.771 for ampl, but NOT phases;  where from?
\begin{table*}
{\fontsize{6}{7.2}\selectfont  %  <--- works and better; 2nd (skip) should be 1.2 times 1st (size)
\begin{threeparttable}
\fontsize{6}{7.2}\selectfont  %  <--- works and better; 2nd (skip) should be 1.2 times 1st (size)

\caption{Analysis of the \texttt{OGLE} $I$-photometry (2001-2017) for the 42
pRRd stars listed in Table~1 of Prudil et al. (2017), sorted according to
decreasing amplitude ratio, $A_{1,D}$/$A_{1,A}$.  The analyses were made using
our non-linear least-squares (Levenberg--Marquardt) multi-frequency program,
which, in addition to solving for the dominant and additional amplitudes and
phases, also solved for the two independent frequencies (with error estimates).
The identifications in the first column are the \texttt{OGLE-BLG-RRLYR} star
numbers, \underline{underlined} for the three stars observed by {\it K2}, and
followed by an asterisk if either $P_D$, $P_A$ or both exhibits Blazhko
modulations (see Table 2 of P17).  Column (2) contains the epochs of the
analyzed photometry and gives, in parentheses, the number of $I$-magnitude
measurements included in the analysis, and the rms error of the fit, $\sigma$,
in units of mmag.  Column (3) contains the derived mean magnitudes, columns
(4)-(6) give the dominant and additional pulsation periods and their ratio, and
columns (7)-(9) contain the Fourier first-term amplitudes and their ratios. The
$A_{1,D}$ amplitudes for the six stars with the lowest amplitudes have been
{\it italicized}. The bottom four stars were classified by the \texttt{OGLE}
survey as `RRc'; otherwise the classification was `RRab'.}

\begin{tabular}{llcllcccc}
\toprule
    \multicolumn{1}{c}{\texttt{OGLE}} &  \multicolumn{1}{c}{Epochs (N$_I$,$\sigma$)} &  $<I>$ & \multicolumn{1}{c}{$P_D$} & \multicolumn{1}{c}{$P_A$} & $P_A$/$P_D$  & $A_{1,D}$ & $A_{1,A}$ & \multicolumn{1}{c}{ $A_{1,D}$/$A_{1,A}$} \\
    \multicolumn{1}{c}{ no.    }      &                                     & [mag]  & \multicolumn{1}{c}{[d]}  & \multicolumn{1}{c}{[d]}  &              & [mmag]    &  [mmag]   & \multicolumn{1}{c}{ }                    \\
\multicolumn{1}{c}{(1)}   & \multicolumn{1}{c}{(2)}  & (3)  & \multicolumn{1}{c}{(4)}  & \multicolumn{1}{c}{(5)} & (6) & (7)  & (8) & (9)   \\
\midrule
02943               &  2010-17\,(1665,35) & 17.460$\pm$0.001 & 0.3367005(1)  & 0.2297799(7) & 0.682446(2) &     189.0$\pm$1.2  & ~\,6.9$\pm$1.3  & \!\!\!27.5$\pm$5.3   \\ % C11p1142 (2010-17, Pest, 6-3)
15842               &  2001-17\,(1021,8)  & 15.004$\pm$0.001 & 0.36543510(2) & 0.2534121(2) & 0.693453(1) &     171.4$\pm$0.4  & ~\,7.9$\pm$0.4  & \!\!\!21.7$\pm$1.0   \\ % C11p1630-31 (OGLE3_2001-4+OGLE4_2010-17,  6-3, Pest)
01401               &  2010-16\,(601,16)  & 16.907$\pm$0.001 & 0.3130437(1)  & 0.2139031(4) & 0.683301(1) &     219.7$\pm$1.0  &  13.4$\pm$1.0   & \!\!\!16.4$\pm$1.3   \\ % C11p1136-37 (2010-13, N=426, 6-3, Pest)
05851               &  2001-17\,(7330,52) & 18.124$\pm$0.001 & 0.35468745(5) & 0.2456307(2) & 0.692527(1) &     195.6$\pm$0.9  &  17.1$\pm$0.8   & \!\!\!11.4$\pm$0.6   \\ % C11p1637-38 (6-3, Pest, FMMv2)
14135               &  2010-17\,(1847,38) & 17.583$\pm$0.001 & 0.3008045(1)  & 0.2153329(4) & 0.715857(2) &     176.3$\pm$1.3  &  15.5$\pm$1.3   & \!\!\!11.4$\pm$0.9   \\ % C11p1676  (6-3-3, Pest, FMMv2)
                    &                     &                  & 0.3008045(1)  & 0.2244637(5) & 0.746211(2) &     176.3$\pm$1.3  &  26.8$\pm$1.3   & 6.58$\pm$0.93  \\ %
12880$*$            &  2010-17\,(6586,29) & 16.692$\pm$0.001 & 0.34004076(4) & 0.2390558(2) & 0.703021(1) &     210.1$\pm$0.5  &  27.1$\pm$0.5   & 7.74$\pm$0.17  \\ % C11p1640-42 (6-3, Pest, FMMv2); Prudil: BL(D),nonStat,harm(A),OGLEsec
08621               &  2010-13\,(457,22)  & 16.255$\pm$0.001 & 0.3300141(2)  & 0.2294011(7) & 0.695125(3) &     195.4$\pm$1.6  &  25.3$\pm$1.7   & 7.71$\pm$0.57  \\ % C11p1643 (2010-13, N=457, 6-3, FMMv2, Pest)
30418               &  2010-16\,(556,33)  & 18.002$\pm$0.002 & 0.3640956(3)  & 0.259372(1)  & 0.712373(2) &     185.3$\pm$2.1  &  27.5$\pm$2.1   & 6.75$\pm$0.59  \\ % C11p1645-46 (6-3, Pest, FMMv2)
14063               &  2010-17\,(2013,13) & 16.245$\pm$0.001 & 0.3753218(1)  & 0.2634306(2) & 0.701879(1) &     155.9$\pm$0.4  &  23.9$\pm$0.4   & 6.52$\pm$0.13  \\ % C11p1652 (6-3, Pest, FMMv2)
30470               &  2010-14\,(159,22)  & 17.262$\pm$0.002 & 0.3811721(7)  & 0.263756(2)  & 0.691960(6) &     193.3$\pm$2.7  &  29.9$\pm$3.3   & 6.46$\pm$0.79  \\ % C11p1649  (5-2,Pest, FMMv2)
28160               &  2010-17\,(927,38)  & 17.056$\pm$0.001 & 0.3268868(2)  & 0.2329577(4) & 0.712656(2) &     160.8$\pm$1.8  &  25.2$\pm$1.8   & 6.38$\pm$0.54  \\ % C11p1376 (6-3, Pest)
24500               &  2010-17\,(776,46)  & 18.340$\pm$0.002 & 0.3385730(3)  & 0.2404183(6) & 0.710093(2) &     165.9$\pm$2.4  &  26.5$\pm$2.4   & 6.27$\pm$0.67  \\ % C11p1368 (6-3, Pest)
29286               &  2010-17\,(681,42)  & 18.142$\pm$0.002 & 0.3882796(3)  & 0.2707951(5) & 0.697423(2) &     180.2$\pm$2.4  &  31.1$\pm$2.4   & 5.79$\pm$0.53  \\ % C11p1380-82 (N=681, 6-3 Pest FMMv2)
19058               &  2010-17\,(568,29)  & 17.724$\pm$0.001 & 0.3060771(1)  & 0.2140382(3) & 0.699295(1) &     178.7$\pm$1.8  &  31.8$\pm$1.9   & 5.61$\pm$0.40  \\ % C11p1350  (6-3, Pest)
32721$*$            &  2010-16\,(287,31)  & 17.060$\pm$0.002 & 0.3261655(3)  & 0.2301606(5) & 0.705656(2) &     178.4$\pm$3.2  &  32.6$\pm$3.2   & 5.47$\pm$0.63  \\ % C11p1390 (6-3, Pest))
02990               &  2001-17\,(1003,45) & 17.832$\pm$0.001 & 0.3530475(1)  & 0.2476270(3) & 0.701399(1) &     161.1$\pm$2.1  &  29.7$\pm$2.0   & 5.42$\pm$0.48  \\ % C11p1316-19 (6-3, Pest), 2001-17?
02217               &  2010-17\,(833,18)  & 16.392$\pm$0.001 & 0.3655628(2)  & 0.2569971(4) & 0.703018(1) &{\it 115.8$\pm$0.9} &  22.4$\pm$0.9   & 5.18$\pm$0.26  \\ % C11p1139 (6-3, FMMv2, Pest)
30974               &  2010-16\,(556,33)  & 17.660$\pm$0.002 & 0.2897323(1)  & 0.2042718(3) & 0.705036(1) &     208.0$\pm$2.3  &  40.2$\pm$2.3   & 5.17$\pm$0.33  \\ % C11p1385 (6-3, Pest, FMMv2)
02147               &  2010-17\,(834,17)  & 16.723$\pm$0.001 & 0.3522055(1)  & 0.2506667(2) & 0.711706(1) &     178.5$\pm$0.9  &  35.2$\pm$0.9   & 5.08$\pm$0.15  \\ % C11p1312-15,1625-26 (6-3)
03090$*$            &  2010-17\,(1758,52) & 17.221$\pm$0.001 & 0.3243208(2)  & 0.2308657(3) & 0.711844(1) &     171.1$\pm$1.7  &  33.7$\pm$1.7   & 5.07$\pm$0.28  \\ % C11p1320-23,1623-24 (7-4)
24360               &  2010-17\,(648,37)  & 17.933$\pm$0.002 & 0.3265030(2)  & 0.2300291(4) & 0.704524(2) &     179.1$\pm$2.3  &  36.4$\pm$2.1   & 4.92$\pm$0.38  \\ % C11p1364 (6-3, Pest))
01529$*$            &  2001-17\,(1048,29) & 16.934$\pm$0.001 & 0.3633225(1)  & 0.2545659(2) & 0.700661(1) &     173.5$\pm$1.3  &  35.3$\pm$1.3   & 4.91$\pm$0.22  \\ % C11p1311 (6-3, Pest, FMMv2, 2001-17 combined)
\underline{19121}$*$&  2010-17\,(815,35)  & 17.462$\pm$0.002 & 0.3575967(2)  & 0.2481799(4) & 0.694022(1) &     158.4$\pm$1.8  &  33.1$\pm$1.8   & 4.79$\pm$0.31  \\ % C11p1288 (6-3, FMMv2, Pest)
09117$*$            &  2010-17\,(14998,23)& 16.626$\pm$0.001 & 0.30967722(2) & 0.2190973(1) & 0.707502(1) &     185.4$\pm$0.3  &  39.4$\pm$0.3   & 4.71$\pm$0.04  \\ % C11p1328 (2010-17, 6-3, Pest); C11p1694 re-look
20645               &  2011-17\,(426,32)  & 17.752$\pm$0.002 & 0.3564776(2)  & 0.2462885(5) & 0.690895(2) &     171.8$\pm$3.1  &  37.2$\pm$2.7   & 4.62$\pm$0.38  \\ % C11p1352,1627-28 (6-3, Pest, FMMv2)
13745$*$            &  2010-17\,(1868,32) & 16.386$\pm$0.001 & 0.3567739(1)  & 0.2476187(2) & 0.694049(1) &     175.7$\pm$1.1  &  39.3$\pm$1.0   & 4.47$\pm$0.15  \\ % C11p1332 (6-3, FMMv2 Pest)
22305$*$            &  2010-17\,(780,67)  & 18.514$\pm$0.003 & 0.3406534(3)  & 0.2381711(5) & 0.699160(2) &     202.6$\pm$3.6  &  46.9$\pm$3.7   & 4.32$\pm$0.42  \\ % C11p1360 (6-3, Pest)
\underline{16999}$*$&  2010-17\,(223,33)  & 16.226$\pm$0.003 & 0.3258227(3)  & 0.2314176(6) & 0.710256(2) &     175.6$\pm$3.8  &  40.8$\pm$3.6   & 4.31$\pm$0.47  \\ % C11p1635 (5-2, Pest, FMMv2)
35838               &  2010-16\,(406,13)  & 16.082$\pm$0.001 & 0.31626814(7) & 0.2185164(2) & 0.690921(1) &     206.4$\pm$1.2  &  48.6$\pm$1.0   & 4.25$\pm$0.11  \\ % C11p1396 (6-3, Pest FMMv2)
07688               &  2010-17\,(1549,27) & 17.610$\pm$0.001 & 0.31543654(8) & 0.2181543(2) & 0.691595(1) &     195.1$\pm$1.0  &  47.2$\pm$1.0   & 4.13$\pm$0.11  \\ % C11p1324 (6-3, Pest, FMMv2)
14481               &  2010-16\,(542,12)  & 16.285$\pm$0.001 & 0.4097129(1)  & 0.2820237(3) & 0.688345(1) &     173.7$\pm$0.8  &  45.4$\pm$0.8   & 3.82$\pm$0.08  \\ % C11p1336 (6-3, Pest, FMMv2)
27203$*$            &  2010-16\,(404,36)  & 17.701$\pm$0.002 & 0.3211947(2)  & 0.2233162(4) & 0.695267(2) &     162.5$\pm$3.1  &  43.6$\pm$2.8   & 3.72$\pm$0.31  \\ % C11p1372-73 (6-3, Pest, FMMv2)
32923               &  2010-17\,(526,46)  & 18.270$\pm$0.002 & 0.3492864(3)  & 0.2450249(4) & 0.701501(2) &     162.8$\pm$3.1  &  44.9$\pm$3.1   & 3.63$\pm$0.32  \\ % C11p1395 (N=526, 6-3, FMMv2, Pest)
\underline{00595}   &  2010-17\,(824,12)  & 16.771$\pm$0.001 & 0.2818316(1)  & 0.1997130(1) & 0.708625(1) &{\it 112.5$\pm$0.7} &  31.7$\pm$0.6   & 3.55$\pm$0.09  \\ % C11p1655 (6-3, Pest, FMMv2,)
21400               &  2010-17\,(893,32)  & 17.940$\pm$0.001 & 0.3527255(2)  & 0.2538858(3) & 0.719783(1) &     152.6$\pm$1.6  &  43.5$\pm$1.6   & 3.50$\pm$0.16  \\ % C11p1356-57 (6-3, Pest, FMMv2)
15657               &  2010-16\,(156,28)  & 16.010$\pm$0.007 & 0.3665869(5)  & 0.2540171(7) & 0.692925(3) &     180.0$\pm$3.6  &  55.1$\pm$5.0   & 2.81$\pm$0.22  \\ % C11p1340; c17p594-596; C11p1657
00617               &  2001-17\,(933,52)  & 18.442$\pm$0.003 & 0.3381064(1)  & 0.2417129(2) & 0.714902(1) &     168.8$\pm$2.6  &  61.2$\pm$2.6   & 2.76$\pm$0.16  \\ % C11p1666 (6-3, Pest, fmmv2)
18798               &  2010-15\,(114,35)  & 18.174$\pm$0.013 & 0.3686818(9)  & 0.260384(1)  & 0.706255(5) &    ~\,181$\pm$12   & ~\,54$\pm$13      & 2.5$\pm$0.6    \\ [0.1cm] % C17p590 (see also C11p1344-47)
30908               &  2010-17\,(5849,88) & 18.635$\pm$0.001 & 0.2874647(2)  & 0.2031463(2) & 0.706683(1) &{\it ~\,99.5$\pm$1.6} &  38.0$\pm$1.6   & 2.62$\pm$0.16  \\ % C11p1663 (6-3 Pest FMMv2)
19847               &  2010-14\,(111,31)  & 18.264$\pm$0.005 & 0.3177105(9)  & 0.224449(1)  & 0.706457(6) &{\it 136.1$\pm$6.6} &  59.0$\pm$6.5   & 2.31$\pm$0.37  \\ % C11p1660 (5-2, Pest, FMMv2)
34006               &  2010-17\,(1852,34) & 17.988$\pm$0.001 & 0.3120725(2)  & 0.2194600(2) & 0.703234(1) &{\it ~\,90.4$\pm$1.2} &  43.7$\pm$1.2   & 2.07$\pm$0.08  \\ % C11p1669 (6-3, Pest, FMMv2)
31754               &  2010-17\,(6707,41) & 18.022$\pm$0.001 & 0.3173163(1)  & 0.2259796(1) & 0.712159(1) &{\it ~\,88.9$\pm$0.7} &  77.9$\pm$0.7   & 1.14$\pm$0.02  \\ % C11p1672 (3periods, Pest, FMMv2)
\bottomrule
\end{tabular}

\end{threeparttable}
}
\end{table*}

% Section 3.2.4
\subsubsection{Reanalysis of the \texttt{OGLE} photometry}

% See C11p1233 for 2nd star

The  photometry analyzed by P17 included all the available \texttt{OGLE}
observations made up to and including the 2013 observing season.  Since then
the \texttt{OGLE-IV} team has acquired four more years of photometry.
Primarily because of the extra observations, but also as a check on the
previously derived periods, amplitudes and their ratios, the photometric
observations made through 2017 for the 42 known pRRd stars were analyzed using
our non-linear least-squares multi-frequency fitting program.  The fits to the
data were conducted with six harmonics for the dominant mode and three
harmonics for the additional mode (including linear combination frequencies),
usually with no adjustments for trends but when necessary with additional
trend-fitting.  For those stars with \texttt{OGLE-II, -III} and \texttt{-IV}
$I$-band photometry the data sets were merged.  For several such stars
discontinuities in the photometric time series were found, in which case only
the \texttt{OGLE-IV} data were analyzed.

The resultant mean $I$-magnitudes, periods, first-term Fourier amplitudes and
their ratios are given in Table~8.  In general there is agreement with the
values given in Table\,1 of P17, and with the more recent values given in the
\texttt{OGLE-IV} on-line catalogue (http://ogledb.astrouw.edu.pl/ogle/OCVS/).
Ten of the 42 stars were found by P17 to exhibit Blazhko modulations of either
the dominant mode, the additional mode, or both modes (the ten stars are
identified with an asterisk following the star number in the first column of
Table\,8).  Except for the three stars with K2 and \texttt{OGLE} photometry
(see $\S$3.2.3) no attempt was made to identify additional Blazhko modulations
over and above those identified by P17.  In the discussion that follows the
newly derived periods and amplitudes, as well as additional Fourier
decomposition parameters, are used for the comparison of pRRd stars with
classical (cRRd) and anomalous RRd (aRRd) stars.

\section{DISCUSSION}

% Section 4.1
\subsection {Pulsation Periods}

% see p25 of PM comments see C19p501 for 1st page of Kunder et al. (2020)
In the Petersen diagram (Fig.\,1) the `peculiar' RRd stars lie {\it
very roughly} along the extension of the curve defined by the
`classical' RRd stars. While cRRd stars form a well-defined {\it
narrow} sequence in the diagram, the pRRd and aRRd stars form large
clouds with no apparent structure. The period ratios, $P_S$/$P_L$,
for the pRRd stars vary from 0.682 to 0.720, the range of which,
0.038, is almost twice as big as the range seen for cRRd stars,
0.022 (from 0.726 to 0.748), and all but four of the shortest-period
pRRd stars lie {\it below} the extrapolated cRRd sequence. While the
fundamental-mode periods of the shortest-period cRRd stars extend
down to $P_L$=0.35\,d, the longest-period pRRd star has
$P_L$=0.41\,d; thus, in the period range 0.35$<$$P_L$$<$0.41\,d the
two groups coexist. The 17 pRRd stars in this range amount to
40 per cent of the P17 sample and all have significantly smaller
period ratios than do the stars on the cRRd sequence.

% FIGURE 15 -  P1 vs P0  and AL-AS Diagrams for the K2 and Prudil RRd stars
\begin{figure*}
\begin{center}
\vskip0.1truecm
\begin{overpic}[width=8.2cm]{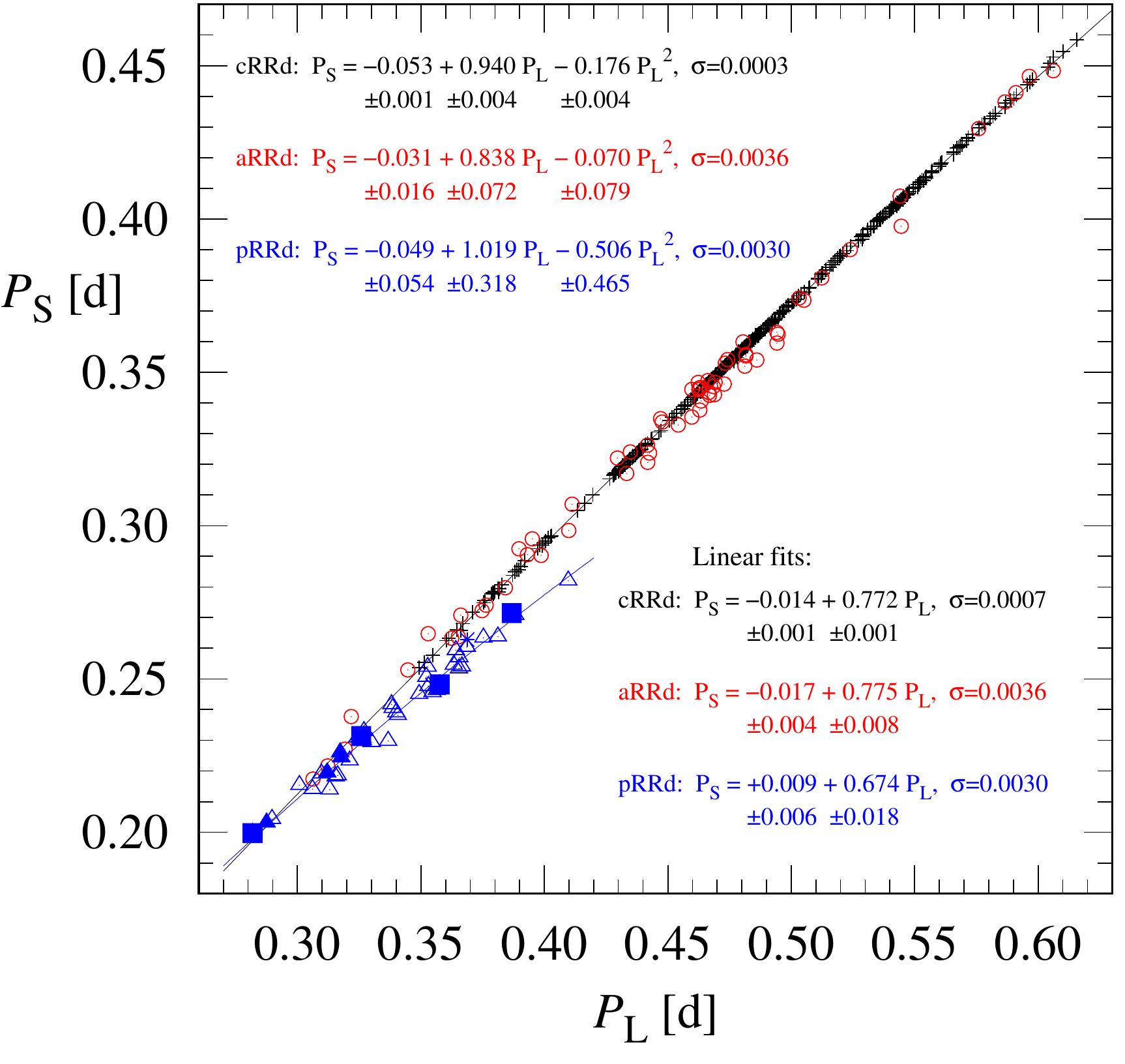} \put(80,87){(a)}   \end{overpic}  %C17Vol5p1037 (old:%C17Vol5p932,945,946);  renamed "Prudil_P1vsP0_2020Nov2.pdf"
\begin{overpic}[width=9.0cm]{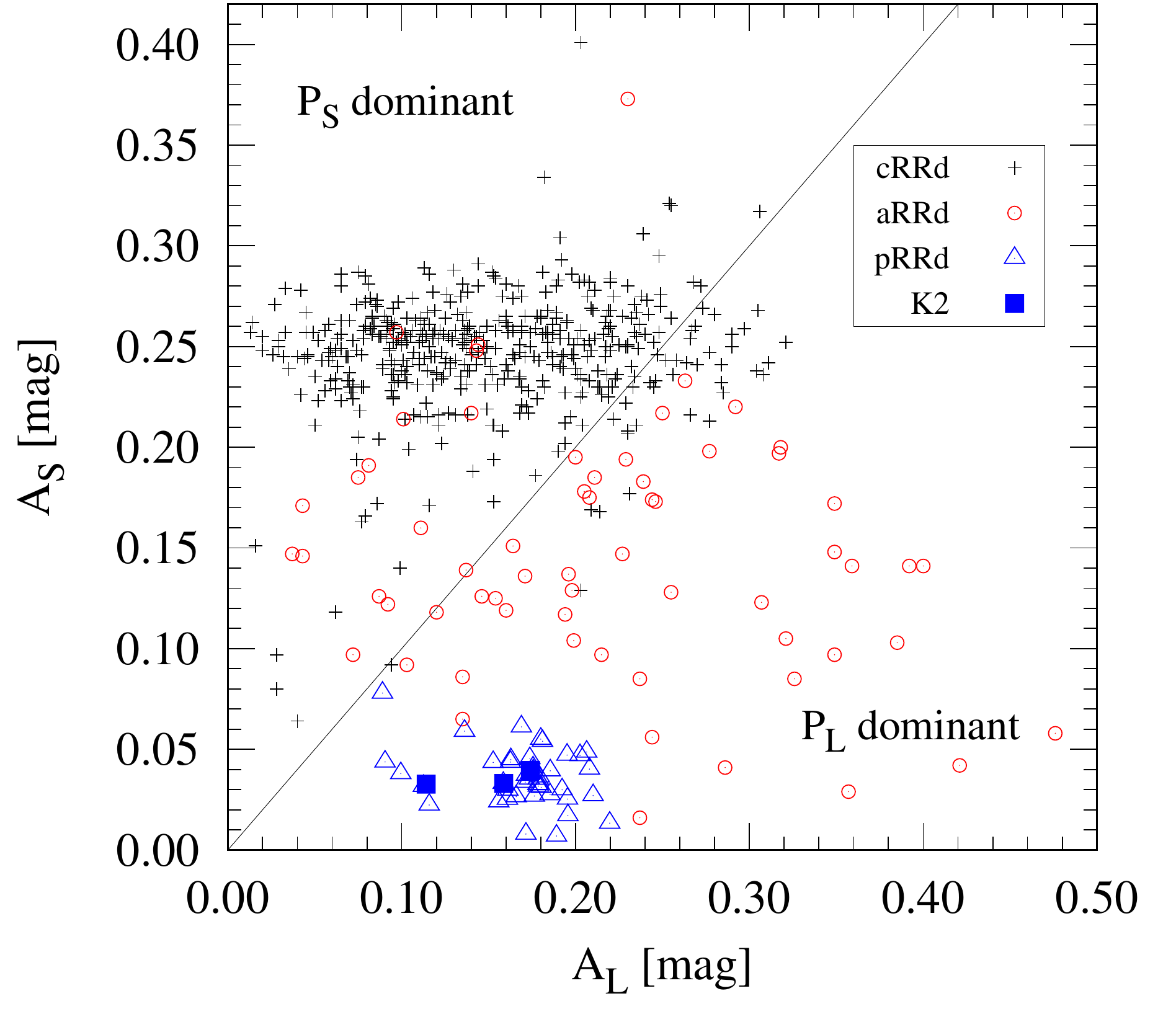} \put(83,78){(b)}   \end{overpic}  %C17Vol6p1040;  renamed "AvsA_2020Nov21.pdf"
\end{center}

\caption{(a) $P_L$-$P_S$ graph comparing the longer and shorter
pulsation periods for the RRd stars in Figs.\,1-2.  The four pRRd
stars observed by {\it K2} are plotted with large blue squares, and
the blue and black `lines' through the pRRd and cRRd points are
fitted least-squares quadratic curves, the equations of which (along
with linear fit equations) are given on the graph. (b) $A_L$-$A_S$
graph comparing the Fourier first-term $I$-amplitudes of the longer-
and shorter-period modes for the RRd stars in Figs.\,1-2.  The
diagonal 1:1 line separates the $P_S$-dominant from the
$P_L$-dominant RRd stars. EPIC\,216764000 is absent because no
\texttt{OGLE} $I$-passband photometry exists for it.}

\label{fig:PetersenDiagram}
\end{figure*}

% See K2C17p1014 for Fig. 10 of Prudil et al.
Assuming that the pRRd variable stars are single stars,
pulsationally unstable in the fundamental and first-overtone {\it
radial} modes as they evolve through the Instability Strip, and that
they have physical characteristics within the ranges seen for
`normal' RR~Lyrae stars, i.e., M\,$\in$\,(0.50--0.70)M$_{\odot}$,
L\,$\in$\,(40--70)L$_{\odot}$, $T_{\rm eff}$\,$\in$\,(5000--8000)K,
and [Fe/H] in the range $-2.5$\,dex (very metal poor) to 0.0\,dex
(solar), P17 were unable, using the linear nonadiabatic
pulsation models of Smolec et al. (2016), to reproduce the periods
and period ratios as low as those observed. The closest match was
with the most extreme model, that with the highest mass
M=0.70M$_{\odot}$, lowest luminosity L=40L$_{\odot}$, highest
metallicity [Fe/H]=0.0\,dex, and hottest effective temperature.
However, for the $T_{\rm eff}$ range where both modes are unstable
none of the models matched the observed periods and period ratios.
As a result, P17 rejected the conclusion that the additional
low-amplitude mode detected in the pRRd stars can be identified with
the first-overtone radial pulsation, i.e., that the pRRd stars are
simply very short-period cRRd stars.

% See K2V17p1023; see C17p741-751 for complete Chadid+(2010) paper
% Furthermore, analysis of V1127\,Aql by Szab\'o et al. (2014) revealed
% period doubling and a possible presence of nonradial pulsations.

A similar grid of linear models was calculated by Chadid et
al. (2010).  Their models, which are graphically summarized in their
fig.\,8, assumed L in the range 40--70L$_{\odot}$, M in the range
0.50--0.80M$_{\odot}$, and $T_{\rm eff}$ in the range 5000--8000\,K,
with metallicities ranging from metal poor, [Fe/H]=--2.3\,dex (i.e.,
Z=0.0001), to solar, 0.0\,dex (i.e., Z=0.0200). The models were
computed for fundamental-mode periods as short as 0.22\,d, i.e., for
the entire range of the observed longer-periods of the pRRd stars.
The most metal-rich of the models predicted period ratios that are
similar to those of the pRRd stars, but the match could be achieved
only for the highest period ratios observed, 0.71--0.72.  If models
more metal-rich than solar had been calculated by Chadid et al. the
locations in the Petersen diagram of the solutions probably would
have extended further into the pRRd domain. The possibility that
pRRd stars are more metal-rich than the Sun needs to be tested with
spectra, EPIC\,216764000 with $<${\it Kp}$>$=13.74\,mag being a good
candidate for such spectroscopy.

Comparing the models of Chadid et al. and the models of Smolec
et al. we see that they agree remarkably well. For [Fe/H]=0.0\,dex
(red dots in fig.\,10 of P17 and black dots in fig.\,8 of Chadid et
al.) the period ratios, $P_S$/$P_L$, are almost the same in
both computations. An excellent agreement of the computed period
ratios is also seen for [Fe/H]=--1.0\,dex. Both sets of models show
that the period ratio has very little sensitivity to mass or
luminosity and depends mostly on [Fe/H]. Where there is a difference
is that P17 considered in their analysis also the growth rates
of the modes.  This information was used as an additional
constraint, requiring that both radial modes have to be linearly
unstable, thus limiting the period range of the acceptable models. It
is for this reason that the P17 models do not extend to
very short periods. For such short-period models at least one
of the radial modes becomes damped and so the models must be
rejected. Chadid et al. did not use the information about
stability of the modes and no such additional constraint was
imposed in their fig.\,8.

To sum up, the linear models of Chadid et al. and Smolec et
al. fail to describe the pRRd stars in two ways: (1) they do not
reproduce period ratios below 0.71; and (2) the pulsation modes are
not unstable for the short periods of the pRRd stars, i.e., the
models lie outside of the instability strip. The period
ratio discrepancy can be easily fixed by assuming
[Fe/H]$>$0.0\,dex, but increasing [Fe/H] most likely will not fix
the growth-rates discrepancy at the short periods. In fact, it
could make it worse -- judging from fig.\,10 of the P17 paper the
shortest period at which both radial modes are unstable becomes even
longer as the [Fe/H] increases.

The Chadid et al. (2010) models were computed for their analysis of
the 15th-magnitude star V1127\,Aql.  With dominant period
$P_0$=0.3560\,d, additional period $P_1$=0.2480\,d, and period ratio
$P_1$/$P_0$=0.6966,  its location in the Petersen diagram is in the
zone occupied by pRRd stars (see Fig.\,1 where it is represented by
a blue cross). Based on the observed periods and the calculated
models V1127\,Aql was considered by Chadid et al. to be a possible
double-mode RR~Lyrae star, but was rejected as such on the grounds
that the period ratio does not fit to the relation defined by the
cRRd stars of our Galaxy. Finally, V1127\,Aql also lies in the
direction of the Galactic Bulge ($l$=38$^\circ\,\!\!\!\!.\;\!$1804,
$b$=--6$^\circ\,\!\!\!\!.\;\!$458), and like many of the known pRRd
stars it too exhibits a Blazhko modulation ($P_{\rm BL}$=26.9\,d).
Thus, the amplitudes for both modes are variable, as is the case for
some of the pRRd stars, e.g., \texttt{OGLE-BLG-RRLYR-09117}. We
conclude that V1127\,Aql almost certainly is another member of the
group of `peculiar' RRd stars.

Another possible pRRd star is AH\,Cam. With dominant period $P_{\rm
L}$=0.368716\,d, secondary period $P_{\rm S}$=0.2628\,d, and $P_{\rm
S}$/$P_{\rm L}$=0.7130, it too lies in the pRRd region of the Petersen diagram
(see Fig.\,1). It is metal-rich with [Fe/H] $>$ --0.5\,dex and like many pRRd
stars it exhibits Blazhko variations ($P_B$=10.8289$\pm$0.0002\,d -- see Smith
et al. 1994; Le\,Borgne et al. 2012). The amplitude of the dominant mode,
$A_{1,D}$=201.3\,mmag, too, is similar to that for pRRd stars.

Also of interest are the three short-period aRRd stars (red circles) that
reside in Fig.\,1 among the pRRd stars.  The three stars were discovered and
classified by Soszy\'nski et al. (2019) and have \texttt{OGLE} identifications
\texttt{OGLE-BLG-RRLYR-60063}, \texttt{OGLE-GD-RRLYR-02968} and
\texttt{OGLE-GD-RRLYR-08752}. Unlike the pRRd stars for which the additional
amplitudes are always much smaller than the dominant amplitudes (see $\S$4.2)
the primary and secondary amplitudes for the three stars are of comparable
value and relatively large (see fig.\,4 of Soszy\'nski et al. 2019). In this
sense all three are more like very-short-period cRRd stars than pRRd stars,
which would make them the shortest-period cRRd stars yet identified and thus of
considerable interest. With mean $I$ magnitudes ranging from 14.6 to 16.2 their
metallicities might be measurable using spectra taken with a large telescope.

In Figure~15(a) the Petersen diagram (Fig.\,1) is viewed from
another perspective. In the $P_L$-$P_S$ diagram the two periods for
all three types of RRd stars appear to be linearly correlated, the
equations of which are given on the graph. Inspection of the
residuals revealed that for the cRRd stars the correlation is
non-linear, a result confirmed when the cRRd points were fit with a
quadratic formula, the rms-error derived from the quadratic fit
being less than half the size of the rms-error derived from the
linear fit, i.e., $\sigma$=0.0003 vs.  0.0007.  When quadratics were
fit to the pRRd and aRRd points their $\sigma$ values remained
unchanged, suggesting that linear fits are sufficient for describing
their $P_L$-$P_S$ relations.  The aRRd relation resembles the cRRd
relation but with larger scatter. Given the large cloud-like nature
of the aRRd distribution in the Petersen diagram this result was
somewhat surprising. Fig.\,15(a) also shows that for the pRRd stars
the slope of the $P_L$-$P_S$ relation, 0.674$\pm$0.018, is
significantly shallower than that for the cRRd (and aRRd) stars,
0.772$\pm$0.001.  If the pRRd stars are simply short-period cRRd
stars this would suggest that for a given $P_L$ the `additional'
period of the pRRd stars is shorter than expected. We regard the
different slopes as strong evidence that there is a fundamental
difference between the pRRd and the cRRd (or aRRd) stars.

In the Petersen diagram (Fig.\,1) the mean curve (black) plotted
through the cRRd points follows from the Fig.\,15(a) fitted
quadratic $P_L$-$P_S$ relation. Extrapolation of the trendline to
periods shorter than 0.35\,d shows that almost all of the pRRd stars
lie below the extrapolated cRRd curve.  The corresponding mean
trendline through the pRRd data points was not plotted in Fig.\,1
(owing to the highly uncertain first and third coefficients for the
pRRd quadratic fit).

%  TABLE 9
\begin{table*}
{\fontsize{6}{7.2}\selectfont  %  <--- works and better; 2nd (skip) should be 1.2 times 1st (size)
\begin{threeparttable}
\fontsize{6}{7.2}\selectfont  %  <--- works and better; 2nd (skip) should be 1.2 times 1st (size)

\caption{Fourier decomposition amplitude-ratio, $R_{\rm 21}$, and
phase-difference, $\phi^c_{\rm 21}$, parameters for the long-period ($P_L$) and
short-period ($P_S$) pulsation modes of the 42 pRRd stars with
\texttt{OGLE}-survey $I$-photometry.  Also given are the derived
light-curve-shape indices, $\Delta$$R_{\rm 21}$ and $\Delta$$\phi_{\rm 21}$
(see text for definitions, and Fig.16).  As in Table\,8 the star numbers are
\underline{underlined} for the three stars observed by {\it K2}, and followed by
an asterisk if either mode exhibits Blazhko modulations.}

\begin{tabular}{llclcrccr}
\toprule
\multicolumn{1}{c}{\texttt{OGLE}} & \multicolumn{1}{c}{$P_L$ [d]} & \multicolumn{1}{c}{ $P_S$/$P_L$}  & \multicolumn{1}{c}{ $R_{\rm 21,L}$} &
    \multicolumn{1}{c}{ $R_{\rm 21,S}$ } & \multicolumn{1}{c}{ $\Delta$$R_{\rm 21}$ } & \multicolumn{1}{c}{ $\phi^c_{\rm 21,L}$ } & \multicolumn{1}{c}{ $\phi^c_{\rm 21,S}$ } &
    \multicolumn{1}{c}{ $\Delta$$\phi_{\rm 21}$ }    \\
    \multicolumn{1}{c}{(1)}   & \multicolumn{1}{c}{(2)}  & (3)  & \multicolumn{1}{c}{(4)}  & \multicolumn{1}{c}{(5)} & \multicolumn{1}{c}{(6)} & (7)  & (8) & \multicolumn{1}{c}{(9)}  \\
\midrule
\underline{595}      & 0.2818316 & 0.708625 & 0.387 $\pm$ 0.006 & 0.134 $\pm$ 0.021 & 0.25 $\pm$ 0.02  & 4.477 $\pm$   0.018 & 4.71    $\pm$   0.16  & -0.23    $\pm$     0.16  \\
617                  & 0.3381064 & 0.714902 & 0.320 $\pm$ 0.016 & 0.083 $\pm$ 0.042 & 0.24 $\pm$ 0.04  & 4.530 $\pm$   0.056 & 4.29    $\pm$   0.50  &  0.24    $\pm$     0.50  \\
1401                 & 0.3130437 & 0.683301 & 0.386 $\pm$ 0.005 & 0.193 $\pm$ 0.077 & 0.19 $\pm$ 0.08  & 4.485 $\pm$   0.015 & 3.44    $\pm$   0.42  &  1.05    $\pm$     0.42  \\
1529$*$              & 0.3633225 & 0.700661 & 0.352 $\pm$ 0.008 & 0.064 $\pm$ 0.037 & 0.29 $\pm$ 0.04  & 4.592 $\pm$   0.025 & 5.25    $\pm$   0.58  & -0.66    $\pm$     0.58   \\
2147                 & 0.3522055 & 0.711706 & 0.324 $\pm$ 0.005 & 0.093 $\pm$ 0.026 & 0.23 $\pm$ 0.03  & 4.620 $\pm$   0.018 & 4.35    $\pm$   0.29  &  0.27    $\pm$     0.29  \\
2217                 & 0.3655628 & 0.703018 & 0.319 $\pm$ 0.009 & 0.135 $\pm$ 0.043 & 0.18 $\pm$ 0.04  & 4.584 $\pm$   0.029 & 4.37    $\pm$   0.31  &  0.22    $\pm$     0.31   \\
2943                 & 0.3367005 & 0.682446 & 0.340 $\pm$ 0.007 & 0.255 $\pm$ 0.190 & 0.09 $\pm$ 0.19  & 4.559 $\pm$   0.023 & 2.91    $\pm$   0.79  &  1.65    $\pm$     0.79  \\
2990                 & 0.3530475 & 0.701399 & 0.291 $\pm$ 0.014 & 0.123 $\pm$ 0.070 & 0.17 $\pm$ 0.07  & 4.588 $\pm$   0.051 & 3.96    $\pm$   0.58  &  0.62    $\pm$     0.58  \\
3090$*$              & 0.3243208 & 0.711844 & 0.317 $\pm$ 0.010 & 0.036 $\pm$ 0.050 & 0.28 $\pm$ 0.05  & 4.580 $\pm$   0.037 & 5.41    $\pm$   1.37  & -0.83    $\pm$     1.38  \\
5851                 & 0.3546875 & 0.692527 & 0.364 $\pm$ 0.005 & 0.086 $\pm$ 0.050 & 0.28 $\pm$ 0.05  & 4.562 $\pm$   0.015 & 5.15    $\pm$   0.59  & -0.59    $\pm$     0.59  \\
7688                 & 0.3154365 & 0.691595 & 0.359 $\pm$ 0.005 & 0.026 $\pm$ 0.021 & 0.33 $\pm$ 0.02  & 4.521 $\pm$   0.017 & 3.85    $\pm$   0.82  &  0.67    $\pm$     0.82  \\
8621                 & 0.3300003 & 0.695144 & 0.314 $\pm$ 0.023 & 0.246 $\pm$ 0.144 & 0.07 $\pm$ 0.15  & 4.537 $\pm$   0.080 & 6.56    $\pm$   0.61  & -2.02    $\pm$     0.61  \\
9117$*$              & 0.3096772 & 0.707502 & 0.379 $\pm$ 0.002 & 0.102 $\pm$ 0.007 & 0.28 $\pm$ 0.01  & 4.468 $\pm$   0.005 & 5.04    $\pm$   0.07  & -0.57    $\pm$     0.07  \\
12880$*$             & 0.3400408 & 0.703021 & 0.380 $\pm$ 0.003 & 0.168 $\pm$ 0.019 & 0.21 $\pm$ 0.02  & 4.534 $\pm$   0.008 & 3.93    $\pm$   0.12  &  0.60    $\pm$     0.12  \\
13745$*$             & 0.3567739 & 0.694049 & 0.340 $\pm$ 0.006 & 0.044 $\pm$ 0.027 & 0.30 $\pm$ 0.03  & 4.620 $\pm$   0.022 & 6.19    $\pm$   0.62  & -1.57    $\pm$     0.62  \\
14063                & 0.3753218 & 0.701879 & 0.307 $\pm$ 0.003 & 0.125 $\pm$ 0.017 & 0.18 $\pm$ 0.02  & 4.679 $\pm$   0.010 & 4.31    $\pm$   0.14  &  0.37    $\pm$     0.14  \\
14135                & 0.3008045 & 0.715857 & 0.339 $\pm$ 0.008 & 0.066 $\pm$ 0.083 & 0.27 $\pm$ 0.08  & 4.599 $\pm$   0.026 & 5.31    $\pm$   1.27  & -0.71    $\pm$     1.27  \\
14481                & 0.4097129 & 0.688345 & 0.333 $\pm$ 0.005 & 0.075 $\pm$ 0.017 & 0.26 $\pm$ 0.02  & 4.680 $\pm$   0.016 & 4.09    $\pm$   0.23  &  0.59    $\pm$     0.23  \\
15657                & 0.3665869 & 0.692925 & 0.407 $\pm$ 0.044 & 0.175 $\pm$ 0.091 & 0.23 $\pm$ 0.10  &  4.81 $\pm$   0.10  & \multicolumn{1}{c}{$\dots$} & \multicolumn{1}{c}{$\dots$} \\
15842                & 0.3654351 & 0.693453 & 0.320 $\pm$ 0.002 & 0.052 $\pm$ 0.046 & 0.27 $\pm$ 0.05  & 4.570 $\pm$   0.008 & 5.02    $\pm$   0.87  & -0.45    $\pm$     0.87  \\
\underline{16999}$*$ & 0.3258227 & 0.710256 & 0.388 $\pm$ 0.024 & 0.125 $\pm$ 0.092 & 0.26 $\pm$ 0.10  & 4.566 $\pm$   0.068 & 3.90    $\pm$   0.68  &  0.67    $\pm$     0.69  \\
18798                & 0.3686818 & 0.706255 & 0.483 $\pm$ 0.140 & $\dots$ & \multicolumn{1}{c}{$\dots$} & 4.82 $\pm$ 0.30  & \multicolumn{1}{c}{$\dots$} & \multicolumn{1}{c}{$\dots$} \\
19058                & 0.3060771 & 0.699295 & 0.362 $\pm$ 0.011 & 0.039 $\pm$ 0.057 & 0.32 $\pm$ 0.06  & 4.598 $\pm$   0.068 & 5.41    $\pm$   1.57  & -0.81    $\pm$     1.58  \\
\underline{19121}$*$ & 0.3575967 & 0.694022 & 0.342 $\pm$ 0.012 & 0.042 $\pm$ 0.054 & 0.30 $\pm$ 0.07  & 4.577 $\pm$   0.040 & 3.51    $\pm$   1.28  &  1.06    $\pm$     1.28  \\
19847                & 0.3177105 & 0.706457 & 0.275 $\pm$ 0.052 & 0.281 $\pm$ 0.128 &-0.01 $\pm$ 0.14  &  4.36 $\pm$   0.18  & 4.51    $\pm$   0.33  & -0.15    $\pm$     0.38  \\
20645                & 0.3564776 & 0.690895 & 0.367 $\pm$ 0.018 & 0.161 $\pm$ 0.071 & 0.21 $\pm$ 0.07  & 4.675 $\pm$   0.056 & 4.21    $\pm$   0.50  &  0.47    $\pm$     0.50  \\
21400                & 0.3527255 & 0.719783 & 0.357 $\pm$ 0.011 & 0.051 $\pm$ 0.037 & 0.31 $\pm$ 0.04  & 4.556 $\pm$   0.035 & 4.01    $\pm$   0.73  &  0.54    $\pm$     0.73  \\
22305$*$             & 0.3406534 & 0.699160 & 0.315 $\pm$ 0.019 & 0.049 $\pm$ 0.078 & 0.27 $\pm$ 0.08  & 4.598 $\pm$   0.068 & \multicolumn{1}{c}{$\dots$} & \multicolumn{1}{c}{$\dots$} \\
24360                & 0.3265030 & 0.704524 & 0.337 $\pm$ 0.013 & 0.059 $\pm$ 0.059 & 0.28 $\pm$ 0.06  & 4.506 $\pm$   0.044 & 4.13    $\pm$   1.08  &  0.37    $\pm$     1.09  \\
24500                & 0.3385730 & 0.710093 & 0.385 $\pm$ 0.016 & 0.220 $\pm$ 0.097 & 0.17 $\pm$ 0.10  & 4.546 $\pm$   0.048 & 2.99    $\pm$   0.45  &  1.55    $\pm$     0.45  \\
27203$*$             & 0.3211947 & 0.695267 & 0.328 $\pm$ 0.019 & 0.143 $\pm$ 0.068 & 0.19 $\pm$ 0.07  & 4.582 $\pm$   0.074 & 4.90    $\pm$   0.47  & -0.32    $\pm$     0.48  \\
28160                & 0.3268868 & 0.712656 & 0.323 $\pm$ 0.012 & 0.129 $\pm$ 0.075 & 0.19 $\pm$ 0.08  & 4.617 $\pm$   0.043 & 4.49    $\pm$   0.58  &  0.13    $\pm$     0.59  \\
29286                & 0.3882796 & 0.697423 & 0.302 $\pm$ 0.014 & 0.119 $\pm$ 0.082 & 0.18 $\pm$ 0.08  & 4.693 $\pm$   0.050 & 3.63    $\pm$   0.64  &  1.06    $\pm$     0.65  \\
30418                & 0.3640956 & 0.712373 & 0.337 $\pm$ 0.012 & 0.166 $\pm$ 0.079 & 0.17 $\pm$ 0.08  & 4.534 $\pm$   0.042 & 4.21    $\pm$   0.49  &  0.32    $\pm$     0.50  \\
30470                & 0.3811710 & 0.691950 & 0.266 $\pm$ 0.036 & 0.104 $\pm$ 0.132 & 0.16 $\pm$ 0.14  &  4.48 $\pm$   0.14  & \multicolumn{1}{c}{$\dots$} & \multicolumn{1}{c}{$\dots$} \\
30908                & 0.2874647 & 0.706683 & 0.276 $\pm$ 0.017 & 0.236 $\pm$ 0.044 & 0.04 $\pm$ 0.05  & 4.296 $\pm$   0.068 & 5.06    $\pm$   0.20  & -0.76    $\pm$     0.21  \\
30974                & 0.2897323 & 0.705036 & 0.367 $\pm$ 0.012 & 0.039 $\pm$ 0.053 & 0.33 $\pm$ 0.05  & 4.466 $\pm$   0.035 & 3.92    $\pm$   1.42  &  0.54    $\pm$     1.42  \\
31754                & 0.3173163 & 0.712159 & 0.181 $\pm$ 0.008 & 0.131 $\pm$ 0.009 & 0.05 $\pm$ 0.01  & 4.403 $\pm$   0.047 & 5.02    $\pm$   0.07  & -0.61    $\pm$     0.09  \\
32721$*$             & 0.3261655 & 0.705656 & 0.330 $\pm$ 0.019 & 0.076 $\pm$ 0.101 & 0.25 $\pm$ 0.10  & 4.406 $\pm$   0.068 & \multicolumn{1}{c}{$\dots$} & \multicolumn{1}{c}{$\dots$} \\
32923                & 0.3492864 & 0.701501 & 0.320 $\pm$ 0.020 & 0.095 $\pm$ 0.070 & 0.23 $\pm$ 0.07  & 4.590 $\pm$   0.072 & 5.00    $\pm$   0.73  & -0.41    $\pm$     0.73  \\
34006                & 0.3120725 & 0.703234 & 0.221 $\pm$ 0.013 & 0.131 $\pm$ 0.028 & 0.09 $\pm$ 0.03  & 4.363 $\pm$   0.062 & 5.11    $\pm$   0.22  & -0.74    $\pm$     0.23  \\
35838                & 0.3162681 & 0.690921 & 0.374 $\pm$ 0.006 & 0.062 $\pm$ 0.022 & 0.31 $\pm$ 0.02  & 4.552 $\pm$   0.017 & 3.75    $\pm$   0.36  &  0.81    $\pm$     0.36  \\
\bottomrule
\end{tabular}
\end{threeparttable}
}
\end{table*}

% Section 4.2
\subsection {Light Curve Amplitudes and Shapes}

% Unfortunately the linear pulsation models of Chadid et al. and Smolec
% et al. give no information on amplitudes or light curve shapes.

% $A_{1,A}/A_{1,D} \sim 0.05$-$0.90$, with typical values $\sim$0.2 (see fig.\,2
% of P17 and our Fig.\,2).

The original 42 pRRd stars were found by P17 to divide into two groups based on
the {\it Fourier first-term} amplitudes of the {\it dominant} oscillations: 36
stars with amplitudes in the range 140$<$$A_{1,D}$$<$220\,mmag, and six stars
with amplitudes $<$140\,mmag (see fig.\,2 of P17).  Our analysis of the larger
2001-2017 \texttt{OGLE} data set confirms this result, in particular the low
$A_{1,D}$ amplitudes for the six low-amplitude stars (see column 7 of
Table\,8).  P17 also found that when {\it total} amplitudes and Fourier
amplitude parameters $R_{\rm 21}$ and $R_{\rm 31}$ for the dominant
oscillations are plotted vs. $P_L$ the division into two groups persists (see
figs.2 and 3 of P17), although amplitude ratios are low only in some of the
low-amplitude pRRd stars.  However, the apparent division went away in the
amplitude-independent $P_L$-$\phi^s_{\rm 21}$ and $P_L$-$\phi^s_{\rm 31}$
phase-difference graphs shown in the bottom two panels of the same figure,
where both indices form tight progressions with all the pRRd stars having
similar $\phi^s_{\rm 21}$ and $\phi^s_{\rm 31}$ characteristics.  In fact, in
all the various fig.\,3 graphs of P17 the locations of the higher-amplitude
pRRd stars lie between those of the {\it single-mode} RRc and RRab pulsators.
Turning to the phased light curves for the dominant mode, after filtering out
the contributions from the additional mode, all 42 stars were seen to have
asymmetric dominant light curves (see fig.\,4 of P17). Based on the entirety of
the observations P17 concluded that despite their very short periods,
$P_L$=0.28--0.41\,d, the {\it dominant} oscillations of {\it all} 42 pRRd stars
correspond to fundamental-mode pulsation.

On the other hand the pulsation modes of the {\it additional} oscillations were
left unexplained.  Also uncertain were the shapes of the secondary light
curves. Due in large part to the relatively low amplitudes (see fig.\,2 of P17,
and column~8 of our Table~8), and in the absence of high-precision photometry,
P17 assumed sinusoidal light curve shapes for the additional pulsations. We can
now say, based on our analysis of the high-precision {\it Kepler} photometry,
that none of the four pRRd stars observed by {\it K2} has a sinusoidal
secondary light curve.  Perhaps the strongest evidence of the asymmetries is
given by the risetimes summarized in Table~4. While a sinusoidal light curve
would have RT=0.50, the four {\it K2} stars have risetimes for the `additional'
oscillations, RT$_A$, ranging from 0.421$\pm$0.002 to 0.472$\pm$0.001. By way
of comparison the light curves of the {\it dominant} longer-period oscillations
for the same four stars are much more asymmetric and have RT$_D$ values ranging
from 0.235$\pm$0.001 to 0.289$\pm$0.001.  The non-sinusoidal nature of the
secondary oscillations is also evident in the low but non-zero Fourier $R_{\rm
21}$ values, which for the four {\it K2} stars range from 0.048$\pm$0.005 to
0.15$\pm$0.04, and are significantly smaller than the values for the dominant
oscillations, which range from 0.324$\pm$0.008 to 0.398$\pm$0.001 (see
Table~4).

Not considered by P17 was how the amplitudes and amplitude ratios of the pRRd
stars compare with those for cRRd (and aRRd) stars. The $P_L$-$A_S/A_L$ diagram
plotted in Fig.\,2 shows that the amplitude ratios of the cRRd stars tend to
decrease as $P_L$ decreases, with a possible discontinuity occuring around
$P_L$$\sim$0.45\,d, and that this trend appears to continue through to the pRRd
stars.  If the sample were limited to classical RRd stars in globular clusters
one would see a clear trend in going from the long-period low-metallicity OoII
cRRd stars to the intermediate-period intermediate-metallicity OoI cRRd stars
(e.g., fig.\,2 of Clement, Ferance \& Simon 1993).

% FIGURE 16  R21(I) vs P_L diagram   (see C17p976 for source)
\begin{figure*}
\begin{center}
\vskip0.1truecm
\begin{overpic}[width=8.5cm]{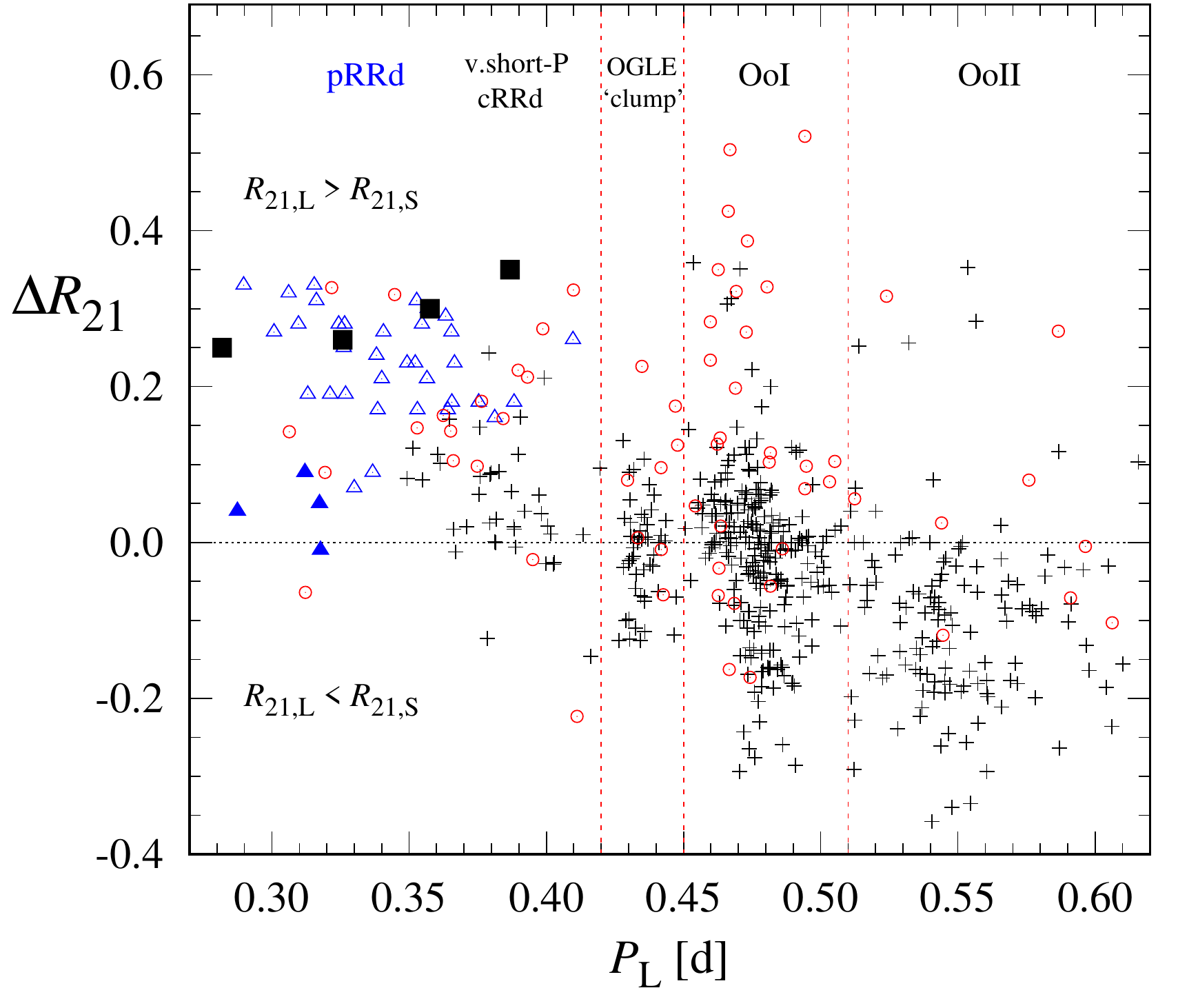}  \put(20,79){(a)}   \end{overpic}  % C17p1011, p1000, K2C17vol6p1051 renamed from DeltaR21_vs_PL_2021May14.pdf"
\begin{overpic}[width=8.4cm]{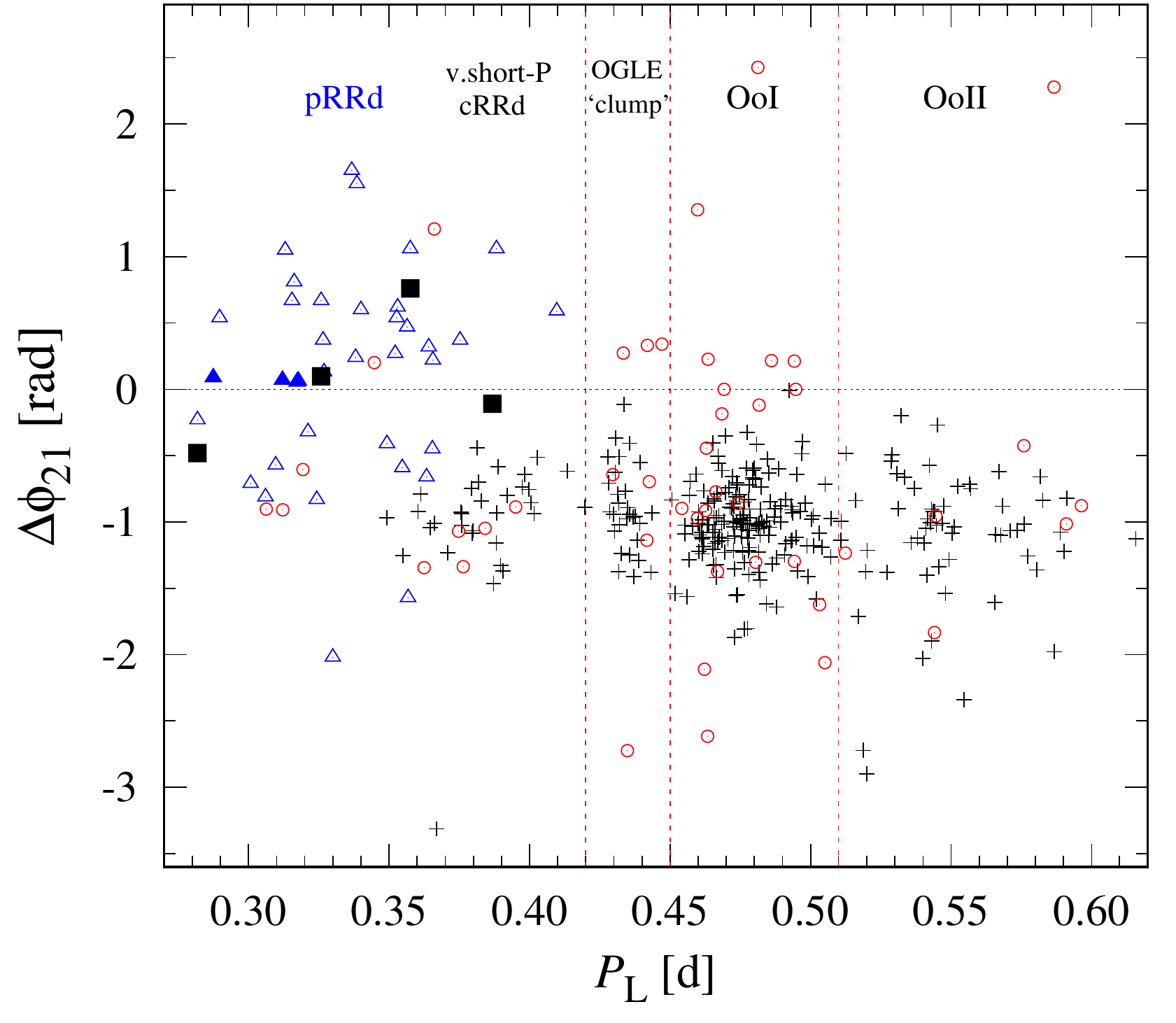}  \put(18,80){(b)}   \end{overpic}  % C17p976, C17p1006-7,  C17p1012  renamed "PHI21_vs_PL_2020Sept3.pdf"
\end{center}

\caption{(a) $\Delta R_{\rm 21}$ vs. $P_{\rm L}$ diagram, and (b)
$\Delta \phi_{\rm 21}$ vs. $P_{\rm L}$ diagram, for the pRRd stars
(blue triangles, and large black-filled squares for the four stars
observed by K2), cRRd stars (black plus signs) and aRRd stars (red
open circles) in Figs.\,1-2. The Fourier amplitude ratios and phase
differences are all derived from \texttt{OGLE} $I$-photometry
(except the points for EPIC\,216764000 which were derived from the
{\it K2} photometry). Typical measurement rms-errors for the
pRRd stars are $\pm$0.06 for $\Delta R_{\rm 21}$ and $\pm$0.66\,rad
for $\Delta \phi_{\rm 21}$; the \texttt{OGLE} online catalog does
not report measurement errors for cRRd and aRRd stars.}

\label{fig:R21-P_diagram}
\end{figure*}

Another perspective on the information contained in Fig.\,2  is given by the
$A_L$-$A_S$ diagram plotted in Figure~15(b). According to this graph the
fundamental difference between the pRRd and cRRd stars is that the
shorter-period {\it additional} amplitudes, $A_S$, of all the pRRd stars are
significantly lower than the $A_S$ amplitudes of almost all of the cRRd stars.
Even the shortest-period cRRd stars which have periods and period ratios most
similar to the pRRd stars, many of which have $P_L$ dominant like the pRRd
stars (see Fig.\,2),  have significantly larger $A_S$ amplitudes
($\sim$0.24\,mag) than the $A_S$ amplitudes of the pRRd stars
($\sim$0.03\,mag). Turning to the amplitudes of the longer-period (fundamental)
modes, the $A_L$ values of the pRRd stars and of the cRRd stars are on average
not too different, and this is so despite the range of the $A_L$ values being
twice as narrow for the former group of stars. This result again supports the
P17 conclusion that the dominant oscillations probably are fundamental-mode
radial pulsations.

Fig.\,15(b) also shows that the main cause of the wide range of amplitude
ratios for the cRRd stars (see Fig.\,2) is the large range of amplitudes for
the longer-period oscillations, $A_L$=0.01--0.32\,mag, the range of amplitudes
for the shorter-period first-overtone mode being fairly small for most of the
cRRd stars ($A_S$=0.2--0.3\,mag). Almost all of the cRRd stars with
$A_S<0.2$\,mag also tend to have low $A_L$ values, i.e., a dominant
first-overtone.

Table~9 contains Fourier decomposition  $R_{\rm 21}$ and  $\phi^c_{\rm 21}$
parameters (including error estimates) for the dominant longer-period and
additional shorter-period pulsation modes of the 42 pRRd stars, all derived
from \texttt{OGLE} $I$-photometry;  neither parameter was computed by P17 for
either mode.
% (and the corresponding parameters given in the \texttt{OGLE} catalog were
% computed assuming only a single mode).   $R_{\rm 31}$ and  $\phi^c_{\rm 31}$
% parameters were also computed, but owing to the low amplitudes of the
% additional mode they are much less accurate and therefore are not reported.
Columns (6) and (9) contain two additional indices, both of which relate to the
shapes of the light curves: $\Delta$$R_{\rm 21}$, defined as the difference
between the $R_{\rm 21}$ amplitude ratio for the two modes, i.e., $R_{\rm 21,L}
- R_{\rm 21,S}$;   and $\Delta$$\phi_{\rm 21}$ defined as the difference
between the   $\phi^c_{\rm 21}$ phase difference for the two modes, i.e.,
$\phi^c_{\rm 21,L} - \phi^c_{\rm 21,S}$.   The uncertainties given in Table\,9
for the indices are only approximate (since they do not include possible
covariance contributions).

% see C17p995 for R21 vs PL graph
The $P_L$-$\Delta R_{\rm 21}$ graph shown in Figure\,16(a) compares  the light
curve shapes of the pRRd stars with those of cRRd and aRRd stars.   The $\Delta
R_{\rm 21}$ values for the cRRd and aRRd stars were computed from the $R_{\rm
21,1}$ and $R_{\rm 21,2}$ amplitude ratios reported in the \texttt{OGLE} online
catalogue (note that since the \texttt{OGLE} catalog does not give error estimates for
the reported Fourier parameters  the uncertainties of the $\Delta R_{\rm 21}$
and $\Delta \phi_{\rm 21}$ values for the aRRd and cRRd stars in Fig.16 are
unknown).
% Typical measurement rms-errors for the pRRd stars are $\pm$0.06 for $\Delta
% R_{\rm 21}$ and 0.66 radians for  $\Delta \phi_{\rm 21}$; the \texttt{OGLE}
% online catalog does not report measurement errors.
The horizontal line at $\Delta R_{\rm 21}$=0 separates the RRd stars into those
stars where $R_{\rm 21,L}$$<$$R_{\rm 21,S}$, which appears to be the case for
almost all the OoII cRRd stars and half the OoI and \texttt{OGLE}-`clump' cRRd
stars, and  those with $R_{\rm 21,L}$$>$$R_{\rm 21,S}$, which is the case for
the (metal-rich) very-short-period cRRd stars, the aRRd stars, and the pRRd
stars.  Of particular interest here is the clear trend when moving from the
longest- to the shortest-period RRd stars, which, if the pRRd and cRRd stars
have distinctly different pulsation types, still requires a physical
explanation.

Fourier phase-difference indices for the three types of RRd stars are compared
in Figure\,16(b).  Almost all of the cRRd stars, regardless of the $P_L$ value,
have negative differences, typically $\sim-1.0$.  This is in striking contrast
to two thirds of the pRRd stars and 11 of the aRRd stars that have positive
$\Delta  \phi_{\rm 21}$ values.  Again this argues for the dissimilarity of the
pRRd and very short-period cRRd stars.

\section{SUMMARY}

% \subsection{\rm EPIC\,216764000}
% see C7p250 for Pawel summary, repeated here

Four `peculiar' RRd (pRRd) stars  observed by the {\it Kepler} space telescope
during NASA's {\it K2} Mission have been identified.   All four stars are
located in the direction of the Galactic Bulge. Three of the stars are
relatively faint (with $<$$V$$>$=18--20\,mag) and were found to be among the
pRRd stars discovered by P17. The fourth star, EPIC\,216764000 (=\,V1125\,Sgr),
which is several magnitudes brighter than the other three stars, is a newly
discovered pRRd star. Detailed frequency analyses of the high-precision
long-cadence {\it K2} photometry for all four stars were performed and the
results used to study cycle-to-cycle light variations.

% Dianna Ross conclusions:
% For EPIC\,216764000 the dominant mode appears to be modulated: the
% Fourier spectrum can be fit with triplets, corresponding to an
% amplitude modulation period of $\sim$60\,d. Unfortunately this
% conclusion is tentative given that the modulation multiplets are not
% resolved.  In the time domain the residuals after prewhitening with
% these two frequencies and their harmonics are non-Gaussian,
% suggesting incomplete prewhitening for both modes which have been
% traced to possibly-periodic amplitude and frequency (phase)
% modulations. No additional frequencies are seen in the star

Both pulsation modes of EPIC\,216764000 are observed to be non-stationary.  Variations of the
amplitudes and phases of the modes appear to be irregular and, at least within
the length of the {\it K2} data, are not repetitive. Besides the two modes, no
additional frequencies are detected in the star, in particular, the 0.61 mode
seen in all well-observed cRRd stars (Moskalik et al. 2018b;  Nemec, Moskalik
et al., in preparation) is not present. Owing to the relative brightness of
EPIC\,216764000 we urge that high-resolution spectra be taken to determine its
chemical abundances, in particular, [Fe/H].

% Campaign 11 conclusions
For the three fainter pRRd stars our analysis of the {\it K2} photometry  and
the \texttt{OGLE} photometry through 2017 has confirmed the dominant periods
and Fourier parameters derived by Soszy\'nski et al. (2017b).  Also confirmed
are the periods and amplitudes (both dominant and additional) derived by P17.
The derived amplitudes and phases were used to calculate Fourier decomposition
amplitude and phase parameters.

% Comparison with cRRd and aRRd stars,
While P17 chose to compare the pRRd stars with single-mode RRc and RRab stars
we have compared their periods, amplitudes and phases with those for classical
and anomalous double-mode RR~Lyrae stars.  By viewing the Petersen diagram from
the perspective of a $P_L$-$P_S$ graph we show that all three RRd star types
obey roughly linear correlations between the two periods, and that the slope of
the correlation for the pRRd stars is shallower than that for the cRRd and aRRd
stars.  In fact, the cRRd $P_L$-$P_S$ relation is better fit with a quadratic
rather than a line.  Various graphs are presented in support of the conclusion
that the pRRd stars are not simply short-period cRRd stars.  Finally, the
galactic field stars V1127\,Aql and AH\,Cam are identified as other probable
members of the class of pRRd stars.

\section*{Acknowledgements} Funding for the {\it Kepler}/{\it K2} Mission was
provided by the NASA Science Mission directorate.   We thank Emese Plachy and
Laszlo Moln\'ar for kindly supplying \texttt{EAP}-pipeline photometry for
EPIC\,216764000.  We also thank Emese, Laszlo, Robert Szab\'o and Katrien
Kolenberg for discussions of RRd stars observed during the {\it Kepler}/{\it
K2} Missions.   JN acknowledges assistance from Dr. Amanda F. Linnell Nemec,
and thanks the Camosun College Faculty Association for supporting his travel to
various {\it Kepler} conferences.  Interesting discussions were had with
Rados\l{}aw Poleski at the {\it Kepler}/{\it K2} SciConIV meeting at the
NASA-Ames Research Center in Mountain View, California.  Finally we thank Prof.
Johanna Jurcsik for her careful review of the paper and for her interesting
suggestions for future work.

% This research has made use of "Aladin sky atlas" developed at CDS, Strasbourg
% Observatory, France.

Data availability: the data underlying this article are available
from the \texttt{MAST} website, at https://archive.stsci.edu/k2/,
and all the datasets were derived from sources in the public domain.


% \bibliographystyle{mnras}

\end{document}